\documentclass[twocolumn,twocolappendix]{aastex631}

\def \zha {$z_{\rm H\alpha}$}
\def \mkms {\rm ~km~s^{-1}}

\accepted{June 20, 2022}

\shorttitle{The Kinematics of Cold, Extra-galactic Fountain Flows}
\shortauthors{Rubin et al.}
\graphicspath{{./}{figures/}}

\begin{document}

\title{On the Kinematics of Cold, Metal-enriched Galactic Fountain Flows in Nearby Star-forming Galaxies}

\correspondingauthor{Kate H. R. Rubin}
\email{krubin@sdsu.edu}

\author[0000-0001-6248-1864]{Kate H. R. Rubin}
\affiliation{Department of Astronomy, San Diego State University, San Diego, CA 92182 USA}
\affiliation{Center for Astrophysics and Space Sciences, University of California, San Diego, La Jolla, CA 92093, USA}

\author{Christian Juarez}
\affiliation{Department of Astronomy, San Diego State University, San Diego, CA 92182 USA}

\author[0000-0001-5810-5225]{Kathy L.~Cooksey}
\affiliation{Department of Physics and Astronomy, University of Hawai`i at Hilo, Hilo, HI 96720, USA}

\author[0000-0002-0355-0134]{Jessica K. Werk}
\affiliation{Department of Astronomy, University of Washington, Seattle, WA 98195-1580, USA}

\author[0000-0002-7738-6875]{J. Xavier Prochaska}
\affiliation{Department of Astronomy \& Astrophysics, University of California, 1156 High Street, Santa Cruz, CA 95064, USA}
\affiliation{University of California Observatories, Lick Observatory, 1156 High Street, Santa Cruz, CA 95064, USA}
\affiliation{Kavli Institute for the Physics and Mathematics of the Universe (WIP), 5-1-5 Kashiwanoha, Kashiwa, 277-8583, Japan}

\author[0000-0002-7893-1054]{John M. O'Meara}
\affiliation{W. M. Keck Observatory, 65-1120 Mamalahoa Highway, Kamuela, HI 96743, USA}

\author[0000-0002-1979-2197]{Joseph N. Burchett}
\affiliation{New Mexico State University, Department of Astronomy, Las Cruces, NM 88001, USA}

\author[0000-0001-9719-4080]{Ryan J. Rickards Vaught}
\affiliation{Center for Astrophysics and Space Sciences, University of California, San Diego, La Jolla, CA 92093, USA}

\author{Varsha P. Kulkarni}
\affiliation{Department of Physics and Astronomy, University of South Carolina, Columbia, SC 29208, USA}

\author{Lorrie A. Straka}
\affiliation{Leiden Observatory, Leiden University, PO Box 9513, NL-2300 RA Leiden, The Netherlands}

\begin{abstract}

We use medium-resolution Keck/Echellette Spectrograph and Imager spectroscopy of bright quasars to study cool gas traced by \ion{Ca}{2} $\lambda\lambda3934,3969$ and \ion{Na}{1} $\lambda\lambda5891,5897$ absorption in the interstellar/circumgalactic media of 21 foreground star-forming galaxies at redshifts $0.03<z<0.20$ with stellar masses $7.4\le\log M_{*}/M_{\odot}\le 10.6$.  The quasar-galaxy pairs were drawn from a unique sample of Sloan Digital Sky Survey quasar spectra with intervening nebular emission, and thus have exceptionally close impact parameters ($R_{\perp}<13$\,kpc).  The strength of this line emission implies that the galaxies' star formation rates (SFRs) span a broad range, with several lying well above the star-forming sequence.  We use Voigt profile modeling to derive column densities and component velocities for each absorber, finding that
column densities $N(\mbox{\ion{Ca}{2}})>10^{12.5}~\rm cm^{-2}$ ($N(\mbox{\ion{Na}{1}})>10^{12.0}~\rm cm^{-2}$) occur with an incidence $f_{\rm C}(\mbox{\ion{Ca}{2}})=0.63^{+0.10}_{-0.11}$ ($f_{\rm C}(\mbox{\ion{Na}{1}})=0.57^{+0.10}_{-0.11}$).  We find no evidence for a dependence of $f_{\rm C}$ or the rest-frame equivalent widths $W_r$(\ion{Ca}{2} K) or $W_r$(\ion{Na}{1} 5891) on $R_{\perp}$ or $M_*$.  Instead, $W_r$(\ion{Ca}{2} K) is correlated with local SFR at $>3\sigma$ significance, suggesting that \ion{Ca}{2} traces star formation-driven outflows.
While most of the absorbers have velocities within $\pm50\mkms$ of the host redshift, their velocity widths (characterized by $\Delta v_{90}$) are universally   30--$177\mkms$ larger than that implied by tilted-ring modeling of the velocities of interstellar material.  These kinematics must trace galactic fountain flows and demonstrate that they persist at  $R_{\perp}>5$\,kpc.  Finally, we assess the relationship between dust reddening and $W_r$(\ion{Ca}{2} K) ($W_r$(\ion{Na}{1} 5891)), finding that 33\% (24\%) of the absorbers are inconsistent with the best-fit Milky Way $E(B-V)$-$W_r$ relations  at $>3\sigma$ significance.

\end{abstract}

\keywords{}

\section{Introduction} \label{sec:intro}

The disk-halo interface is host to diverse baryonic processes that regulate the buildup of stellar mass in star-forming galaxies \citep{ShapiroField1976,Bregman1980,Norman1989,deAvillez2000}.  Supernova activity in galactic disks generate wind-blown bubbles in the interstellar medium (ISM; \citealt{MacLowMcCray1988,TenorioTagle1988,Korpi1999}), some of which are sufficiently powerful to evacuate thermalized supernova ejecta and entrain cold material through this interface into the circumgalactic medium \citep[CGM; e.g.,][]{TomisakaIkeuchi1986,Veilleux1995,Cooper2008,Fielding2018}.  At the same time, the material required to feed ongoing star formation must likewise pass through this region, originating either beyond or in distant regions of the galactic halo, or condensing from previously ejected stellar and interstellar material \citep[e.g.,][]{LehnerHowk2011,Marasco2012,KimOstriker2018}.  

In the Milky Way, disk-halo material is observed in emission across a broad range of phases, including hot, diffuse gas traced by X-ray emission \citep{Egger1995,Kerp1999,KuntzSnowden2000}, a warm, denser phase traced by H$\alpha$ emission \citep[e.g.,][]{WeinerWilliams1996,Bland-Hawthorn1998,Haffner2003}, the cool, neutral material that emits at 21 cm \citep[e.g.,][]{Bajaja1985,WakkervanWoerden1991,Kalberla2005,McClure-Griffiths2009}, and a subdominant cold phase arising in molecular clouds \citep{Gillmon2006,Heyer2015,Rohser2016}.  Gas with temperatures spanning much of this range has likewise been observed in metal-line absorption toward distant QSOs or UV-bright Galactic stars \citep[e.g.,][]{Richter2001a,Richter2001c,Wakker2001,Howk2003,Yao2009,LehnerHowk2011,Werk2019}.  Indeed, detection of absorption due to the \ion{Ca}{2} $\lambda \lambda 3934, 3969$ and \ion{Na}{1} $\lambda \lambda 5891, 5897$ transitions toward stars in the Galactic disk and halo provided the first evidence for the existence of the ISM \citep[e.g.,][]{Hartmann1904,Hobbs1969,Hobbs1974}, and for the presence of interstellar material above the Galactic plane \citep{MunchZirin1961}.  
The propensity of these transitions to arise in warm (temperature $T < 10,000$ K) and cold ($T<1000$ K) gas phases, respectively, make them effective tracers of the neutral ISM \citep{Crawford1992,Welty1996,Richter2011,Puspitarini2012}.

Over the past several decades, detailed study of these absorption transitions have, e.g., provided important constraints on the distances and temperatures of massive \ion{H}{1} cloud complexes \citep{Wakker2001,BenBekhti2008,BenBekhti2012}; revealed the small-scale structure of neutral material in the Milky Way halo \citep{Smoker2015,Bish2019}; and placed novel constraints on the physics and composition of interstellar dust \citep[e.g.,][]{Phillips1984,Sembach1994,Welty1996,Murga2015}.  The comprehensive analysis of \ion{Ca}{2} and \ion{Na}{1} transitions in several hundred QSO spectra by \citet{BenBekhti2012} demonstrated that this absorption has comparable Milky Way sky coverage to that of \ion{H}{1} detected in emission, and that approximately half of the absorber sample have positions and velocities consistent with those of known \ion{H}{1} complexes.



Absorption from \ion{Ca}{2} is also known to trace cool circumgalactic material in the halos of external galaxies.  Several studies have used spectroscopy of background QSO sightlines to identify this transition in association with known foreground systems, reporting detections within projected separations $R_{\perp} \lesssim 30$ kpc \citep[e.g.,][]{BoksenbergSargent1978,Boksenberg1980,Blades1981,Bergeron1987,Zych2007}.  Taking advantage of Sloan Digital Sky Survey (SDSS) spectroscopy of more than 100,000 quasars, \citet{ZhuMenard2013} analyzed the mean \ion{Ca}{2} signal induced as a function of projected separation from nearly one million foreground galaxies, tracing significantly detected absorption from $R_{\perp} \sim 7$ to $200$\,kpc. 
Detections of circumgalactic \ion{Na}{1}, on the other hand, have been rarer: prior to the advent of the SDSS, fewer than ten galaxy-absorber pairs, all within
$R_{\perp} < 15$ kpc, were reported in the literature \citep[e.g.,][]{Bergeron1987,Womble1990,Stocke1991,Richter2011}. The mining of SDSS QSO spectra for individual \ion{Na}{1} systems increased this sample by a modest factor \citep[e.g.,][]{Cherinka2011,York2012,Straka2015}; however, the limited signal-to-noise and spectral resolution of these data are ill-suited to detailed study of either of these transitions in individual QSO sightlines.  Instead, \ion{Na}{1} absorption has long been leveraged to study ISM kinematics in ``down-the-barrel" galaxy spectroscopy, revealing ubiquitous, large-scale outflows in massive, starbursting and active galactic nucleus (AGN)-host systems \citep[e.g.,][]{Heckman2000,Martin2005,Rupke2005,Rupke2017,Veilleux2020,Rupke2021}, as well as in more typical star-forming galaxies with stellar masses $10 < \log M_*/M_{\odot} < 11$ \citep{ChenTremonti2010,Concas2019,RobertsBorsani2020}.

In this work, we use medium-resolution ($\mathcal{R} \approx 8000$) optical spectroscopy of 21 bright quasars confirmed to lie exceptionally close to known foreground systems at redshifts $0.03 < z < 0.20$ to study cold ($T\lesssim10,000$\,K) disk-halo material traced by \ion{Ca}{2} $\lambda \lambda 3934,3969$ and \ion{Na}{1} $\lambda \lambda 5891, 5897$ absorption.
These sightlines were drawn from a unique sample of quasars surveyed by the SDSS, for which unassociated foreground nebular emission lines were identified in their SDSS fiber spectra.
These systems, called Galaxies on Top of Quasars (or GOTOQs), were first discovered by \citet{Noterdaeme2010}, and later studies have since uncovered a sample of 103 such objects \citep{York2012,Straka2013,Straka2015}. 
\citet{Kulkarni2022} recently presented {\it HST}/COS spectroscopy of eight GOTOQs (including five in the present study), confirming that these systems give rise  to damped or subdamped Ly$\alpha$ absorption in all cases.
\citet{Straka2015} performed photometric analysis of the SDSS imaging of the full sample of 103 pairs, constraining galaxy luminosities, stellar masses, and impact parameters ($R_{\perp}$), and used emission-line fluxes measured from the SDSS spectroscopy to assess the galaxies' star formation activity at the location of the fiber.
Here we combine these measurements with our sensitive follow-up optical spectroscopy to explore the incidence and kinematics of \ion{Ca}{2} and \ion{Na}{1} absorption within $R_{\perp} < 13$ kpc of a sample of external galaxies for the first time.  We use our sample to trace the dependence of the absorption strengths of these transitions on the stellar masses ($M_*$) and local star formation rates (SFRs) of the foreground host systems, as well as their relationship to the dust reddening along the sightlines.  The relatively high (echellette) spectral resolution of our dataset, in combination with the uniquely small impact parameters of the QSOs we target, permit novel insights into the ubiquity of galactic fountain flows in the nearby star-forming galaxy population.

We describe our sample selection and echellette spectroscopy in Section~\ref{sec:obs}, and describe salient properties of the foreground host galaxies in our sample as measured by \citet{Straka2015} in Section~\ref{sec:fg_galaxies}.
We detail our methods of measuring foreground galaxy redshifts and absorption-line equivalent widths, column densities, and kinematics in Section~\ref{sec:analysis}.  Section~\ref{sec:results} presents our results on the relationship between these absorption-line properties and $R_{\perp}$, dust reddening, foreground galaxy $M_*$, and local SFR.  In Section~\ref{sec:model}, we develop a simple model of the \ion{Ca}{2}- and \ion{Na}{1}-absorbing properties of the Milky Way's ISM and demonstrate that such a model fails to explain the large column densities and kinematic widths we measure.  We discuss the implications of these findings in light of complementary studies of \ion{Ca}{2} and \ion{Na}{1} absorption detected toward background QSO sightlines and in down-the-barrel galaxy spectroscopy in Section~\ref{sec:discussion}.  We adopt a $\Lambda$CDM cosmology with $H_0 = 70~\rm km~s^{-1}~Mpc^{-1}$, $\Omega_{\rm M} = 0.3$, and $\Omega_{\rm \Lambda}=0.7$.  Magnitudes quoted are in the AB system.

\section{Sample Selection and Observations} \label{sec:obs}

Our target quasar sample is drawn from the parent sample of 103 GOTOQs discovered in SDSS spectra by \citet{Noterdaeme2010}, \citet{York2012}, \citet{Straka2013}, and \citet{Straka2015}.  The latter study performed photometric analysis of the SDSS imaging of all QSO-galaxy pairs, measuring galaxy luminosities, impact parameters, and stellar masses.  They also calculated SFRs from the extinction-corrected H$\alpha$ and [\ion{O}{2}] luminosities measured within the SDSS fiber spectroscopy of the background QSOs.  The intervening galaxies in this parent sample span a range of redshifts $0 < z < 0.84$, have impact parameters 0.4\,kpc $< R_{\perp} <$ 12.7 kpc, and span a wide range in stellar mass ($7.3 < \log M_*/M_{\odot} < 11.5$).  

\begin{figure*}
\includegraphics[clip,width=1.0\textwidth,trim={0 1cm 0 1cm}]{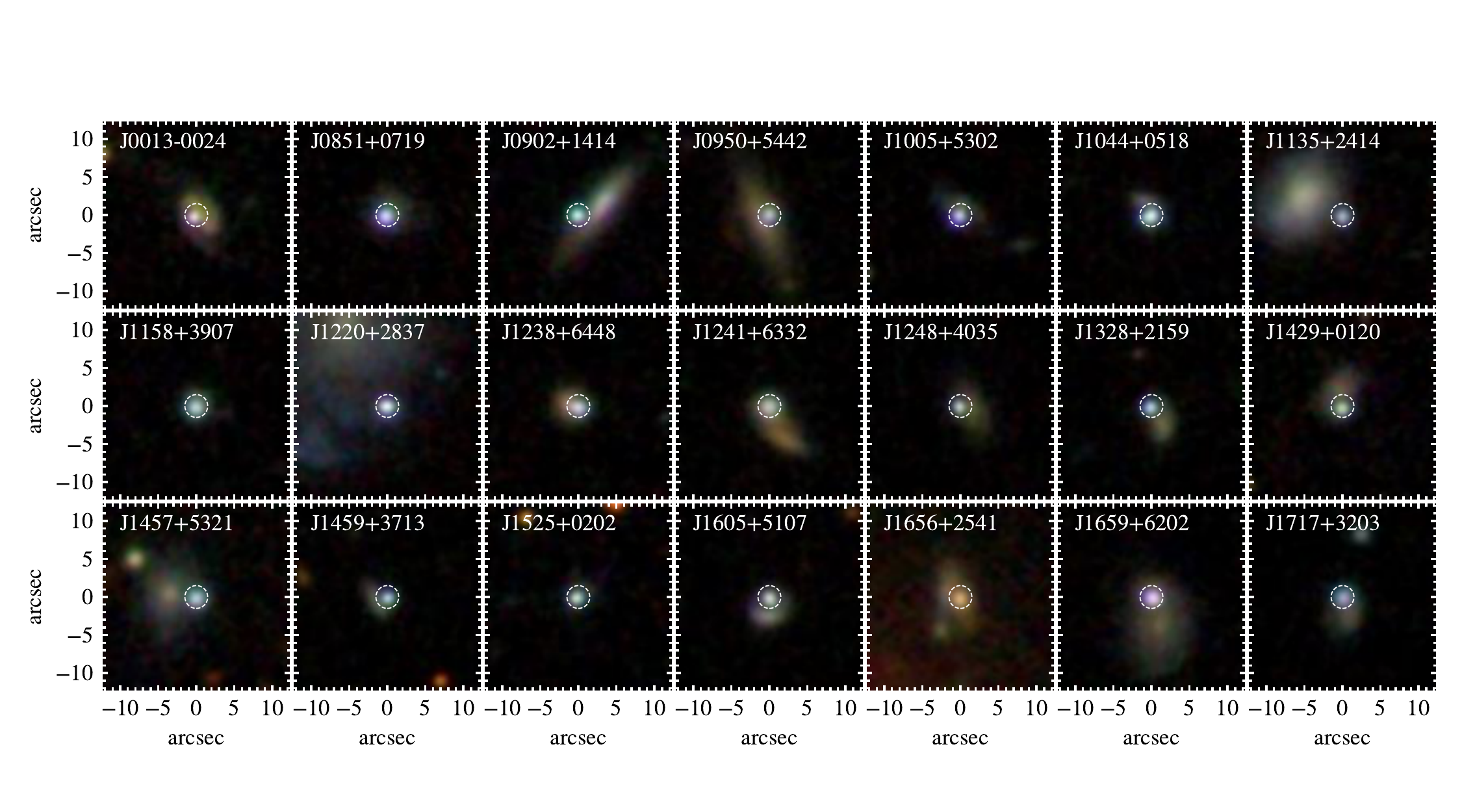}
\caption{SDSS $gri$ color imaging of all GOTOQs for which we have obtained ESI spectroscopy \citep{York2000}.  Each panel is $25\arcsec \times 25\arcsec$.  The images are labeled with the corresponding GOTOQ ID at the upper left.  The dashed white circle indicates the size of the $3\arcsec$ diameter fiber used for the SDSS spectroscopy of each system.  \label{fig:ims}}
\end{figure*}

We used the following criteria to select targets for follow-up echellette-resolution spectroscopy: (1) a continuum-emitting counterpart to the foreground system was identified by \citet{Straka2015}; (2) the foreground galaxy redshift must be such that the \ion{Na}{1} D doublet falls outside of  spectral regions with significant atmospheric absorption (at observed wavelengths $\lambda_{\rm obs} = 6850$--6950 \AA\ and 7580--7710 \AA); and (3) the quasar must be sufficiently bright to yield a $2\sigma$ rest equivalent width ($W_r$) detection limit of $\approx 0.02$ \AA\ at $\lambda_{\rm obs} = 6000$--7500 \AA\ in an exposure time of $\leq 1$\,hour.  This latter constraint corresponds to an $r$-band magnitude limit of $m_r \lesssim 19.1$ for the background quasar.  Approximately 36 GOTOQs in the \citet{Straka2015} parent sample satisfy all of these criteria.  We completed follow-up spectroscopy of 21 of these targets. Table~\ref{tab.gotoqs} lists their coordinates, as well as the QSO redshifts, impact parameters, and other properties of the foreground galaxies as reported in \citet{Straka2015}.  
SDSS color images of each system are included in Figure~\ref{fig:ims}.

Our observations were carried out using the Echellette Spectrograph and Imager \citep[ESI;][]{Sheinis2002} on the Keck II Telescope on 2017 March 6 UT and 2017 June 22-23 UT.  Seeing conditions ranged between FWHM $\sim 0.4\arcsec$--$0.8\arcsec$ over the course of the program.  We used the $0.5\arcsec$ wide longslit with ESI, which affords an FWHM resolution of $\mathcal{R} \approx 8000$ ($37.3~\rm km~s^{-1}$), a spectral dispersion of $10~\rm km~s^{-1}$, and a typical wavelength coverage of $\rm 3990-\!10130\,\AA$.
We exposed for between 20 and 70 minutes total per object, dividing each observation into two to four individual exposures.  

The data were reduced using the XIDL ESIRedux data reduction pipeline\footnote{\url{https://www2.keck.hawaii.edu/inst/esi/ESIRedux/}}.  The pipeline includes bias subtraction, flat-fielding, wavelength calibration, the tracing of order curvature, object identification, sky subtraction, cosmic ray rejection, and relative flux calibration.  
We also used it to apply a vacuum and  heliocentric correction to each spectrum.

\begin{figure*}
\includegraphics[clip,width=1.0\textwidth]{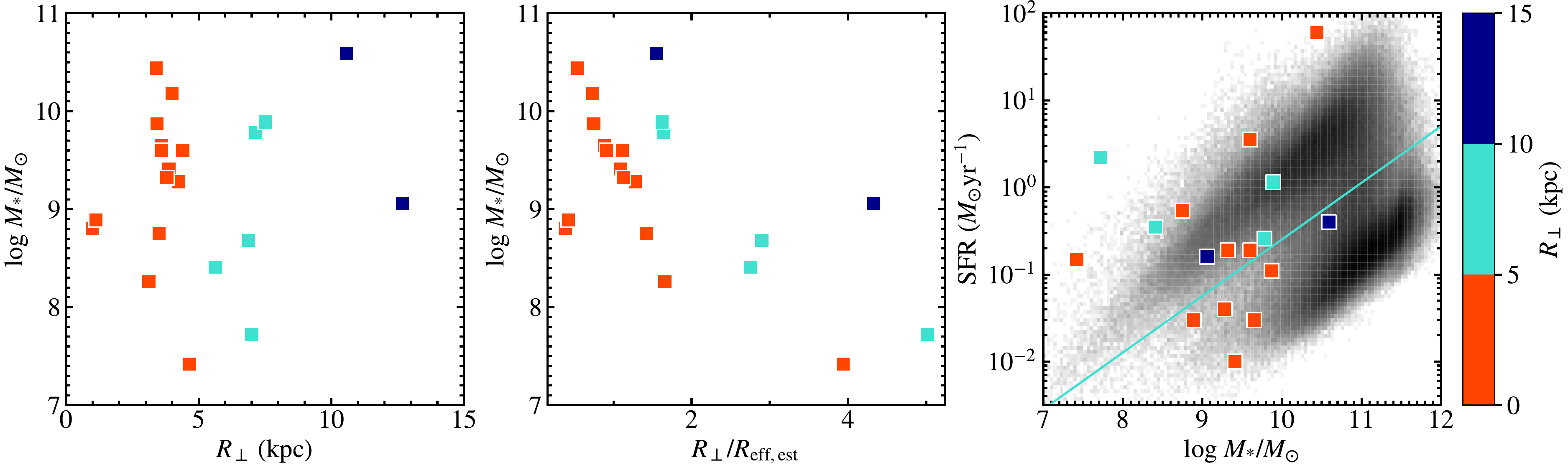}
\caption{ {\it Left:} Distribution of $\log M_*/M_{\odot}$ vs.\ $R_{\perp}$ for our sample.  Points are color-coded by the $R_{\perp}$ value for each system across all panels, as indicated by the color bar at right. {\it Middle:}  Distribution of $\log M_*/M_{\odot}$ vs.\ $R_{\perp}/R_{\rm eff,est}$ for our foreground GOTOQ sample.  {\it Right:} Distribution of $\rm SFR_{local}$ vs.\ $\log M_*/M_{\odot}$ for our foreground GOTOQ sample.  The SFR values shown here should be considered lower limits on the total SFR of each system due to fiber losses.  The grayscale histogram shows the distribution of total SFR vs.\ $M_*$ for all galaxies included in the MPA-JHU catalog of these values for SDSS DR7 \citep{Brinchmann2004}.  The turquoise line shows a linear fit to the minimum in the galaxy distribution between star-forming and quiescent systems by \citet{Moustakas2013} for $z=0$.  
\label{fig:rperp_mstar_sfr}}
\end{figure*}

\begin{deluxetable*}{lccccccccc}
\tablewidth{0pt}
\tablecaption{Observed GOTOQ Sample\label{tab.gotoqs}}
\tablehead{
\colhead{Sight Line} & \colhead{R.A.} & \colhead{Decl.} & \colhead{$z_{\rm QSO}$\tablenotemark{a}} & \colhead{$z_{\rm H\alpha}$\tablenotemark{b}} & \colhead{$R_{\perp}$\tablenotemark{a}} & \colhead{$m_r$(QSO)\tablenotemark{a}} & \colhead{$\log M_*/M_{\odot}$\tablenotemark{a}} & \colhead{SFR(H$\alpha$)\tablenotemark{a}} & \colhead{$E(B-V)_{(g-i)}$\tablenotemark{a}}
\\
 & \colhead{(J2000)} & \colhead{(J2000)} & & & \colhead{\hfill(kpc)} & \colhead{(mag)} & & \colhead{($M_{\odot}~\rm yr^{-1}$)} & \colhead{(mag)}
}
\startdata
GOTOQJ0013--0024  & 00:13:42.45 & --00:24:12.60 & 1.641 & 0.15537 & 3.40 & 18.58 & 10.4 & 60.30 & 0.24 \\
GOTOQJ0851+0719  & 08:51:13.74 & +07:19:59.80 & 1.650 & 0.13010 & 5.63 & 17.97 & 8.4 & 0.35 & $-0.06$ \\
GOTOQJ0902+1414* & 09:02:50.47 & +14:14:08.29 & 0.980 & 0.05044 & 3.59 & 18.43 & 9.7 & 0.03 & 0.05 \\
GOTOQJ0950+5442* & 09:50:13.74 & +54:42:54.65 & 0.700 & 0.04586 & 0.98 & 18.33 & 8.8 & \nodata & $-0.09$ \\
GOTOQJ1005+5302  & 10:05:14.21 & +53:02:40.04 & 0.560 & 0.13547 & 3.60 & 18.79 & 9.6 & 0.19 & $-0.15$ \\
GOTOQJ1044+0518* & 10:44:30.26 & +05:18:57.32 & 0.900 & 0.10781 & 3.51 & 17.77 & 8.8 & 0.54 & 0.21 \\
GOTOQJ1135+2414* & 11:35:55.66 & +24:14:38.10 & 1.450 & 0.03426 & 3.88 & 19.22 & 9.4 & 0.01 & $-0.10$ \\
GOTOQJ1158+3907* & 11:58:22.85 & +39:07:12.96 & 1.160 & 0.18337 & 4.66 & 18.04 & 7.4 & 0.15 & 0.07 \\
GOTOQJ1220+2837  & 12:20:37.23 & +28:37:52.03 & 2.200 & 0.02762 & 6.88 & 17.91 & 8.7 & \nodata & 0.04 \\
GOTOQJ1238+6448  & 12:38:46.68 & +64:48:36.60 & 1.560 & 0.11859 & 7.00 & 17.93 & 7.7 & 2.21 & 0.10 \\
GOTOQJ1241+6332  & 12:41:57.55 & +63:32:41.63 & 2.620 & 0.14270 & 10.57 & 17.96 & 10.6 & 0.40 & 0.22 \\
GOTOQJ1248+4035  & 12:48:14.43 & +40:35:35.13 & 2.110 & 0.15132 & 4.00 & 19.11 & 10.2 & \nodata & 0.12 \\
GOTOQJ1328+2159  & 13:28:24.33 & +21:59:19.66 & 0.330 & 0.13524 & 12.68 & 18.96 & 9.1 & 0.16 & $-0.16$ \\
GOTOQJ1429+0120  & 14:29:17.69 & +01:20:58.93 & 1.130 & 0.08395 & 3.43 & 18.70 & 9.9 & 0.11 & 0.13 \\
GOTOQJ1457+5321* & 14:57:19.00 & +53:21:59.27 & 1.200 & 0.06594 & 4.24 & 18.16 & 9.3 & 0.04 & 0.00 \\
GOTOQJ1459+3713  & 14:59:38.50 & +37:13:14.70 & 1.220 & 0.14866 & 4.40 & 19.02 & 9.6 & 3.55 & $-0.04$ \\
GOTOQJ1525+0202* & 15:25:14.08 & +02:02:54.68 & 1.220 & 0.09019 & 3.12 & 18.73 & 8.3 & \nodata & 0.14 \\
GOTOQJ1605+5107  & 16:05:21.26 & +51:07:40.95 & 1.230 & 0.09899 & 3.80 & 18.58 & 9.3 & 0.19 & 0.19 \\
GOTOQJ1656+2541  & 16:56:43.35 & +25:41:36.80 & 0.243 & 0.03451 & 1.13 & 18.16 & 8.9 & 0.03 & 0.42 \\
GOTOQJ1659+6202  & 16:59:58.94 & +62:02:18.14 & 0.230 & 0.11026 & 7.15 & 17.80 & 9.8 & 0.26 & $-0.03$ \\
GOTOQJ1717+3203  & 17:17:04.14 & +32:03:20.93 & 0.660 & 0.20016 & 7.51 & 18.68 & 9.9 & 1.15 & 0.00 \\
\enddata
\tablenotetext{a}{These quantities are drawn from the analysis of \citet{Straka2015}.  Values of $\log M_*/M_{\odot}$ and SFR(H$\alpha$) were calculated for the foreground galaxy.   As discussed in Section~\ref{sec:fg_galaxies}, the latter estimates should be considered lower limits due to the likelihood of fiber losses, and are referred to as $\rm SFR_{\rm local}$ throughout the text.  Values of $E(B-V)_{(g-i)}$ refer to the background QSO and are estimated by comparing each QSO's $(g-i)$ color to the median $(g-i)$ color for QSOs at the same redshift as reported in \citet{Schneider2007}.}
\tablenotetext{b}{This is the foreground galaxy redshift calculated as described in Section~\ref{subsec:redshifts}.}
\tablenotetext{*}{For sight lines marked with an asterisk, we use SDSS spectra rather than ESI spectra to determine a precise emission-line redshift. }
\end{deluxetable*}

\section{Foreground Galaxy Properties} \label{sec:fg_galaxies}

For this analysis, we draw on stellar mass estimates reported by \citet{Straka2015} for the parent GOTOQ sample.  Stellar masses were determined via spectral energy distribution (SED) model fits to photometry of the host galaxies measured in the five SDSS passbands with the photometric redshift code \texttt{HYPERZ} \citep{Bolzonella2000,Straka2015}.  
The left panel of Figure~\ref{fig:rperp_mstar_sfr} shows the $\log M_*/M_{\odot}$ distribution of our foreground host sample vs.\ $R_{\perp}$.  
These systems span 
an overall wide range of stellar masses ($7.4 \le \log M_*/M_{\odot} \le 10.6$), with a median $\log M_*/M_{\odot} = 9.3$.  Our sightlines sample this parameter space relatively thoroughly within $R_{\perp} < 9$ kpc; however, we caution that our constraints beyond $R_{\perp} > 10$ kpc are sparse.

Under the assumption that the absorption strength of our transitions of interest at a given $R_{\perp}$ may depend on the relative extent of a galaxy's stellar component, we use the observed relation between $M_*$ and effective radius ($R_{\rm eff}$) for late-type galaxies to estimate $R_{\rm eff}$ for each host.  We use the best-fit $R_{\rm eff}$-$M_*$ relation estimated by \citet{vanderWel2014} for systems having $0 < z < 0.5$:
\begin{equation}
    R_{\rm eff, est} = 10^{0.86} \left (\frac{M_*}{5 \times 10^{10} M_{\odot}} \right )^{0.24}{\rm kpc}.
\end{equation}
These values fall in the range $1.2~\mathrm{kpc} \le R_{\rm eff,est} \le 6.8~\mathrm{kpc}$ for our sample.
These authors also assess the intrinsic scatter in this relation, estimating $\sigma(\log R_{\rm eff}) = 0.16$.  The true size of any given galaxy in our sample may therefore differ from this best-fit estimated size by a few kiloparsecs; however, we note that estimates of galaxy halo virial radii (often used in CGM studies in the same way we will use $R_{\rm eff, est}$ below) are typically subject to a greater degree of uncertainty.  We normalize the $R_{\perp}$ value for each system by its $R_{\rm eff, est}$ estimate and compare this quantity to $\log M_*/M_{\odot}$ in the middle panel of Figure~\ref{fig:rperp_mstar_sfr}. Due to the correlation between $M_*$ and $R_{\rm eff,est}$, the sightlines with $R_{\perp}/R_{\rm eff,est} > 2$ tend to probe the lower-$M_*$ hosts in our sample (i.e., those with $\log M_*/M_{\odot} \lesssim 9$).

We likewise make use of the SFRs estimated by \citet{Straka2015} for these systems
from the extinction-corrected H$\alpha$ luminosities measured in the SDSS fiber spectra. 
Extinction corrections were determined from the ratio of H$\alpha$ to H$\beta$ line luminosities and adopted an SMC extinction curve \citep{Straka2015}.  The \citet{Kennicutt1998} empirical calibration was then applied to the intrinsic H$\alpha$ luminosities.  As noted by \citet{Straka2015}, because the SDSS fibers used to observe these galaxies were typically placed such that a significant fraction of their H$\alpha$ line emission was lost (see Figure~\ref{fig:ims}), these SFRs should be considered lower limits on the total star formation activity of the hosts.  
Moreover, the fraction of line emission missed by the fiber is likely larger for systems with larger impact parameters.  We explore this effect in Appendix~\ref{sec:appendix_SFRfrac}, modeling the distribution of star formation in each foreground system as an exponential disk with a scale radius consistent with $R_{\rm eff,est}$.  This simple analysis implies that the SDSS fibers capture $\gtrsim10\%$ of the H$\alpha$ emitted by the majority of the galaxies probed within $R_{\perp} < 5$ kpc, but may miss $\gtrsim 90-99\%$ of the H$\alpha$ emission from systems at larger spatial offsets.  
The measured H$\alpha$ luminosities and SFRs instead provide accurate assessments of the star formation activity close to the absorbing material detected along our QSO sightlines (i.e., the ``local" SFR). For this reason, we refer to this quantity as $\rm SFR_{local}$ below.

The distribution of $\rm SFR_{local}$ and $\log M_*/M_{\odot}$ values for our foreground galaxy sample is shown in the rightmost panel of Figure~\ref{fig:rperp_mstar_sfr} with colored points.  The grayscale histogram shows the distribution of {\it total} SFR and $\log M_*/M_{\odot}$ for the SDSS DR7 galaxy population \citep{Brinchmann2004}.  The turquoise curve shows a linear fit to the minimum in the bimodal galaxy distribution estimated by \citet{Moustakas2013} and extrapolated to $z=0$.  Several of our foreground galaxies lie below this line, in the parameter space primarily occupied by non-star-forming, early-type systems.  
This is likely because we have not measured their total, integrated SFRs (as described above).  The modeling we perform in Appendix~\ref{sec:appendix_SFRfrac} implies that all of our systems likely have total SFRs $> 0.1~M_{\odot}~\rm yr^{-1}$, and that those systems with $R_{\perp} > 5$ kpc may have total SFRs $\gtrsim 10~M_{\odot}~\rm yr^{-1}$.  The latter galaxies should therefore be considered starbursting systems.  
The location of our sample in this parameter space may likewise be affected by overestimation of the galaxy stellar masses due to systematics associated with SED modeling of the shallow SDSS photometry.  The $1\sigma$ uncertainty intervals for the $\log M_*/M_{\odot}$ values reported by \citet{Straka2015} for our sample have a mean of 0.54 dex, and range up to 2.0 dex.

\section{Line Profile Analysis} \label{sec:analysis}

\subsection{Foreground Galaxy Redshifts} \label{subsec:redshifts}

Because we are interested in the detailed kinematic structure of absorption detected along our target sight lines, and because \citet{Straka2015} reported redshifts with only four significant figures, we draw on our ESI spectra to measure more precise redshifts for our GOTOQ sample.  We inspected each ESI spectrum for the presence of narrow emission features at the observed wavelengths of H$\alpha$ and [\ion{O}{3}] $\lambda 5008$ for the associated foreground galaxy.  We identified both transitions in eight sight lines, and identified only H$\alpha$ in an additional six systems.  The remaining seven sight lines (indicated with asterisks in Table~\ref{tab.gotoqs}) lack narrow emission features at the expected locations of H$\alpha$ and [\ion{O}{3}]; this is most likely because the ESI slit placement was insufficiently close to the foreground system.  For these sightlines, we use their SDSS DR16 spectra \citep{York2000,Ahumada2020} to re-assess the GOTOQ redshift.  We determine the continuum level of each QSO by fitting a spline function to feature-free spectral regions using 
the \texttt{lt\_continuumfit} GUI, available with the Python package \texttt{linetools}\footnote{\url{https://linetools.readthedocs.io/en/latest/}} \citep{linetools2016}.  
This tool presents the user with an automatically generated continuum spline, fit to a set of knots whose flux levels are determined from the mean flux in a series of spectral ``chunks".  We performed a visual inspection of these knots, adjusting their placement in cases where their location was unduly affected by nearby absorption or emission features.  

We subtracted this continuum level from each spectrum and performed a Gaussian fit to the residual flux in a spectral region within either $\pm300~\rm km~s^{-1}$ (for ESI spectra) or $\pm 600~\rm km~s^{-1}$ (for the SDSS spectra) of the observed wavelength of H$\alpha$. 
We used the Levenberg-Marquardt least-squares fitter available within the \texttt{astropy.modeling} package \citep{astropy2018} to determine the best-fit Gaussian wavelength centroid for this line.
The typical magnitude of the redshift uncertainty implied by the covariance matrix for the fitted parameters is 2--$5~\rm km~s^{-1}$ for the ESI spectra and 4--$15~\rm km~s^{-1}$ for the SDSS spectra.  The fitted redshift values are all within a maximum of $\pm112\mkms$ of those published for the foreground systems by \citet{Straka2015}.
We refer to the redshifts determined via this method as \zha\ in the following text.

\subsection{Absorption-line Profile Characterization and Modeling} \label{subsec:abs_modeling}

We then characterized the absorption strength and kinematics of the \ion{Ca}{2} H \& K and \ion{Na}{1} transitions associated with each GOTOQ.  We used the \texttt{XAbsSysGui}, available with \texttt{linetools}, to perform a visual inspection of these transitions.  In cases in which an absorption feature is clearly evident within $\pm300~\rm km~s^{-1}$ of \zha, we use this GUI to manually select the velocity window to be used for the computation of the $W_r$ of each line.  We also noted the occasional presence of blended absorption features that are unassociated with \zha.  In cases of transitions lacking clear absorption features, velocity windows were set to $\pm 150~\rm km~s^{-1}$ by default, but were adjusted as necessary to exclude unassociated blends.  These windows were used to calculate upper limits on $W_r$.
Spectral regions covering the \ion{Ca}{2} H \& K and \ion{Na}{1} doublet transitions in the rest frame of the corresponding foreground galaxies for five systems in our sample are shown in Figure~\ref{fig:velplot1}.  Similar figures showing the remaining sight lines are included in Appendix~\ref{sec:appendix_spectra}.  Our ESI spectra have signal-to-noise ratios (S/Ns) in the range 20--$34~\rm pix^{-1}$ with a median ${\rm S/N} = 24~\rm pix^{-1}$ within $\lesssim 400~\rm km~s^{-1}$ of the GOTOQ \ion{Na}{1} transitions.  The spectral S/N within 200--$400~\rm km~s^{-1}$ of the \ion{Ca}{2} K transitions ranges between 2--$22~\rm pix^{-1}$, with a median ${\rm S/N} = 10~\rm pix^{-1}$.
 
We used the velocity windows mentioned above to compute the $W_r$ for each \ion{Ca}{2} and \ion{Na}{1} transition.   For those sightlines yielding a significantly detected $W_r$ in at least one transition, we refer to these absorbers as ``systems" in the following.  We also used the apparent optical depth method \citep{SavageSembach1991} to compute the column density of each transition and its uncertainty. For those systems in which both doublet lines are significantly detected and  unblended, we computed the mean of the column densities of both doublet lines, weighted by their respective uncertainties, and report this value as $N_{\rm aod}$.  For those systems in which only the transition with the larger oscillator strength (\ion{Ca}{2} K or \ion{Na}{1} 5891) is significantly detected, we adopt its apparent optical depth column density as the value of $N_{\rm aod}$.
For those sightlines in which the stronger line is not detected, we report $3\sigma$ upper limits on the column density computed from the apparent optical depth method for the stronger transition only.
All velocity limits, $W_r$ and $N_{\rm aod}$ values, and the associated uncertainties ($\sigma_{W_r}$ and $\sigma_{N_{\rm aod}}$) are reported in Tables~\ref{tab.CaIIabsinfo} and \ref{tab.NaIabsinfo}.

\begin{figure*}[!ht]
  \includegraphics[width=1.0\textwidth,clip]{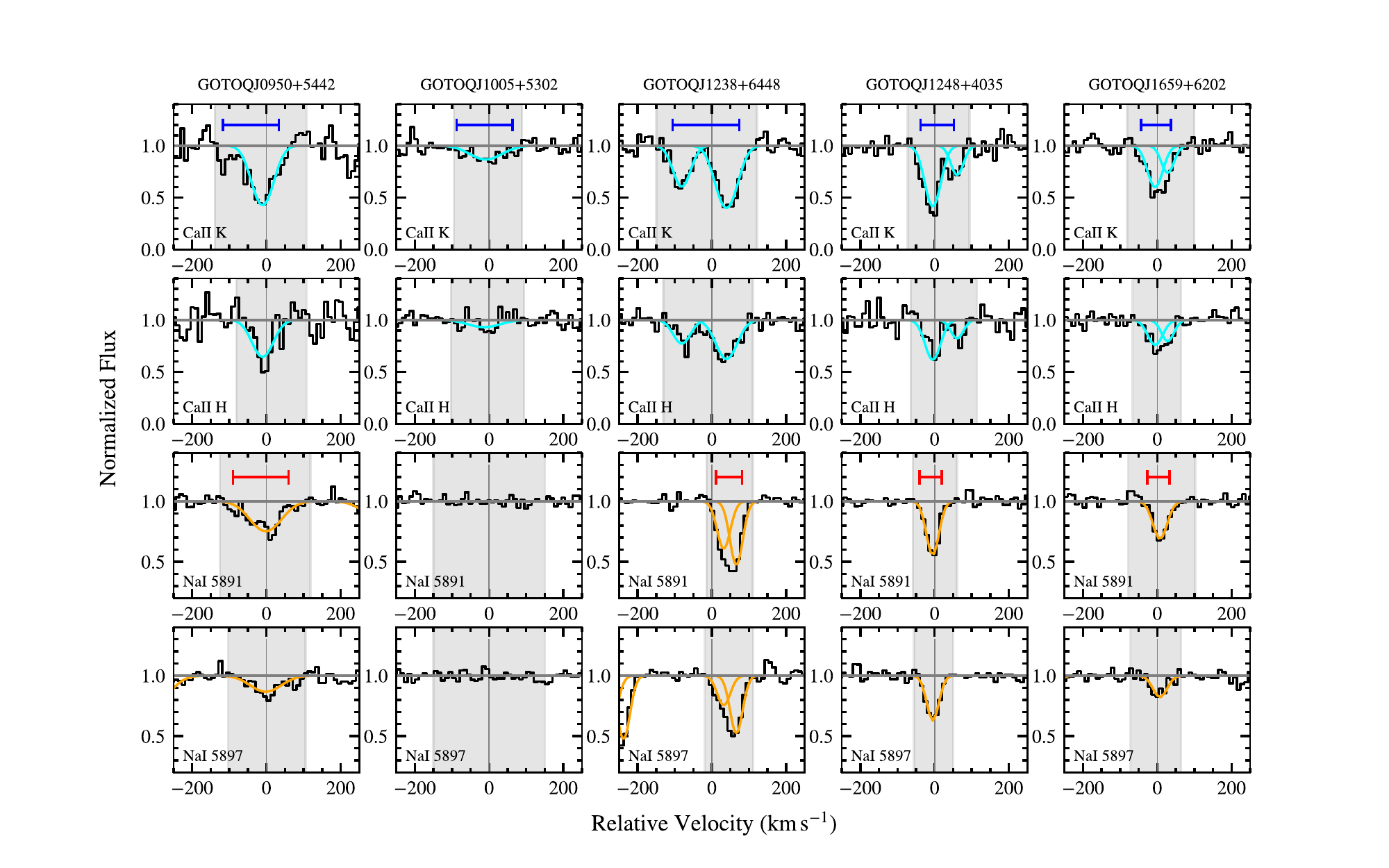}
\caption{Regions of five of our ESI GOTOQ spectra showing the locations of \ion{Ca}{2} H \& K and \ion{Na}{1} $\lambda \lambda 5891, 5897$ transitions associated with the foreground galaxy.  The velocity is defined relative to the GOTOQ redshift estimated from a Gaussian fit to its H$\alpha$ emission as described in Section~\ref{subsec:redshifts} (\zha).  The gray horizontal line indicates the continuum level, and the gray shaded region shows the velocity window selected for computation of $W_r$, $N_{\rm aod}$ and $\Delta v_{90}$.  The blue and red bars show the pixels that contain $>5\%$ of the total apparent optical depth of the line (determined by stepping inward from the profile edges), and the length of these bars corresponds to $\Delta v_{90}$.  Best-fit profile models are shown with cyan (for \ion{Ca}{2}) and orange (for \ion{Na}{1}) curves for systems with significantly detected absorption (see Section~\ref{subsec:abs_modeling} for details).  \label{fig:velplot1}}
\end{figure*}

\citet{Straka2015}
measured unblended $W_r$(\ion{Na}{1} 5891) values using the corresponding SDSS spectra for 16 of our 21 sightlines: 12 of these are upper limits consistent with our constraints; two are detections consistent with our values; and two of the \citet{Straka2015} $W_r$(\ion{Na}{1} 5891) values are larger by $1.4\!-\!2.6\sigma$.  These authors likewise presented measurements of $W_r$(\ion{Ca}{2} K) for each of our sightlines, five of which are upper limits consistent with our constraints.  The remainder are detections that are all larger than our measurements, and 10 of these differ by $>1.0\sigma$. This offset may arise from the use of a larger velocity window by \citet[][although the adopted window is not specified in that work]{Straka2015} and\slash or the inclusion of noise features for some systems.  

We characterize the velocity spread of each significantly detected absorption-line system in a model-independent way using a modified version of the $\Delta v_{90}$ measurement described in \citet{ProchaskaWolfe1997}.  We first smooth the apparent optical depth profile of each system with a boxcar of width $= 37.3\mkms$ and replace any negative apparent optical depth values with a value of zero.  We then step inward from the left and right edges of each profile, summing the apparent optical depths to identify the pixels containing $>5\%$ of the integrated optical depth of the system.  The corresponding value of $\Delta v_{90}$ is the velocity width between these left- (at relative velocity $\delta v_{90,\rm left}$) and rightmost pixels (at relative velocity $\delta v_{90, \rm right}$).  This measurement is listed in Tables~\ref{tab.CaIIabsinfo} and \ref{tab.NaIabsinfo}, and we make use of both these values and our estimates of $\delta v_{90,\rm left}$ and $\delta v_{90,\rm right}$ in the kinematic analyses to follow.

We performed Voigt profile modeling of each significantly detected absorption system using the publicly available \texttt{veeper} Python package\footnote{\url{https://github.com/jnburchett/veeper}}. The \texttt{veeper}, developed by coauthor J.~Burchett, 
determines best-fit values of the column density ($N_{\rm vp}$), Doppler parameter ($b_{\rm D}$), and central velocity relative to \zha\ ($\delta v$) via least-squares minimization.  Parameter space was explored using the iterative \texttt{MPFIT} software, originally written in IDL by C.~Markwardt\footnote{\url{http://cow.physics.wisc.edu/~craigm/idl/idl.html}} and then rewritten in Python by M.~Rivers\footnote{\url{http://cars.uchicago.edu/software}}.  
The user sets initial guesses for each parameter by eye and may then inspect the resulting fit using an interactive GUI.  
The permitted values of $b_{\rm D}$ were limited to the range $1~\mathrm{km~s^{-1}} < b_{\rm D} < 85\mkms$.
Both transitions of each ion were fit simultaneously, and we adopted a Gaussian line spread function with $\sigma = 15.8~\rm km~s^{-1}$ across the full spectral range.  Each absorber was fit twice; once with a single velocity component and, again, with two velocity components initially offset by $\pm10~\rm km~s^{-1}$.  We adopted the best-fit parameters of the two-component fit if it yielded a lower reduced-$\chi^2$ ($\chi^2_r$) value than the one-component fit and reasonable values for the formal 1$\sigma$ parameter uncertainties calculated from the covariance matrix (i.e., $\sigma_{\log N_{\rm vp}} < 0.5$).  While some of these systems may have more than two absorbing structures along the line of sight, we did not attempt more complex profile modeling (e.g., with three or more components) because we generally achieved low $\chi^2$ values with our one- or two-component fits ($\chi^2_r = 0.67\!-\!4.52$), and because our primary findings and conclusions would not be affected by invoking more complex analyses.

There are two absorption-line systems for which both our one-component and two-component \texttt{veeper} fitting fails to yield useful parameter constraints (i.e., $\sigma_{\log N_{\rm vp}} > 1$ or $\sigma_{b_{\rm D}} > 50\mkms$): the \ion{Ca}{2} absorber toward GOTOQJ1328+2159, and the \ion{Na}{1} absorber toward GOTOQJ1429+0120.  We posit that this is due to noise features in these profiles that cause the two doublet lines to exhibit unphysical doublet ratios.  In these cases, we fix the value of the total column density to $N_{\rm aod}$ and perform a one-component \texttt{veeper} fit allowing only the $b_{\rm D}$ and $\delta v$ parameters to vary.  We also note that the 1$\sigma$ parameter uncertainties calculated from the covariance matrix for each absorber fit are formally allowed to overlap regions of parameter space that are excluded from exploration during the fitting process.  This results in values of $\sigma_{b_{\rm D}}>b_{\rm D}$ for a few of the weaker components in our two-component fits, implying 1$\sigma$ confidence intervals that extend to negative values.  The Doppler parameters are thus not well-constrained in these cases; however, the corresponding uncertainties on $\log N_{\rm vp}$ and $\delta v$ should reflect the distribution of each parameter value that corresponds to a $\Delta \chi^2 = 1$ if all other parameters are allowed to vary to keep the $\chi^2$ as low as possible (i.e., they are ``marginalized" uncertainty intervals).

Best-fit Voigt profile models for each securely detected absorber in our sample are shown in Figure~\ref{fig:velplot1} and in Appendix~\ref{sec:appendix_spectra}.  The resulting best-fit values of each model parameter, along with their uncertainties, are listed in Tables~\ref{tab.CaIIabsinfo} and \ref{tab.NaIabsinfo}.  The two systems for which we adopt a fixed column density in our profile fitting are indicated with an asterisk in the $\log N_{\rm vp}$ table columns.  We will primarily use our $N_{\rm vp}$ values where available in the analysis to follow.  We note that while the values of $\log N_{\rm aod}$ and the total $\log N_{\rm vp}$ (summed over all components) are universally within $\pm0.2$ dex for all of our \ion{Ca}{2} absorbers and for the vast majority of our \ion{Na}{1} systems, there are three sightlines for which the total $\log N_{\rm vp}$(\ion{Na}{1}) exceeds $\log N_{\rm aod}$(\ion{Na}{1}) by 0.35--0.66 dex (J1238+6448, J1248+4035, and J1717+3203).  As these absorbers are among the strongest systems in our sample, these offsets are likely due to saturation effects.


\begin{deluxetable*}{lccccccccc}
\tablewidth{700pt}
\tabletypesize{\scriptsize}
\tablecaption{\ion{Ca}{2} Absorption-line Equivalent Widths, Kinematics, and Best-fit Voigt Profile Model Parameters\label{tab.CaIIabsinfo}}
\tablehead{
\colhead{Sight Line} & \colhead{$R_{\perp}$} & \colhead{$W_r$(\ion{Ca}{2} K)\tablenotemark{a}} & \colhead{Velocity Limits} & \colhead{$\log N_{\rm aod}$(\ion{Ca}{2})\tablenotemark{a}} & \colhead{$\Delta v_{90}$(\ion{Ca}{2} K)} & \colhead{$\log N_{\rm vp}$(\ion{Ca}{2})} & \colhead{$b_{\rm D}$(\ion{Ca}{2})} & \colhead{$\delta v$(\ion{Ca}{2})} & \colhead{$\chi_r^{2}$(\ion{Ca}{2})} \\
\colhead{} & \colhead{\hfill(kpc)} & \colhead{(\AA)} & \colhead{($\rm km~s^{-1}$)} & \colhead{($\rm cm^{-2}$)} & \colhead{($\rm km~s^{-1}$)} & \colhead{($\rm cm^{-2}$)} & \colhead{($\rm km~s^{-1}$)} & \colhead{($\rm km~s^{-1}$)} & \colhead{}
}
\startdata
J0013--0024 & 3.4 & $0.96\pm0.07$ & [$-83,131$] & $13.02\pm0.09$ & $130$ & $13.17\pm0.12$ & \nodata & \nodata & $1.58$ \\
 & & \nodata & \nodata & \nodata & \nodata & $12.71\pm0.22$ & $11.4\pm15.9$ & $-28.5\pm6.2$ & \nodata  \\
 & & \nodata & \nodata & \nodata & \nodata & $12.99\pm0.15$ & $47.3\pm17.1$ & $28.1\pm15.7$ & \nodata  \\
J0851+0719 & 5.6 & $ < 0.12$ & [$-150,150$] & $ < 12.14$ & \nodata & \nodata & \nodata & \nodata & \nodata \\
J0902+1414 & 3.6 & $ < 0.20$ & [$-42,70$] & $ < 12.41$ & \nodata & \nodata & \nodata & \nodata & \nodata \\
J0950+5442 & 1.0 & $0.64\pm0.08$ & [$-139,108$] & $12.93\pm0.05$ & $150$ & $12.96\pm0.05$ & $29.6\pm5.1$ & $-6.0\pm3.1$ & $0.99$ \\
J1005+5302 & 3.6 & $0.21\pm0.03$ & [$-95,88$] & $12.40\pm0.06$ & $150$ & $12.35\pm0.07$ & $55.0\pm11.3$ & $-9.7\pm7.5$ & $0.98$ \\
J1044+0518 & 3.5 & $0.32\pm0.03$ & [$-78,105$] & $12.56\pm0.04$ & $100$ & $12.62\pm0.04$ & \nodata & \nodata & $1.00$ \\
 & & \nodata & \nodata & \nodata & \nodata & $11.68\pm0.20$ & $8.9\pm42.8$ & $-21.3\pm11.6$ & \nodata  \\
 & & \nodata & \nodata & \nodata & \nodata & $12.57\pm0.04$ & $20.0\pm5.1$ & $45.3\pm2.3$ & \nodata  \\
J1135+2414 & 3.9 & $ < 0.24$ & [$-75,94$] & $ < 12.49$ & \nodata & \nodata & \nodata & \nodata & \nodata \\
J1158+3907 & 4.7 & $0.15\pm0.04$ & [$-59,94$] & $12.27\pm0.12$ & $100$ & $12.31\pm0.10$ & $54.0\pm17.3$ & $63.1\pm11.6$ & $1.19$ \\
J1220+2837 & 6.9 & $0.33\pm0.07$ & [$-50,47$] & $12.75\pm0.08$ & $50$ & $12.91\pm0.14$ & $13.5\pm6.6$ & $-4.5\pm3.2$ & $1.33$ \\
J1238+6448 & 7.0 & $0.86\pm0.04$ & [$-150,122$] & $13.11\pm0.02$ & $180$ & $13.15\pm0.02$ & \nodata & \nodata & $0.97$ \\
 & & \nodata & \nodata & \nodata & \nodata & $12.65\pm0.04$ & $22.1\pm4.3$ & $-80.0\pm2.2$ & \nodata  \\
 & & \nodata & \nodata & \nodata & \nodata & $12.98\pm0.02$ & $30.9\pm2.5$ & $40.5\pm1.5$ & \nodata  \\
J1241+6332 & 10.6 & $0.65\pm0.03$ & [$-42,113$] & $13.03\pm0.02$ & $80$ & $13.09\pm0.02$ & $35.8\pm2.5$ & $33.9\pm1.6$ & $2.30$ \\
J1248+4035 & 4.0 & $0.58\pm0.03$ & [$-73,94$] & $12.92\pm0.03$ & $90$ & $13.02\pm0.05$ & \nodata & \nodata & $1.65$ \\
 & & \nodata & \nodata & \nodata & \nodata & $12.89\pm0.04$ & $17.6\pm3.1$ & $-3.8\pm1.4$ & \nodata  \\
 & & \nodata & \nodata & \nodata & \nodata & $12.45\pm0.17$ & $7.2\pm4.3$ & $61.7\pm2.7$ & \nodata  \\
J1328+2159 & 12.7 & $0.24\pm0.03$ & [$-42,55$] & $12.53\pm0.05$ & $60$ & \nodata* & $13.3\pm6.0$ & $7.1\pm2.8$ & $1.22$ \\
J1429+0120 & 3.4 & $0.24\pm0.06$ & [$-48,72$] & $12.57\pm0.11$ & $60$ & $12.66\pm0.20$ & $10.0\pm8.9$ & $10.4\pm3.7$ & $1.19$ \\
J1457+5321 & 4.2 & $ < 0.20$ & [$-100,119$] & $ < 12.39$ & \nodata & \nodata & \nodata & \nodata & \nodata \\
J1459+3713 & 4.4 & $ < 1.05$ & [$-150,150$] & $ < 13.47$ & \nodata & \nodata & \nodata & \nodata & \nodata \\
J1525+0202 & 3.1 & $ < 0.21$ & [$-67,150$] & $ < 12.38$ & \nodata & \nodata & \nodata & \nodata & \nodata \\
J1605+5107 & 3.8 & $0.45\pm0.08$ & [$-100,41$] & $12.84\pm0.07$ & $80$ & $12.86\pm0.06$ & $26.6\pm6.9$ & $-16.1\pm3.9$ & $1.50$ \\
J1656+2541 & 1.1 & $ < 0.42$ & [$-59,47$] & $ < 12.80$ & \nodata & \nodata & \nodata & \nodata & \nodata \\
J1659+6202 & 7.2 & $0.40\pm0.03$ & [$-81,99$] & $12.76\pm0.03$ & $80$ & $12.96\pm0.19$ & \nodata & \nodata & $0.83$ \\
 & & \nodata & \nodata & \nodata & \nodata & $12.63\pm0.04$ & $15.5\pm5.1$ & $-4.5\pm2.3$ & \nodata  \\
 & & \nodata & \nodata & \nodata & \nodata & $12.69\pm0.36$ & $4.0\pm2.1$ & $29.2\pm1.6$ & \nodata  \\
J1717+3203 & 7.5 & $0.62\pm0.02$ & [$-114,77$] & $12.99\pm0.02$ & $90$ & $13.18\pm0.06$ & \nodata & \nodata & $1.09$ \\
 & & \nodata & \nodata & \nodata & \nodata & $12.52\pm0.10$ & $19.5\pm8.3$ & $-30.3\pm5.6$ & \nodata  \\
 & & \nodata & \nodata & \nodata & \nodata & $13.08\pm0.08$ & $10.0\pm1.7$ & $11.4\pm1.9$ & \nodata  \\
\enddata
\tablecomments{Best-fit Voigt profile model parameters for \ion{Ca}{2} systems fit with a single absorbing component are listed in a single table row.  For systems fit with two components, we list the total $\log N_{\rm vp}$ of the system in the first row for each sightline and include best-fit $\log N_{\rm vp}$, $b_{\rm D}$, and $\delta v$ values for the individual components in the following two rows.}
\tablenotetext{a}{Upper limits are reported at the $3\sigma$ level.}
\tablenotetext{*}{To model this absorber, we fixed the column density to the measured value of $N_{\rm aod}$ as described in Section~\ref{subsec:abs_modeling}.}
\end{deluxetable*}


\begin{deluxetable*}{lccccccccc}
\tablewidth{700pt}
\tabletypesize{\scriptsize}
\tablecaption{\ion{Na}{1} Absorption-line Equivalent Widths, Kinematics, and Best-fit Voigt Profile Model Parameters\label{tab.NaIabsinfo}}
\tablehead{
\colhead{Sight Line} & \colhead{$R_{\perp}$} & \colhead{$W_r$(\ion{Na}{1} 5891)\tablenotemark{a}} & \colhead{Velocity Limits} & \colhead{$\log N_{\rm aod}$(\ion{Na}{1})\tablenotemark{a}} & \colhead{$\Delta v_{90}$(\ion{Na}{1} 5891)} & \colhead{$\log N_{\rm vp}$(\ion{Na}{1})} & \colhead{$b_D$(\ion{Na}{1})} & \colhead{$\delta v$(\ion{Na}{1})} & \colhead{$\chi_r^{2}$(\ion{Na}{1})} \\
\colhead{} & \colhead{\hfill(kpc)} & \colhead{(\AA)} & \colhead{($\rm km~s^{-1}$)} & \colhead{($\rm cm^{-2}$)} & \colhead{($\rm km~s^{-1}$)} & \colhead{($\rm cm^{-2}$)} & \colhead{($\rm km~s^{-1}$)} & \colhead{($\rm km~s^{-1}$)} & \colhead{}
}
\startdata
J0013--0024 & 3.4 & $0.62\pm0.03$ & [$-60,98$] & $12.60\pm0.02$ & $100$ & $12.78\pm0.07$ & \nodata & \nodata & $0.67$ \\
 & & \nodata & \nodata & \nodata & \nodata & $12.37\pm0.05$ & $25.0\pm5.8$ & $-10.2\pm3.6$ & \nodata \\
 & & \nodata & \nodata & \nodata & \nodata & $12.57\pm0.10$ & $5.4\pm1.9$ & $36.8\pm2.1$ & \nodata \\
J0851+0719 & 5.6 & $ < 0.12$ & [$-150,150$] & $ < 11.80$ & \nodata & \nodata & \nodata & \nodata & \nodata \\
J0902+1414 & 3.6 & $0.21\pm0.03$ & [$-40,90$] & $12.11\pm0.05$ & $60$ & $12.34\pm0.09$ & $6.4\pm1.5$ & $1.4\pm1.6$ & $0.89$ \\
J0950+5442 & 1.0 & $0.50\pm0.05$ & [$-125,119$] & $12.46\pm0.03$ & $150$ & $12.46\pm0.03$ & $52.1\pm5.3$ & $-2.8\pm3.5$ & $0.92$ \\
J1005+5302 & 3.6 & $ < 0.12$ & [$-150,150$] & $ < 11.78$ & \nodata & \nodata & \nodata & \nodata & \nodata \\
J1044+0518 & 3.5 & $ < 0.10$ & [$-150,150$] & $ < 11.69$ & \nodata & \nodata & \nodata & \nodata & \nodata \\
J1135+2414 & 3.9 & $ < 0.10$ & [$-100,47$] & $ < 11.71$ & \nodata & \nodata & \nodata & \nodata & \nodata \\
J1158+3907 & 4.7 & $ < 0.10$ & [$-81,99$] & $ < 11.69$ & \nodata & \nodata & \nodata & \nodata & \nodata \\
J1220+2837 & 6.9 & $0.79\pm0.02$ & [$-61,72$] & $12.74\pm0.01$ & $70$ & $12.87\pm0.02$ & $18.3\pm1.1$ & $-7.8\pm0.5$ & $4.52$ \\
J1238+6448 & 7.0 & $0.73\pm0.03$ & [$-14,111$] & $12.75\pm0.01$ & $70$ & $13.41\pm0.27$ & \nodata & \nodata & $0.80$ \\
 & & \nodata & \nodata & \nodata & \nodata & $12.44\pm0.06$ & $10.5\pm3.4$ & $33.0\pm1.5$ & \nodata \\
 & & \nodata & \nodata & \nodata & \nodata & $13.36\pm0.31$ & $6.0\pm1.1$ & $67.0\pm0.7$ & \nodata \\
J1241+6332 & 10.6 & $0.85\pm0.03$ & [$-61,122$] & $12.76\pm0.01$ & $100$ & $12.84\pm0.01$ & \nodata & \nodata & $2.57$ \\
 & & \nodata & \nodata & \nodata & \nodata & $11.99\pm0.05$ & $16.6\pm3.8$ & $-16.0\pm2.0$ & \nodata \\
 & & \nodata & \nodata & \nodata & \nodata & $12.77\pm0.01$ & $23.6\pm1.0$ & $50.0\pm0.6$ & \nodata \\
J1248+4035 & 4.0 & $0.36\pm0.02$ & [$-59,61$] & $12.40\pm0.02$ & $60$ & $12.75\pm0.05$ & $7.0\pm0.5$ & $-1.3\pm0.5$ & $0.87$ \\
J1328+2159 & 12.7 & $0.11\pm0.04$ & [$-39,97$] & $11.79\pm0.14$ & $70$ & $11.73\pm0.11$ & $13.0\pm15.3$ & $6.3\pm5.7$ & $0.83$ \\
J1429+0120 & 3.4 & $0.18\pm0.02$ & [$-39,63$] & $12.09\pm0.04$ & $50$ & \nodata* & $10.1\pm5.1$ & $-2.4\pm2.0$ & $1.32$ \\
J1457+5321 & 4.2 & $0.26\pm0.04$ & [$-95,86$] & $12.16\pm0.07$ & $100$ & $12.12\pm0.06$ & $36.6\pm8.2$ & $-26.3\pm4.9$ & $0.89$ \\
J1459+3713 & 4.4 & $ < 0.14$ & [$-150,150$] & $ < 11.86$ & \nodata & \nodata & \nodata & \nodata & \nodata \\
J1525+0202 & 3.1 & $ < 0.14$ & [$-150,150$] & $ < 11.85$ & \nodata & \nodata & \nodata & \nodata & \nodata \\
J1605+5107 & 3.8 & $0.26\pm0.04$ & [$-89,99$] & $12.21\pm0.05$ & $90$ & $12.23\pm0.08$ & $9.1\pm5.4$ & $3.3\pm1.9$ & $1.16$ \\
J1656+2541 & 1.1 & $ < 0.12$ & [$-50,150$] & $ < 11.77$ & \nodata & \nodata & \nodata & \nodata & \nodata \\
J1659+6202 & 7.2 & $0.24\pm0.03$ & [$-78,102$] & $12.17\pm0.04$ & $60$ & $12.28\pm0.03$ & $13.8\pm3.5$ & $6.3\pm1.4$ & $0.89$ \\
J1717+3203 & 7.5 & $0.65\pm0.03$ & [$-78,80$] & $12.71\pm0.01$ & $60$ & $13.16\pm0.11$ & \nodata & \nodata & $1.06$ \\
 & & \nodata & \nodata & \nodata & \nodata & $13.11\pm0.11$ & $8.8\pm0.9$ & $8.7\pm1.0$ & \nodata \\
 & & \nodata & \nodata & \nodata & \nodata & $12.19\pm0.43$ & $5.4\pm9.6$ & $-34.0\pm2.6$ & \nodata \\
\enddata
\tablecomments{Best-fit Voigt profile model parameters for \ion{Na}{1} systems fit with a single absorbing component are listed in a single table row.  For systems fit with two components, we list the total $\log N_{\rm vp}$ of the system in the first row for each sightline and include best-fit $\log N_{\rm vp}$, $b_{\rm D}$, and $\delta v$ values for the individual components in the following two rows.}
\tablenotetext{a}{Upper limits are reported at the $3\sigma$ level.}
\tablenotetext{*}{To model this absorber, we fixed the column density to the measured value of $N_{\rm aod}$ as described in Section~\ref{subsec:abs_modeling}.}
\end{deluxetable*}

\begin{figure*}[ht]
  \includegraphics[width=0.5\textwidth]{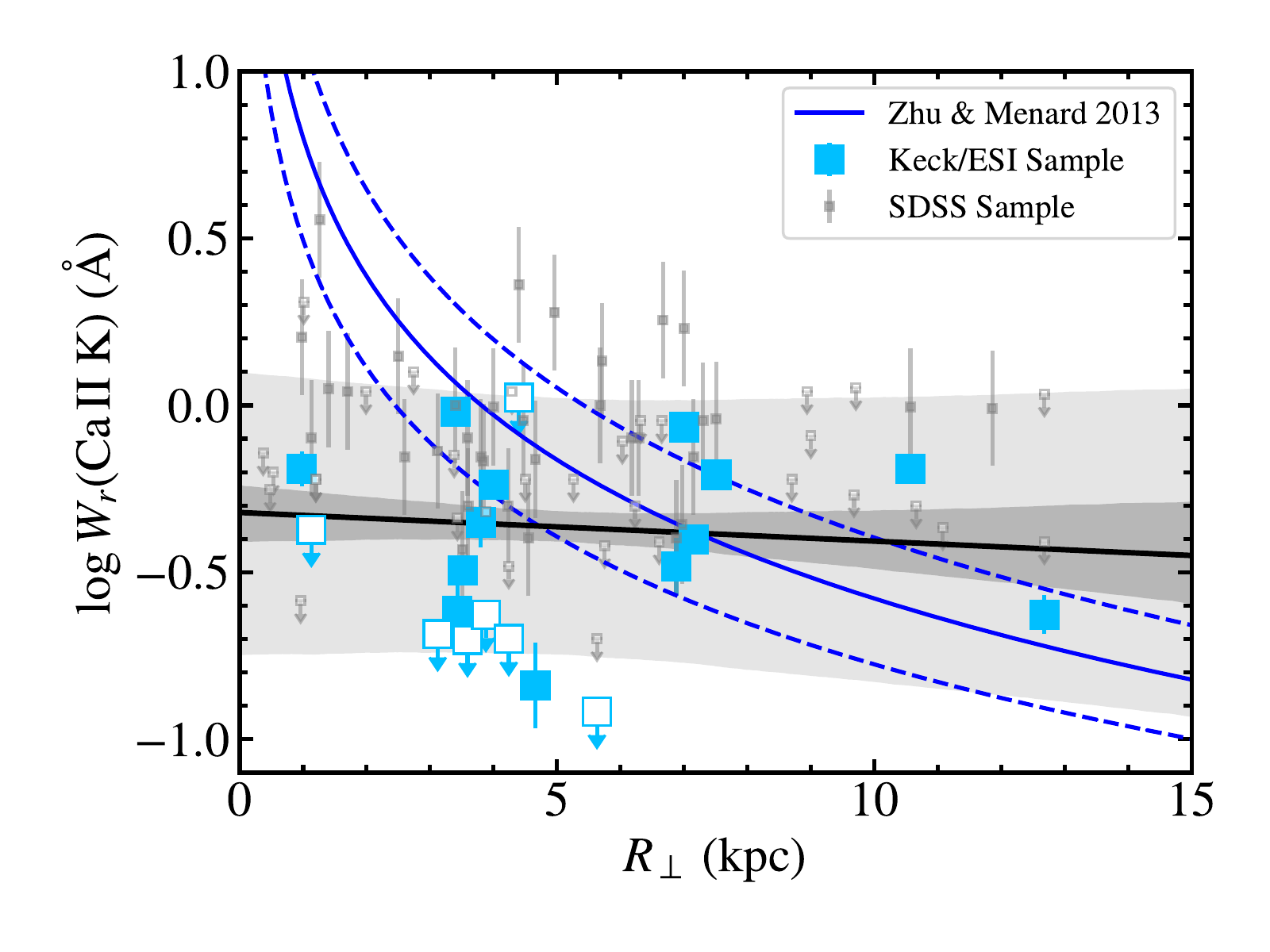}
  \includegraphics[width=0.5\textwidth]{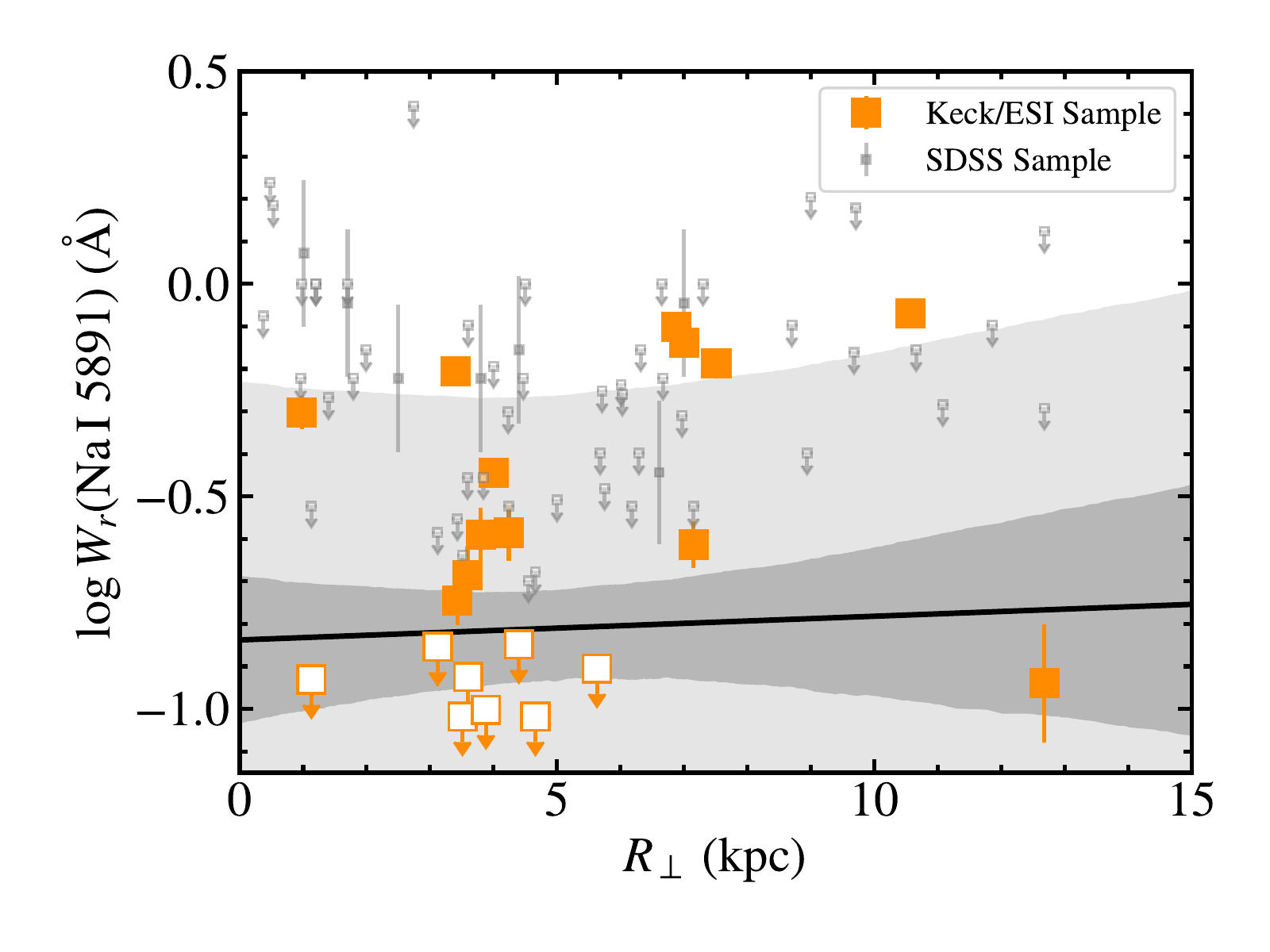}
\caption{Total system $W_r(\mbox{\ion{Ca}{2} K})$ (left) and $W_r$(\ion{Na}{1} 5891) (right) vs.\ projected distance from the associated host galaxy.  Colored points show constraints from our ESI spectroscopy.  Upper limits, indicated with open squares, are shown in cases for which $W_r < 3\sigma_{W_r}$ and represent 3$\sigma$ limits.  Gray points show measurements reported in \citet{Straka2015} from their analysis of SDSS spectroscopy probing the parent sample of GOTOQs.  We exclude absorbers that were flagged as blended by \citet{Straka2015}.  We also exclude any systems in which the \ion{Ca}{2} K transition falls more than 20 \AA\ blueward of the Ly$\alpha$ emission line of the corresponding QSO to avoid blending from the Ly$\alpha$ forest. Black solid lines show best-fit linear relations between $\log W_r$ and $R_{\perp}$ (see Section~\ref{subsec:Wr_Rperp}), and medium gray contours show the inner $\pm34$\% of the locus of fits drawn at random from the posterior probability density function of each linear model.  The light gray region extends the boundaries of the medium gray $1\sigma$ region by the best-fit value of $\sigma_C$ to approximately indicate the degree of intrinsic scatter implied by the data.
Our $W_r$ measurements exhibit no apparent anticorrelation with increasing projected distance from the foreground host.
\label{fig:ew_rperp}}
\end{figure*}

\begin{figure*}[ht]
  \includegraphics[width=0.5\textwidth]{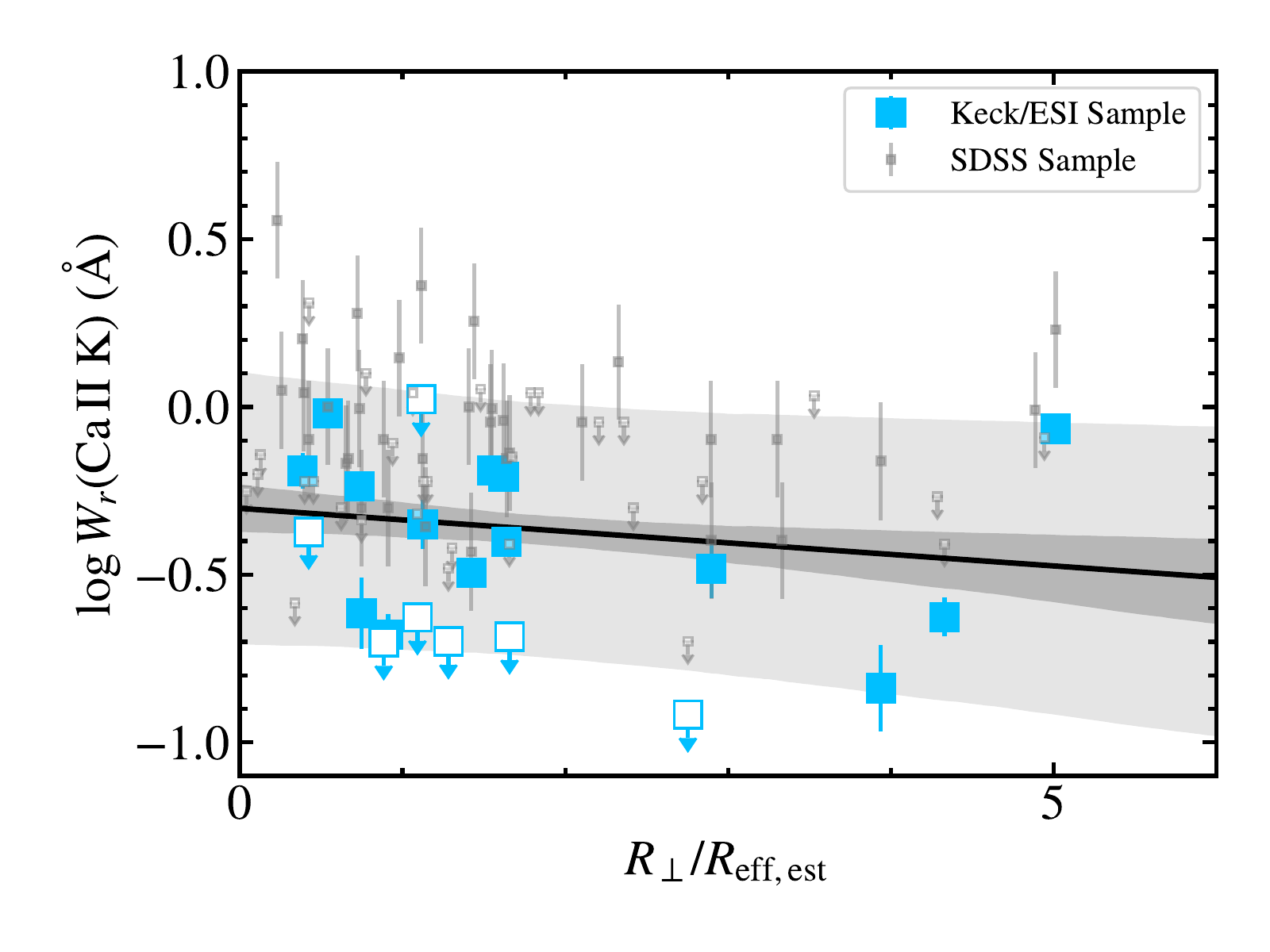}
  \includegraphics[width=0.5\textwidth]{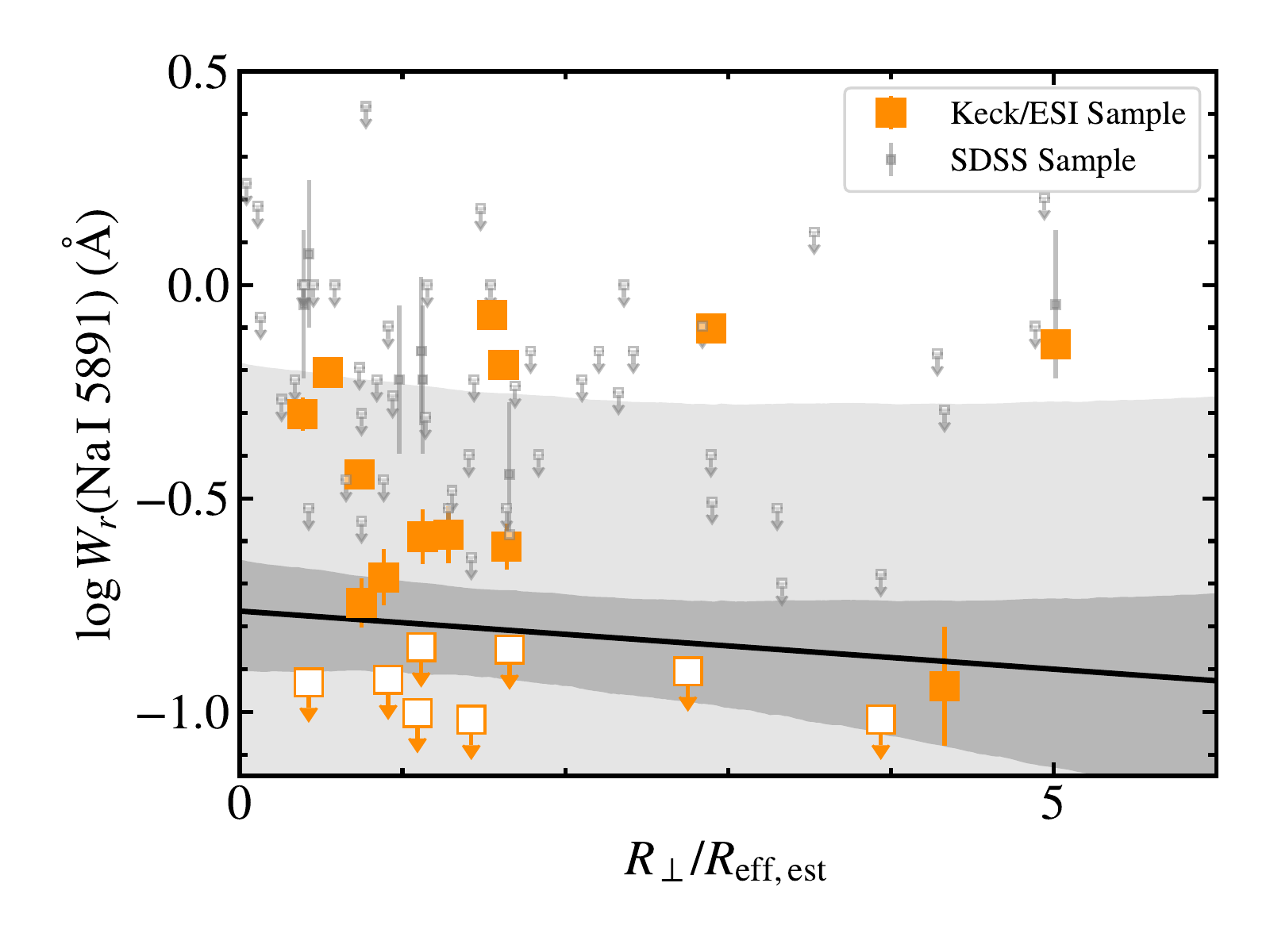}
\caption{Total system $W_r$(\ion{Ca}{2} K) (left) and $W_r$(\ion{Na}{1} 5891) (right) vs.\ projected distance from the associated host galaxy, normalized by the galaxy's estimated effective radius.  Symbols, lines and contours are as described in the Figure~\ref{fig:ew_rperp} caption.  The relation between $\log W_r$(\ion{Ca}{2} K) and $R_{\perp}/R_{\rm eff,est}$ for the combined ESI and SDSS sample has a slope $m = -0.035^{+0.028}_{-0.030}$, indicative of a weak anticorrelation between these quantities.  The relation between $\log W_r$(\ion{Na}{1} 5891) and $R_{\perp}/R_{\rm eff,est}$  exhibits no significant anticorrelation. \label{fig:ew_rperp_renorm}}
\end{figure*}

\section{\ion{Ca}{2} and \ion{Na}{1} Absorption Properties of the Disk-Halo Interface}\label{sec:results}

Here we examine the incidence of \ion{Ca}{2} and \ion{Na}{1} absorption in the disk-halo environment and assess the relation between their absorption strengths and $R_{\perp}$.  

\subsection{$\log W_r$-$R_{\perp}$ Relations}\label{subsec:Wr_Rperp}
We show our measurements of the total system $W_r$(\ion{Ca}{2} K) and $W_r$(\ion{Na}{1} 5891) vs.\ $R_{\perp}$ in Figure~\ref{fig:ew_rperp} with colored points.  Detections and 3$\sigma$ upper limits on $W_r$ for these transitions measured from SDSS spectra of the parent GOTOQ sample by \citet{Straka2015} are shown in gray.  Within our ESI sample, absorption detections span the full range of $R_{\perp}$ probed, with nondetections arising only within $< 6$~kpc. 
 We note here that several of our foreground galaxies observed at $R_{\perp} > 6$~kpc may have higher global SFRs ($\gtrsim10\!-\!100~M_{\odot}~\rm yr^{-1}$; see Appendix~\ref{sec:appendix_SFRfrac}) than those observed at $R_{\perp} < 6$ kpc.  Under the assumption that galaxies that are more actively star-forming will have larger $W_r$(\ion{Ca}{2}) and $W_r$(\ion{Na}{1}) across a broad range of impact parameters, this potential bias may drive an enhancement in our observed $W_r$ values at large $R_{\perp}$.  While we cannot reliably quantify the global SFRs of our galaxy sample with current data, we can instead draw on the measurements of global $M_*$ described in Section~\ref{sec:fg_galaxies} to assess the degree to which an analogous relation between $M_*$ and $W_r$ may impact the distributions of datapoints shown in Figure~\ref{fig:ew_rperp}.  The left-hand panel of Figure~\ref{fig:rperp_mstar_sfr} demonstrates that our sample contains equal numbers of galaxies with stellar masses falling above and below the median value ($\log M_*/M_{\odot} = 9.3$) at $R_{\perp}>6$ kpc. This suggests that the bias described above does not have a major impact on our analysis of the relation between absorber properties and $R_{\perp}$; however, we caution that new data enabling the measurement of the global SFRs in our foreground galaxies are needed to fully disentangle the relationships between $R_{\perp}$, global star-formation activity, and $W_r$.

The blue curves in Figure~\ref{fig:ew_rperp} show the power-law relation (with 1$\sigma$ uncertainties) fit to the mean \ion{Ca}{2} K absorption signal 
measured in SDSS spectra of QSO sightlines vs.\ the projected separation of these QSOs from known foreground systems by \citet{ZhuMenard2013}.  Because this latter analysis included all sightlines having 3 kpc $< R_{\perp} <$ 10 kpc in a single bin, the fitted relation is insensitive to potential changes in the power-law slope at very small separations.  Nevertheless, the absorbers in our dataset do not exhibit 
larger $W_r$ at smaller projected separations as implied by this fit. 
We caution that the foreground galaxy sample identified by \citet{ZhuMenard2013} has higher stellar masses than those we study here (i.e., the median stellar mass in the former sample is $\log M_*/M_{\odot}\sim10.3$), which could explain the larger $W_r$ implied by their fitted relation at $R_{\perp}\sim3$--4 kpc.


Figure~\ref{fig:ew_rperp_renorm} shows the same $W_r$ measurements presented in Figure~\ref{fig:ew_rperp} vs.\ $R_{\perp}/R_{\rm eff, est}$.  
We remind the reader that those galaxies probed at $R_{\perp}/R_{\rm eff, est}>2$ have systematically lower stellar masses than those probed at $R_{\perp}/R_{\rm eff, est}<2$.  Moreover, the modeling described in Appendix~\ref{sec:appendix_SFRfrac} suggests the former systems exhibit a broad range of global SFRs, spanning between ${\sim}0.5~M_{\odot}~\rm yr^{-1}$ and $>100~M_{\odot}~\rm yr^{-1}$.  While absorption nondetections are more evenly distributed across this parameter space than across the range in $R_{\perp}$, the relation between $W_r$ and $R_{\perp}/R_{\rm eff, est}$ does not exhibit a clear anticorrelation for either ion.  

To quantitatively test for correlations (or a lack thereof) in these quantities, we model these datasets assuming a linear relation between $\log W_r$ and either $R_{\perp}$ or $R_{\perp}/R_{\rm eff, est}$:
\begin{equation}\label{eq:linear}
    \log W_r = b + m R_{\perp}.
\end{equation}
We follow \citet{Chen2010a} and \citet{Rubin2018a} to compute the likelihood function for this model.  Briefly, for all securely detected $W_r$ values, the contribution to the logarithm of the likelihood is $\chi^2 / 2$.  For non-detections, each term in the product used to compute the likelihood is the integral from $-\infty$ to the value of the $W_r$ upper limit of a Gaussian function similar in form to that used to calculate $\chi^2$ (see \citealt{Rubin2018a} for the full likelihood function).  We also assume that the relation in Equation~\ref{eq:linear} has an intrinsic cosmic variance, $\sigma_C$, such that the Gaussian variance adopted for each measurement in the likelihood function is $s_i^2 = \sigma_i^2 + \sigma_C^2$, with $\sigma_i$ equal to the measurement uncertainty in each $\log W_r$ value.

We use the Python software package \texttt{emcee} to perform affine-invariant ensemble Markov Chain Monte Carlo sampling of the posterior probability density function (PPDF) for this model \citep{Foreman-Mackey2013}.  We adopt uniform priors for all three parameters within the intervals $-5.0 < m < 5.0$ (with $m$ having units of either $\rm kpc^{-1}$ or being unitless, as appropriate), $-10.0 < b < 10.0$, and $-10.0 < \ln \sigma_C < 10.0$.  We implement 100 ``walkers", each of which take 5000 steps (the first 1000 of which are discarded) to thoroughly sample the PPDF.  We interpret the median and $\pm34$th percentiles of the marginalized PPDF for each parameter as its best value and uncertainty interval.

We show the resulting best-fit relations between $\log W_r$ and either $R_{\perp}$ or $R_{\perp}/R_{\rm eff, est}$ for the combined ESI and SDSS datasets in Figures~\ref{fig:ew_rperp} and \ref{fig:ew_rperp_renorm}, respectively, with solid black lines.  The medium gray contours show the inner $\pm34$\% of the locus of fits for 1000 sets of parameters drawn at random from the PPDF of each data-model comparison.  The light gray contours indicate the boundaries of the inner $\pm34$\% locus, extended on either side by the best-fit value of $\sigma_C$.
We also list the best-fit parameters and their uncertainty intervals for each dataset in Table~\ref{tab:linear_fits}.  Three of the four best-fit values of the slope ($m$) are consistent with zero, confirming a lack of any significant correlation between both $\log W_r(\mbox{\ion{Ca}{2} K})$ and $\log W_r(\mbox{\ion{Na}{1} 5891})$ and $R_{\perp}$, as well as between 
$\log W_r(\mbox{\ion{Na}{1} 5891})$ and $R_{\perp}/R_{\rm eff,est}$.  The $\log W_r(\mbox{\ion{Ca}{2} K})$-$R_{\perp}/R_{\rm eff,est}$ relation has a slope $m=-0.035^{+0.028}_{-0.030}$, weakly suggestive of an anticorrelation between these variables.  

Given our finding in Section~\ref{subsec:abs_modeling} that the \citet{Straka2015} $W_r$ values are frequently larger than those we measure for the same sightlines, we also perform the same modeling including only our ESI dataset.  The resulting best-fit model parameters are listed in Table~\ref{tab:linear_fits}.  Here again, three of the four best-fit slopes are consistent with zero.  Moreover, the $\log W_r(\mbox{\ion{Na}{1} 5891})$-$R_{\perp}$ relation has a slope that is marginally \emph{positive} ($m=+0.058^{+0.046}_{-0.042}\,{\rm kpc}^{-1}$).  All together, we interpret these results as further confirmation of a lack of any anticorrelation between $W_r$ and $R_{\perp}$ or $R_{\perp}/R_{\rm eff,est}$.   

 Keeping in mind the caveat that these findings may be affected by a bias in our galaxy sample toward higher global SFRs at larger $R_{\perp}$ (as discussed toward the beginning of this section), we note that 
the lack of a strong dependence of our $W_r$ values on projected distance is unique among the QSO-galaxy pair literature.  The vast majority of these studies instead have reported a statistically significant decline in the $W_r$ of a wide range of ionic transitions (including transitions of \ion{H}{1}, \ion{C}{2}, \ion{C}{3}, \ion{C}{4}, \ion{Si}{2}, \ion{Si}{3}, \ion{Mg}{2} and \ion{Ca}{2}) with $R_{\perp}$ \citep[e.g.,][]{LanzettaBowen1990,Kacprzak2008,Chen2010a,Nielsen2013,Werk2013,ZhuMenard2013,Burchett2016,Kulkarni2022}.  However, these works have included sight lines over a much larger range of projected separations ($R_{\perp} \gtrsim 100$ kpc) than are included here, and many of them have included few (if any) sightlines with $R_{\perp} < 15$~kpc \citep[e.g.,][]{LanzettaBowen1990,Chen2010a,Werk2013}.
The findings of \citet{Kacprzak2013}, a study of \ion{Mg}{2} absorption along a sample of seven GOTOQ sightlines selected from \citet{Noterdaeme2010} and \citet{York2012}, confirm that sightlines with impact parameters $\gtrsim 10$ kpc drive the well-known anticorrelation between $W_r$(\ion{Mg}{2} 2796) and $R_{\perp}$, while the $W_r$(\ion{Mg}{2} 2796) values for sightlines within this projected distance exhibit no significant dependence on $R_{\perp}$.   On the other hand, \citet{Kulkarni2022} noted that the strong anticorrelation between $N$(\ion{H}{1}) and $R_{\perp}$ exhibited by their sample of 113 galaxies associated with DLAs and sub-DLAs  (assembled from their study of eight GOTOQs and the literature across $0 < z < 4.4$) appears to extend well within $R_{\perp} < 10$ kpc.  This apparent disagreement with both \citet{Kacprzak2013} and the present study may be driven by a variety of factors, including the use of different ionic transitions and quantities characterizing absorption-line strength (i.e., $W_r$ vs.\ $N$), and differing absorber-galaxy pair selection criteria.

\begin{deluxetable*}{llllc}
\tablecaption{Best-fit Parameters for Linear $\log W_r - R_{\perp}$ Models\label{tab:linear_fits}}
\tabletypesize{\footnotesize}
\tablehead{
\colhead{Data Set} & \colhead{Relation} & \colhead{$m$} & \colhead{$b$} & \colhead{$\sigma_C$}\\ 
\colhead{} & \colhead{} & \colhead{} & \colhead{} & \colhead{}}
\startdata
ESI \& \citet{Straka2015} & $\log W_r(\mbox{\ion{Ca}{2} K})$-$R_{\perp}$ & $-0.009\pm 0.015~\rm kpc^{-1}$ & $-0.32\pm0.09$ & $0.34_{-0.04}^{+0.05}$\\
 & $\log W_r(\mbox{\ion{Ca}{2} K})$-$R_{\perp}/R_{\rm eff,est}$ & $-0.035_{-0.030}^{+0.028}$ & $-0.30\pm0.07$ & $0.33_{-0.04}^{+0.05}$\\
 & $\log W_r(\mbox{\ion{Na}{1} 5891})$-$R_{\perp}$ & $+0.006_{-0.027}^{+0.026}~\rm kpc^{-1}$ & $-0.84_{-0.18}^{+0.15}$ & $0.46_{-0.07}^{+0.10}$ \\
 & $\log W_r(\mbox{\ion{Na}{1} 5891})$-$R_{\perp}/R_{\rm eff,est}$ & $-0.028_{-0.053}^{+0.046}$ & $-0.76_{-0.14}^{+0.12}$ & $0.46_{-0.07}^{+0.10}$\\
 \hline
 ESI Only & $\log W_r(\mbox{\ion{Ca}{2} K})$-$R_{\perp}$ & $+0.022_{-0.028}^{+0.031}~\rm kpc^{-1}$ & $-0.64_{-0.20}^{+0.17}$ & $0.35_{-0.07}^{+0.10}$\\
  & $\log W_r(\mbox{\ion{Ca}{2} K})$-$R_{\perp}/R_{\rm eff,est}$ & $-0.006_{-0.061}^{+0.063}$ & $-0.52_{-0.15}^{+0.14}$ & $0.34_{-0.07}^{+0.10}$\\
   & $\log W_r(\mbox{\ion{Na}{1} 5891})$-$R_{\perp}$ & $+0.058_{-0.042}^{+0.046}~\rm kpc^{-1}$ & $-1.05_{-0.30}^{+0.25}$ & $0.52_{-0.11}^{+0.16}$ \\
   & $\log W_r(\mbox{\ion{Na}{1} 5891})$-$R_{\perp}/R_{\rm eff,est}$ & $+0.016_{-0.097}^{+0.098}$ & $-0.78_{-0.23}^{+0.20}$ & $0.54_{-0.11}^{+0.16}$\\
\enddata
\end{deluxetable*}

\subsection{Column Densities and Covering Fractions}\label{subsec:results_cf}

\begin{figure*}[ht]
 \includegraphics[width=0.5\textwidth]{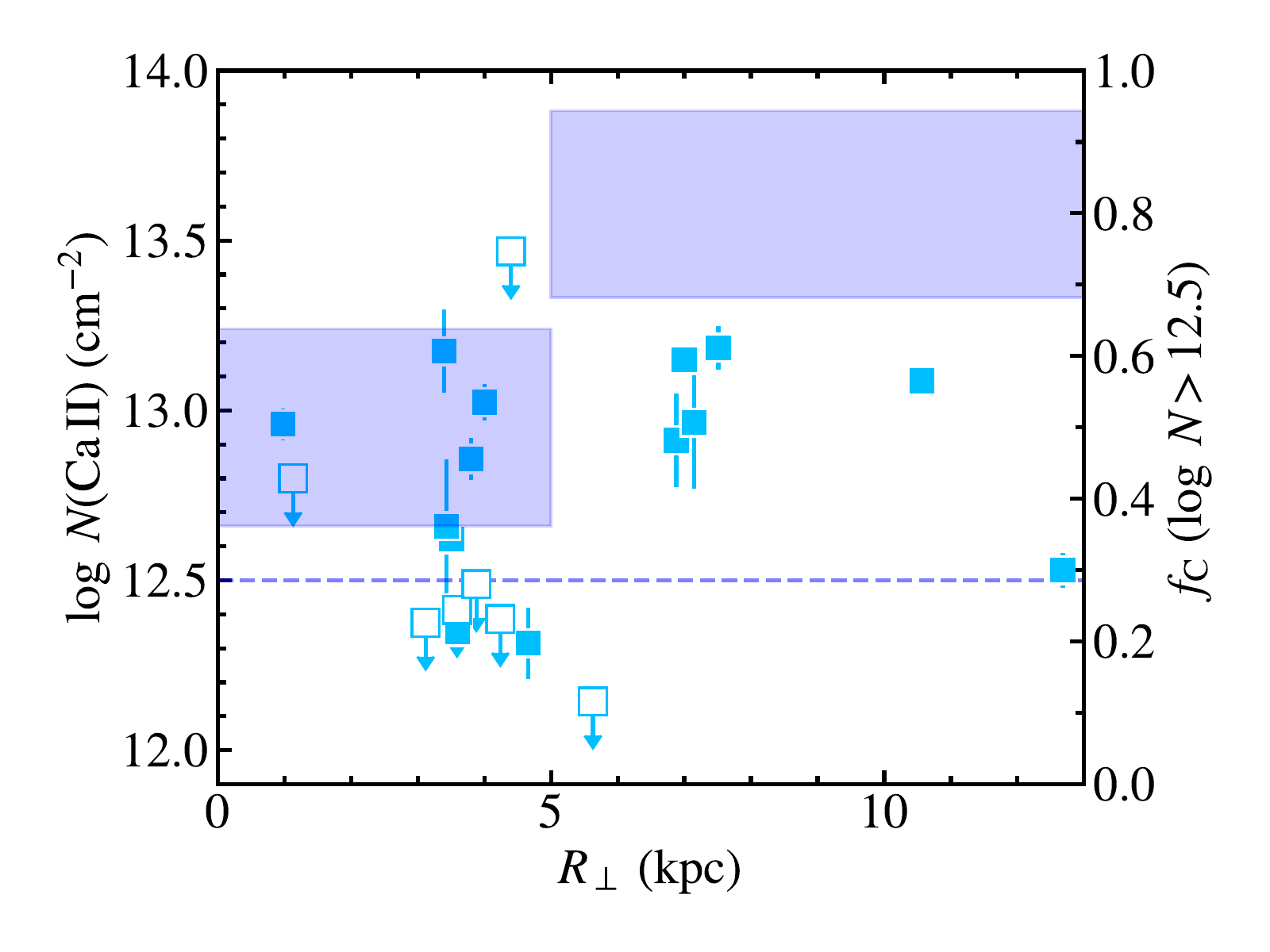}
  \includegraphics[width=0.5\textwidth]{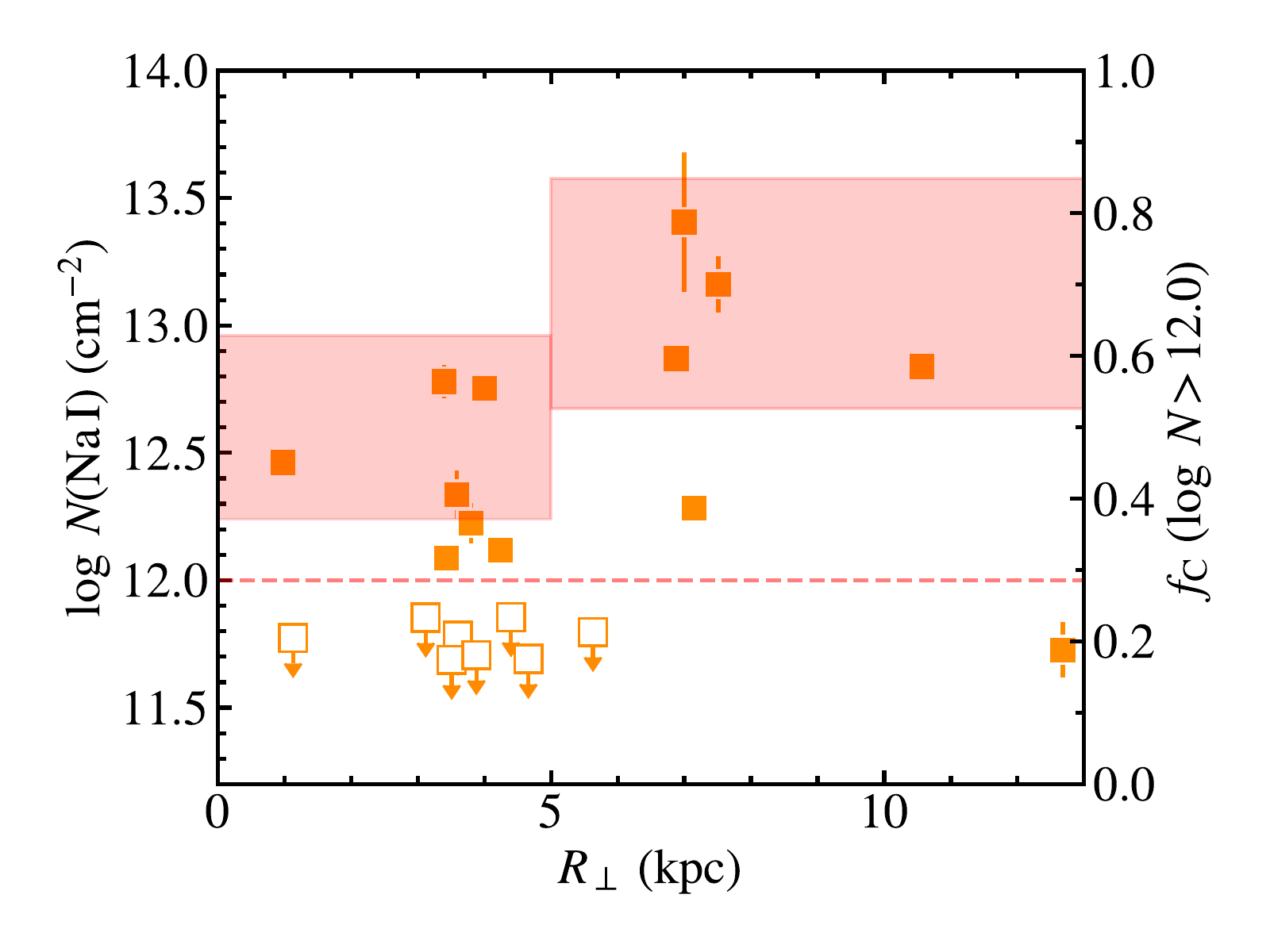}
 \includegraphics[width=0.5\textwidth]{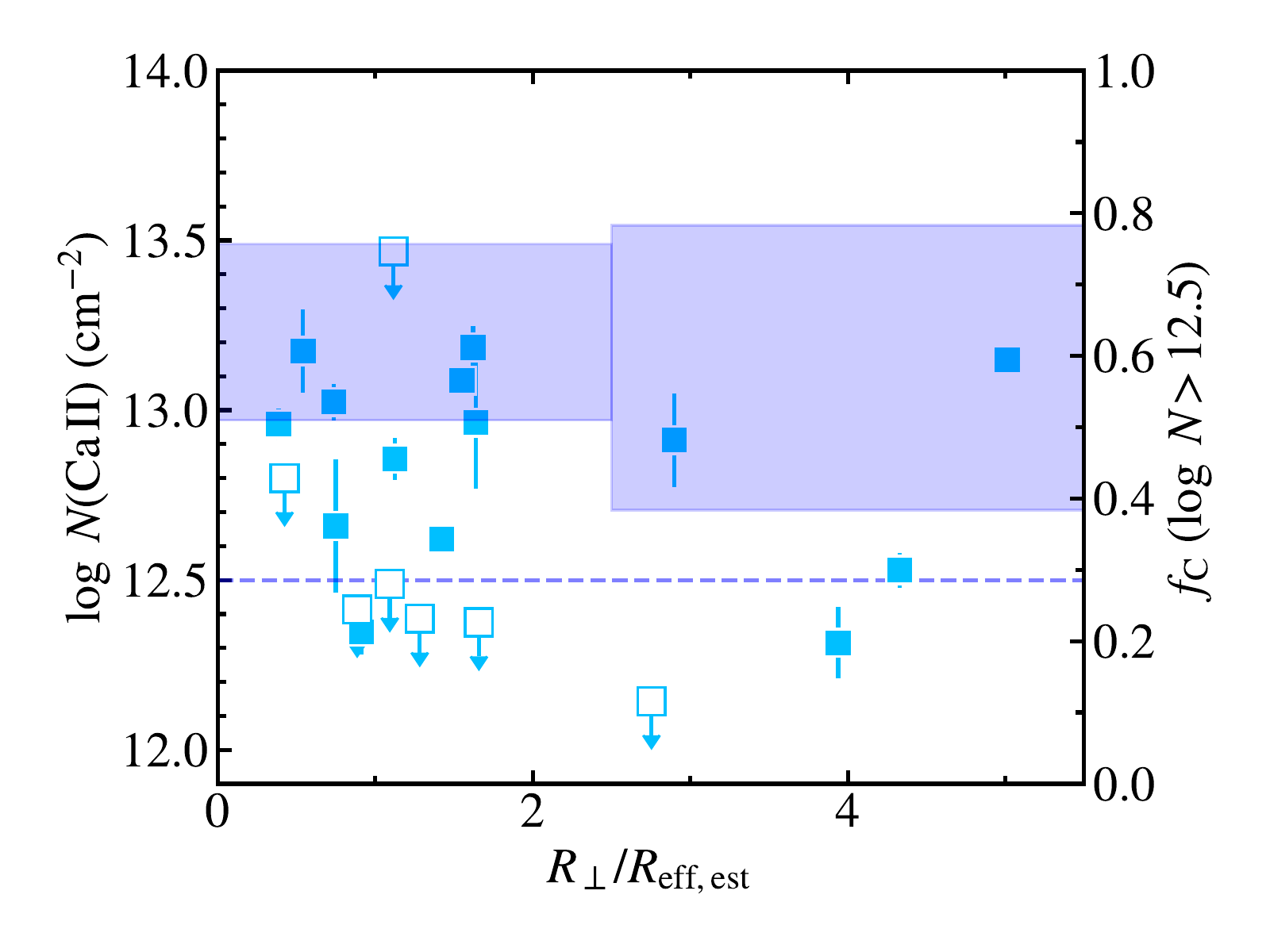}
  \includegraphics[width=0.5\textwidth]{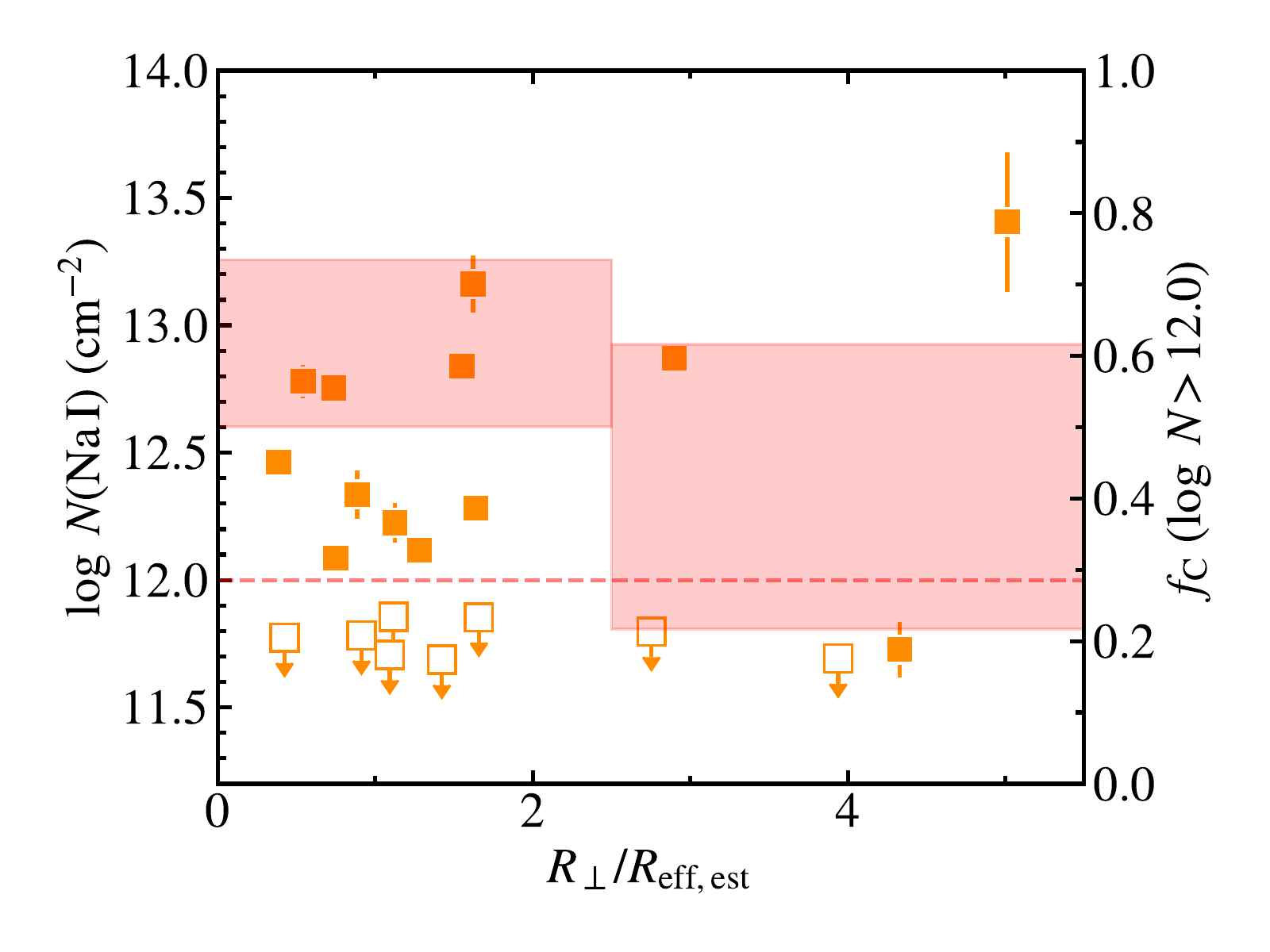}
\caption{{\it Top Row:} Total  system column density of \ion{Ca}{2} (left) and \ion{Na}{1} (right) vs.\ projected distance from the associated GOTOQs.  Open squares with downward arrows represent 3$\sigma$ upper limits calculated using the apparent optical depth method.  The filled boxes indicate the $\pm34$th percentile Wilson score confidence intervals, with respect to the right axes, for the covering fraction of absorbers having $\log N(\mbox{\ion{Ca}{2}})>12.5$ and $\log N(\mbox{\ion{Na}{1}})>12.0$, respectively.
{\it Bottom Row:} Same as above, vs.\ $R_{\perp}/R_{\rm eff,est}$.
  \label{fig:logN_rperp}}
\end{figure*}

Figure~\ref{fig:logN_rperp} shows the total system column densities (including all velocity components) of \ion{Ca}{2} (left) and \ion{Na}{1} (right) in each GOTOQ sightline in our sample vs.\ $R_{\perp}$ (top row) and vs.\ $R_{\perp}/R_{\rm eff,est}$ (bottom row).  As with the $W_r$ values discussed above, the measured column densities do not appear to exhibit any dependence on either $R_{\perp}$ or $R_{\perp}/R_{\rm eff, est}$.  

We assess the covering fraction ($f_{\rm C}$) of these absorbers by dividing the number of systems with column densities above a given threshold by the total number of sightlines (excluding nondetections above the threshold).  These thresholds are chosen to lie just above the majority of 3$\sigma$ upper limits for each ion; i.e., $N(\mbox{\ion{Ca}{2}})>10^{12.5}~\rm cm^{-2}$ and $N(\mbox{\ion{Na}{1}})>10^{12.0}~\rm cm^{-2}$.  We adopt the $\pm34$th percentile Wilson score intervals as uncertainty intervals for each covering fraction.  Overall, we measure covering fractions $f_{\rm C}(\mbox{\ion{Ca}{2}})= 0.63^{+0.10}_{-0.11}$ and $f_{\rm C}(\mbox{\ion{Na}{1}})= 0.57^{+0.10}_{-0.11}$.  We also compute covering fractions within two bins in $R_{\perp}$ and $R_{\perp}/R_{\rm eff,est}$ and show the results with filled boxes in Figure~\ref{fig:logN_rperp}.  These covering fractions do not vary significantly (i.e., by $>2\sigma$) as a function of either of these measures of projected distance.

It is notable that the overall $f_{\rm C}$ values for \ion{Ca}{2} and \ion{Na}{1} are statistically consistent with each other, given that \ion{Ca}{2} is known to trace a wider range of gas densities and temperatures \citep{Phillips1984,Vallerga1993,BenBekhti2012,Murga2015}.  If we instead adopt equivalent column density thresholds for both ions ($N > 10^{12.5}~\rm cm^{-2}$), we find a value $f_{\rm C}(\mbox{\ion{Na}{1}})=0.33^{+0.11}_{-0.09}$ which is $1.9\sigma$ below that of $f_{\rm C}$(\ion{Ca}{2}).  This difference accords with a picture in which \ion{Na}{1}-absorbing structures are smaller in size and/or less abundant than \ion{Ca}{2}-absorbing clouds \citep[e.g.,][]{Bish2019}.  These values are also broadly consistent with the incidence of intermediate and high-velocity \ion{Ca}{2} and \ion{Na}{1} absorbers detected toward a sample of 408 QSO sightlines probing the Milky Way disk-halo interface and halo by \citet{BenBekhti2012}, in spite of their use of more sensitive column density thresholds:  these authors measured $f_{\rm C} = 0.5$ for a threshold $N(\mbox{\ion{Ca}{2}})\ge 10^{11.4}~\rm cm^{-2}$ and $f_{\rm C} = 0.35$ for a threshold $N(\mbox{\ion{Na}{1}})\ge 10^{10.9}~\rm cm^{-2}$.
Similar covering fractions for these ions were measured toward 
multiple stellar sightlines probing intermediate-velocity material $\sim3$ kpc above the Milky Way's disk by \citet{Bish2019} (i.e., $\log N(\mbox{\ion{Ca}{2}})> 11.5) = 0.63^{+0.07}_{-0.14}$ and $f_{\rm C}(\log N(\mbox{\ion{Na}{1}})>11.3) = 0.26^{+0.06}_{-0.08}$).  This implies that our GOTOQ sightlines have overall higher column densities than those measured in both the \citet{BenBekhti2012}  and \citet{Bish2019} samples.

We speculate that this may be due to the limited path through the Milky Way probed by the stellar and QSO sightlines used in these studies.  In particular, because the focus of
these works is on characterizing extraplanar material, they have explicitly
excluded absorbers having velocities consistent with that of the Milky Way's disk rotation curve (i.e., ISM absorbers) from their analyses.  The intermediate- and high-velocity clouds targeted by \citet{BenBekhti2012} are typically found to be located within $<2.5$ kpc and $\sim5$--20 kpc away from the Milky Way's disk, respectively, in cases in which distance information is available (see \citealt{Richter2017} and references therein).
Our GOTOQ sightlines, by contrast, are sensitive to all absorbers above our column density detection threshold ($\log N(\mbox{\ion{Ca}{2}}) \gtrsim 12.1$--12.4 and $\log N(\mbox{\ion{Na}{1}}) \gtrsim 11.9$) regardless of velocity  or location along the line of sight. This bias is compounded by a lack of Milky Way halo sightlines located at low Galactic latitudes: existing sightline samples probe relatively short paths through the disk and extraplanar region due to their height above the disk plane
\citep[e.g,][]{Bish2021}.

Finally, we note that our \ion{Ca}{2} and \ion{Na}{1} covering fractions are significantly lower than the unity covering fraction measured for \ion{Mg}{2} absorbers having $W_r(\mbox{\ion{Mg}{2}}~2796)>1$ \AA\ detected along the seven GOTOQ sightlines studied by \citet{Kacprzak2013}.  These absorbers have larger $W_r$ values than any in our sample and probe a broader range of gas phases that are known to extend well beyond galactic disks into their halos \citep[e.g.,][]{BergeronStasinska1986,Chen2010a,Nielsen2013,Lan2014}.

\begin{figure}[h]
 \includegraphics[width=0.5\textwidth]{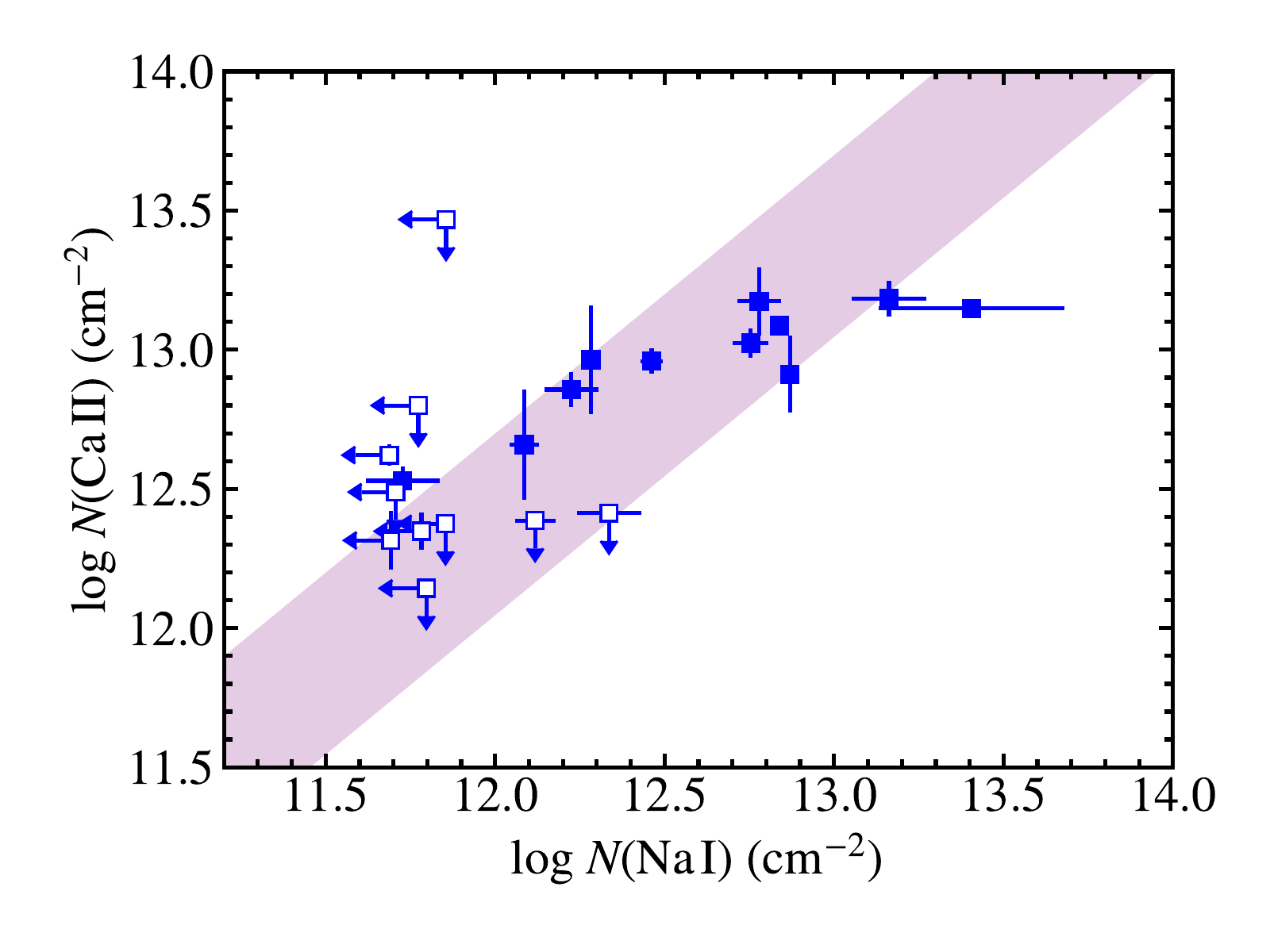}
\caption{Total system column density of \ion{Ca}{2} vs.\ that of \ion{Na}{1} in individual GOTOQ sightlines.  Sightlines along which we do not securely detect one or both of these ions are indicated with open squares placed at the corresponding 3$\sigma$ upper limit.  The purple filled region indicates the range in the average ratio $N$(\ion{Na}{1})/$N$(\ion{Ca}{2}) observed toward high Galactic latitude sightlines probing the Milky Way by \citet{Murga2015}.
\label{fig:logN_CaII_NaI}}
\end{figure}

Figure~\ref{fig:logN_CaII_NaI} compares our total column density constraints for \ion{Na}{1} and \ion{Ca}{2} in individual sightlines.  We find that, in general, larger column densities of \ion{Na}{1} are associated with larger column densities of \ion{Ca}{2}.  The purple filled region in this figure indicates the range in the average ratio $\langle N$(\ion{Na}{1})/$N$(\ion{Ca}{2})$\rangle \approx0.2$--0.9 measured along high-latitude Milky Way halo sightlines by \citet{Murga2015}.   This latter work analyzed the coadded spectra of many thousands of extragalactic sources, and the absorption signal they report arises from material at all velocities along the line of sight (including contributions from both the Milky Way's ISM and CGM).  Our measurements largely fall within this range, suggesting that the gaseous environments probed by our QSO sample are similar to those arising in the Milky Way.

\subsection{Absorption Kinematics}\label{subsec:results_kinematics}

\begin{figure*}[ht]
 \includegraphics[width=\textwidth]{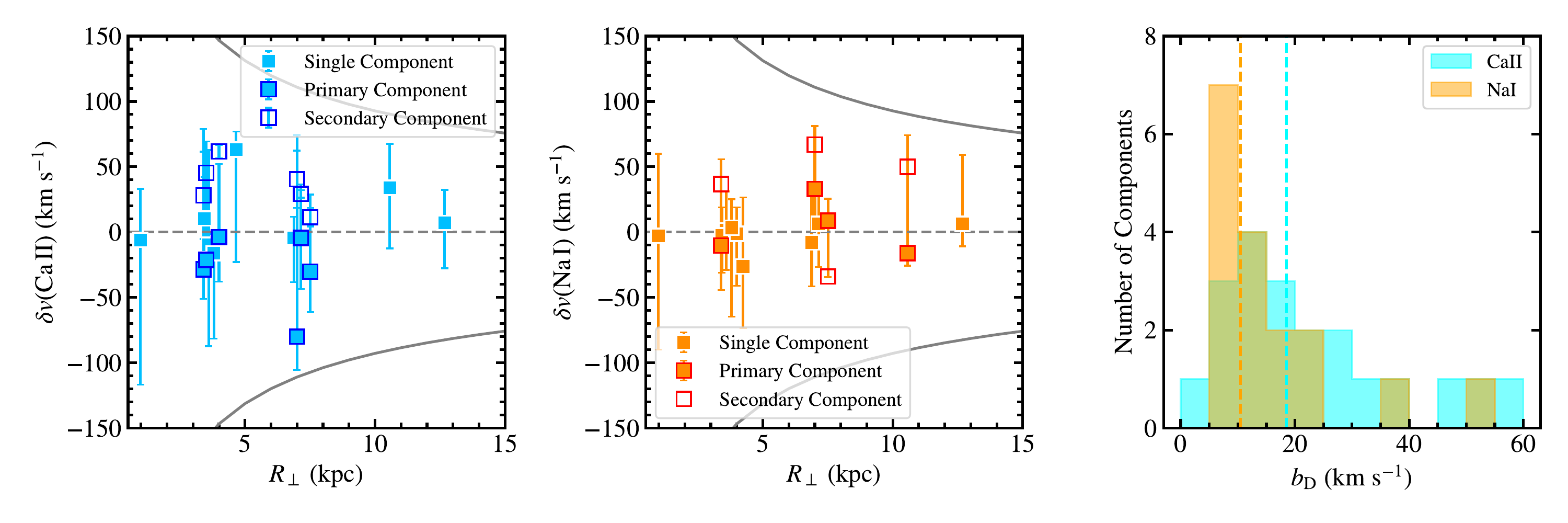}
\caption{Absorption velocity offsets relative to \zha\ for \ion{Ca}{2} (left) and \ion{Na}{1} (middle) for each securely detected absorption system vs.\ projected distance from the galaxy host. Systems fit with single velocity components are shown with solid light blue and orange squares.  The primary and secondary components of systems fit with two velocity components are shown with light blue/orange squares outlined with dark blue/red and open squares outlined with dark blue/red, respectively.  Error bars show 
the span of velocities included in the $\Delta v_{90}$ measurement for the \ion{Ca}{2} K and \ion{Na}{1} 5891 lines.
The gray curves indicate the radial velocity required to escape an $M_h = 10^{10}M_{\odot}$ halo (a conservative minimum escape velocity given the stellar mass distribution of our sample) as a function of total distance from the halo center.  The projected velocity of all detected absorption is well below this threshold.  The rightmost panel shows the distribution of best-fit $b_{\rm D}$ values for all components in our \ion{Ca}{2} (cyan) and \ion{Na}{1} (orange) absorbers.  The median value of each distribution is shown with a vertical dashed line.   \label{fig:vel_rperp}}
\end{figure*}

The best-fit component velocities (relative to \zha) of each absorption system with  a total $W_r > 3\sigma_{W_r}$ are shown in Figure~\ref{fig:vel_rperp}
vs.\ projected distance from the associated galaxy host.  The 
uncertainty interval on each point is set to extend from $\delta v_{\rm 90,left}$ to $\delta v_{\rm 90, right}$ to indicate the velocity space covered by each system.
 We remind the reader that \zha\ need not be equivalent to the average redshift of each host galaxy but rather indicates the redshift of nebular emission along the same sightline as that used to probe the absorbing gas.  We therefore interpret absorption with velocities very close ($|\delta v| \lesssim 10~\rm km~s^{-1}$) to \zha\ as interstellar material lying in the host galaxy's disk and rotating with its \ion{H}{2} regions.   We assume that absorbers with larger velocity offsets (or extents) may be extraplanar in nature and/or part of ongoing bulk outflow from or inflow toward the disk.
This rough velocity criterion is motivated by the theoretical considerations laid out in Section~\ref{sec:model}, where they will be further refined (to account for the $M_*$ and $R_{\perp}$ of each system, as well as uncertainties in foreground galaxy orientation).  
For comparison, the detailed study of the spatially resolved velocity distributions of disk and extraplanar absorbers in the nearby galaxy M33 by  \citet{Zheng2017} identified \ion{Si}{4} components having velocities within $\pm20\mkms$ of the local \ion{H}{1} 21 cm emission peak as ``disk" absorbers, and uncovered numerous extraplanar absorbers at relative velocities $\pm 30\!-\!110\mkms$.

Among the 20 \ion{Ca}{2} velocity components included in Figure~\ref{fig:vel_rperp}, only three (15\%) have 
$|\delta v| > 50~\rm km~s^{-1}$; 11 (55\%) have $|\delta v| > 20~\rm km~s^{-1}$; and 14 (70\%) have $|\delta v| > 10~\rm km~s^{-1}$. 
The remaining six systems have velocity centroids consistent with galactic disk rotation.  The $\Delta v_{90}$ values for the \ion{Ca}{2} absorbers, however, lie in the range $50~\mathrm{km~s^{-1}}\le \Delta v_{90} \le 180\mkms$, and thus imply the presence of outflowing/inflowing absorbing material in every case.
The \ion{Na}{1} absorbers exhibit component velocity offsets at yet lower rates: among the 17 components shown, only one (6\%) has $|\delta v| > 50~\rm km~s^{-1}$; six (35\%) have $|\delta v| > 20~\rm km~s^{-1}$; 
and only eight (47\%) have $|\delta v| > 10~\rm km~s^{-1}$. These profiles are all likewise kinematically broad ($50~\mathrm{km~s^{-1}} \le \Delta v_{90}(\mbox{\ion{Na}{1} 5891}) \le 150\mkms$), suggesting that the ongoing fountain motions traced by \ion{Ca}{2} also include a cold component.

For reference, Figure~\ref{fig:vel_rperp} shows the radial velocity that would be required to escape a dark matter halo having $M_h = 10^{10}M_{\odot}$, assuming that $R_\perp$ is equal to the total distance ($R$) from the halo center (rather than the projected distance), and that $v_{\rm esc} = \sqrt{2GM_h/R}$.  Our foreground systems have a range in stellar mass $7.4 \lesssim \log M_*/M_{\odot} \lesssim 10.6$, implying they range in halo mass over $10.3 \lesssim \log M_h/M_{\odot} \lesssim 12.0$ \citep{Moster2013}; thus, the escape velocity of an $M_h = 10^{10} M_{\odot}$ halo may safely be considered the minimum required for these absorbers to escape from any system in our sample.  With the caveat that our spectroscopy is sensitive only to motion along the line of sight (such that our $\delta v$ values are likely somewhat lower than the three-dimensional velocity of the gas), we find that none of the absorbers in our sample have central velocities close to that required for escape from their host systems.  Moreover, none of the velocity limits of the profiles (indicated by the [$\delta v_{\rm 90, left}$, $\delta v_{\rm 90, right}$] intervals) extend beyond this escape velocity limit.

The rightmost panel of Figure~\ref{fig:vel_rperp}
shows the distribution of the best-fit $b_{\rm D}$ values for our \ion{Ca}{2} and \ion{Na}{1} absorption component sample.  The median value of the former is $18.5~\rm km~s^{-1}$, while the median value of $b_{\rm D}$(\ion{Na}{1}) is $10.5~\rm km~s^{-1}$, close to the resolution of our spectrograph.  
In contrast, the QSO absorption-line study of \ion{Ca}{2} and \ion{Na}{1} absorption in Milky Way disk-halo clouds by \citet{BenBekhti2012} measured median Doppler parameter values of $3.3\mkms$ for \ion{Ca}{2} and $2.1\mkms$ for \ion{Na}{1}, with maximum values of $\approx 10\mkms$ for both ions.  This suggests that the absorbing components in our sample are likely composed of multiple individual ``clouds", and that our $b_{\rm D}$ values are predominantly reflective of turbulent velocity dispersions among these clouds (with a subdominant contribution from thermal broadening).  

\begin{figure}[h]
 \includegraphics[width=\columnwidth]{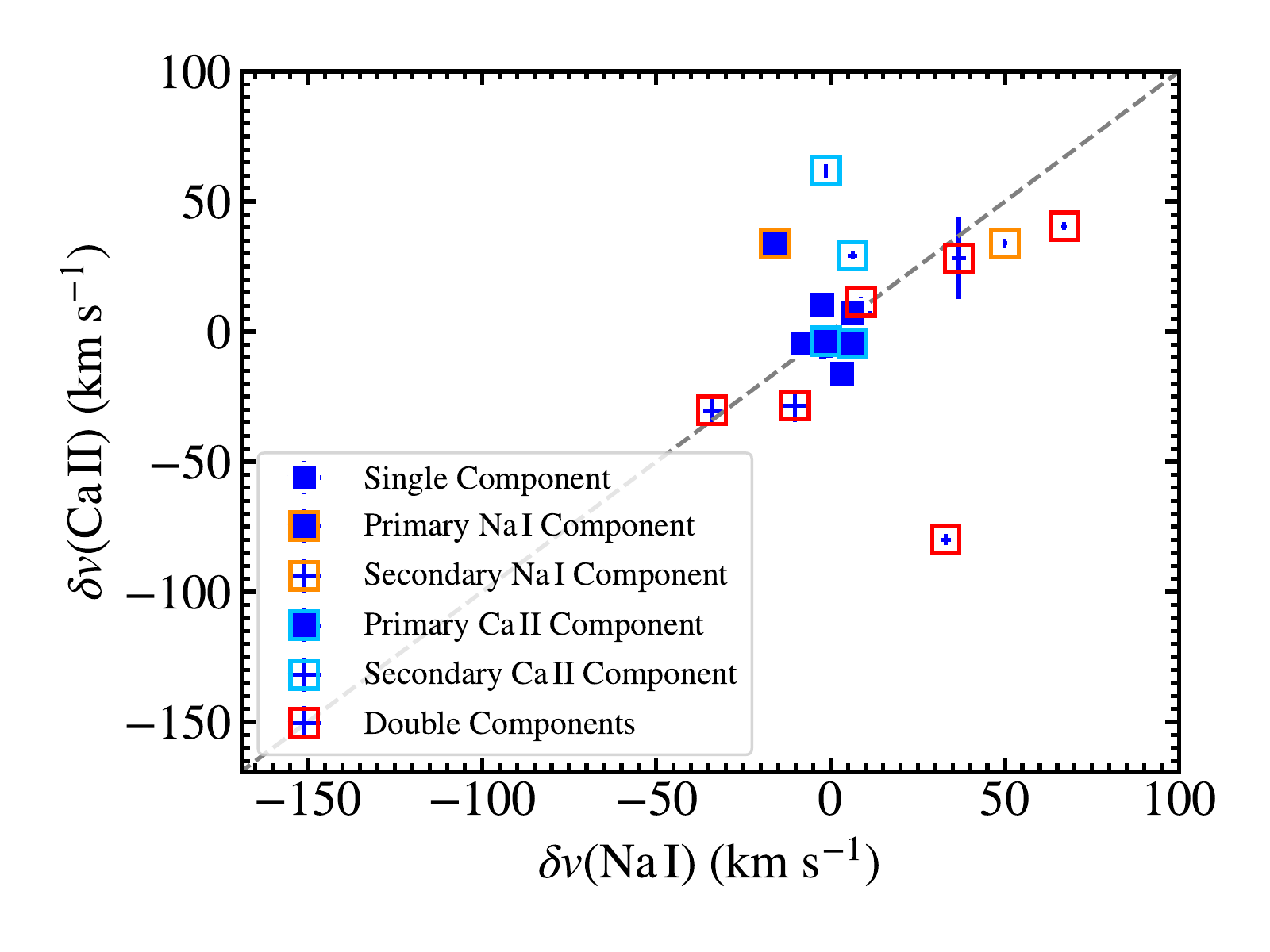}
\caption{Fitted absorption velocity offsets relative to \zha\ for securely detected \ion{Ca}{2} systems vs.\ those for securely detected \ion{Na}{1} systems.  Error bars show the uncertainties in these fitted values.  Absorbers fit with single velocity components in both transitions are shown with solid blue squares.  Absorbers in which \ion{Ca}{2} was fit with one component and \ion{Na}{1} was fit with two components are indicated with solid blue and open squares outlined in orange.  Absorbers in which \ion{Na}{1} was fit with one component and \ion{Ca}{2} was fit with two components are indicated in a similar fashion, as listed in the legend. Velocities for absorbers in which both ions were fit with two components are shown with open red squares.   \label{fig:vel_CaII_NaI}}
\end{figure}

Figure~\ref{fig:vel_CaII_NaI} shows the best-fit $\delta v$ value for each \ion{Ca}{2} component vs.\ the corresponding value of $\delta v$(\ion{Na}{1})  for each system.  Systems for which we have fit \ion{Ca}{2} (or \ion{Na}{1}) with a single component and the other ion with two components appear twice, each with the same $y$- (or $x$-) axis value. 
 There are three systems for which we fit two velocity components to both ions; in these cases, we match components in order of increasing velocity.  We do not require that the $\delta v$ values for \ion{Ca}{2} and \ion{Na}{1} fall within some minimum velocity offset to include them here; instead, we use this figure to assess the degree to which our fitted \ion{Na}{1} and \ion{Ca}{2} components exhibit similar velocities.
The component velocities align closely along many of our sightlines:
the quantity $|\delta v$(\ion{Ca}{2}) $-$ $\delta v$(\ion{Na}{1})$|$ has a median value $10.9\mkms$, and exceeds $25\mkms$ for only four of the 17 component pairs considered.  However, the Pearson correlation coefficient for these measurements is 0.23 with a $P$-value of $37\%$, indicating a relatively high likelihood that uncorrelated data could yield a similar or more extreme coefficient.  If we consider only those systems for which we adopt consistent numbers of components for both \ion{Ca}{2} and \ion{Na}{1}, we measure a Pearson correlation coefficient of 0.35 with a $P$-value of $29\%$.  

Given that our spectroscopy likely cannot resolve the individual absorbing structures producing the observed line profiles, as well as the significant probability that \ion{Na}{1} occurs in fewer of these structures than does \ion{Ca}{2} \citep[e.g.,][]{BenBekhti2012,Bish2019}, our simple approach to modeling these profiles likely obfuscates the velocity alignment of these ions.  Even with this limitation, our dataset points to a relatively high degree of velocity coherence between the two gas phases we trace.  
In Milky Way studies, the kinematics of these ions are typically compared via analysis of the $N(\mbox{\ion{Ca}{2}})/N(\mbox{\ion{Na}{1}})$ ratio as a function of velocity relative to the local standard of rest \citep[LSR; e.g.,][]{RoutlySpitzer1952,Sembach1994,BenBekhti2012}.  This ratio has average values of $N(\mbox{\ion{Ca}{2}})/N(\mbox{\ion{Na}{1}}) \approx 0.69$ at velocities close to the LSR and increases at larger velocity offsets (likely due to the so-called Routly-Spitzer effect; \citealt{RoutlySpitzer1952, Sembach1994}).  While these measurements are not directly analogous to those presented in Figure~\ref{fig:vel_CaII_NaI}, they are similarly suggestive of kinematic coherence of these ions.

\subsection{Relation between $W_r$ and Dust Reddening}

\ion{Na}{1} and \ion{Ca}{2} absorption is known to be correlated with dust across a variety of astrophysical environments, including in the Milky Way ISM and halo \citep[e.g.,][]{Sembach1993,MunariZwitter1997,Poznanski2012,Murga2015} and in external galaxies \citep[e.g.,][]{Wild2005,ChenTremonti2010,Phillips2013,Baron2016,Rupke2021}.  However, current evidence suggests that the strength and form of the relationship between $E(B-V)$ and $W_r$(\ion{Na}{1}) in particular 
depends on the environment probed and/or on the approach to measuring these quantities \citep[e.g.,][]{Rupke2021}.    
Here we investigate the relationship between dust reddening and $W_r$(\ion{Na}{1}) and $W_r$(\ion{Ca}{2}) in our GOTOQ sample, and compare it to that derived for the Milky Way.

We adopt the estimate of $E(B-V)_{(g-i)}$ reported by \citet{Straka2015} for the QSOs in our sample as a proxy for the dust column density associated with each foreground host.  These estimates are based on the observed-frame $(g-i)$ color excess of each QSO relative to the median $(g-i)$ for QSOs at the same redshift in the fourth edition of the SDSS Quasar Catalog \citep{Schneider2007}.  In a study of the relation between QSO colors and the presence and strength of foreground \ion{Mg}{2} absorbers in the SDSS QSO sample, \citet{York2006} found that the QSO color excess $\Delta (g-i)$ is tightly correlated with the dust reddening $E(B-V)$ associated with foreground absorbers and measured from composite QSO spectra shifted into the absorber rest frame.   These authors adopted an SMC reddening law \citep{Prevot1984} to calculate the expected relation $E(B-V)_{(g-i)} = \Delta (g-i)(1+z_{\rm abs})^{-1.2}/1.506$, and found that the average $E(B-V)_{(g-i)}$ in samples of $\gtrsim 100$ objects corresponds closely to the $E(B-V)$ of their composite spectra: $\langle E(B-V)_{(g-i)}\rangle = 0.98 \times E(B-V) - 0.002$.
However, \citet{York2006} also demonstrated that $\Delta (g-i)$ values for individual quasars with no detected foreground absorbers 
exhibit significant scatter with FWHM $\approx 0.27$\,mag\footnote{This quantity is estimated by fitting a Gaussian model to a digitized version of the data in Figure 3 of \citet{York2006}.} (with a mean value $\Delta (g-i) = -0.013$).  This implies an intrinsic dispersion $\sigma_{\rm intr} (\Delta (g-i)) = 0.12$. 

To estimate the total uncertainty in each $E(B-V)_{(g-i)}$ value, we consider both this intrinsic scatter and uncertainty due to measurement error.  \citet{Straka2015} stated that the maximum error in their measurements of apparent magnitudes for both the QSOs and foreground galaxies in their GOTOQ sample is $0.05$ mag.  We therefore assume a measurement error of $\sigma_{\rm meas}(\Delta (g-i)) = 0.07$.  We multiply both $\sigma_{\rm meas}$ and $\sigma_{\rm intr}$ by the quantity $(1+z_{\rm abs})^{-1.2}/1.506$ and add the results in quadrature to compute a total $\sigma_{\rm tot}(E(B-V)_{(g-i)})$ for each GOTOQ sightline.  

\begin{figure*}[ht]
 \includegraphics[width=0.5\textwidth]{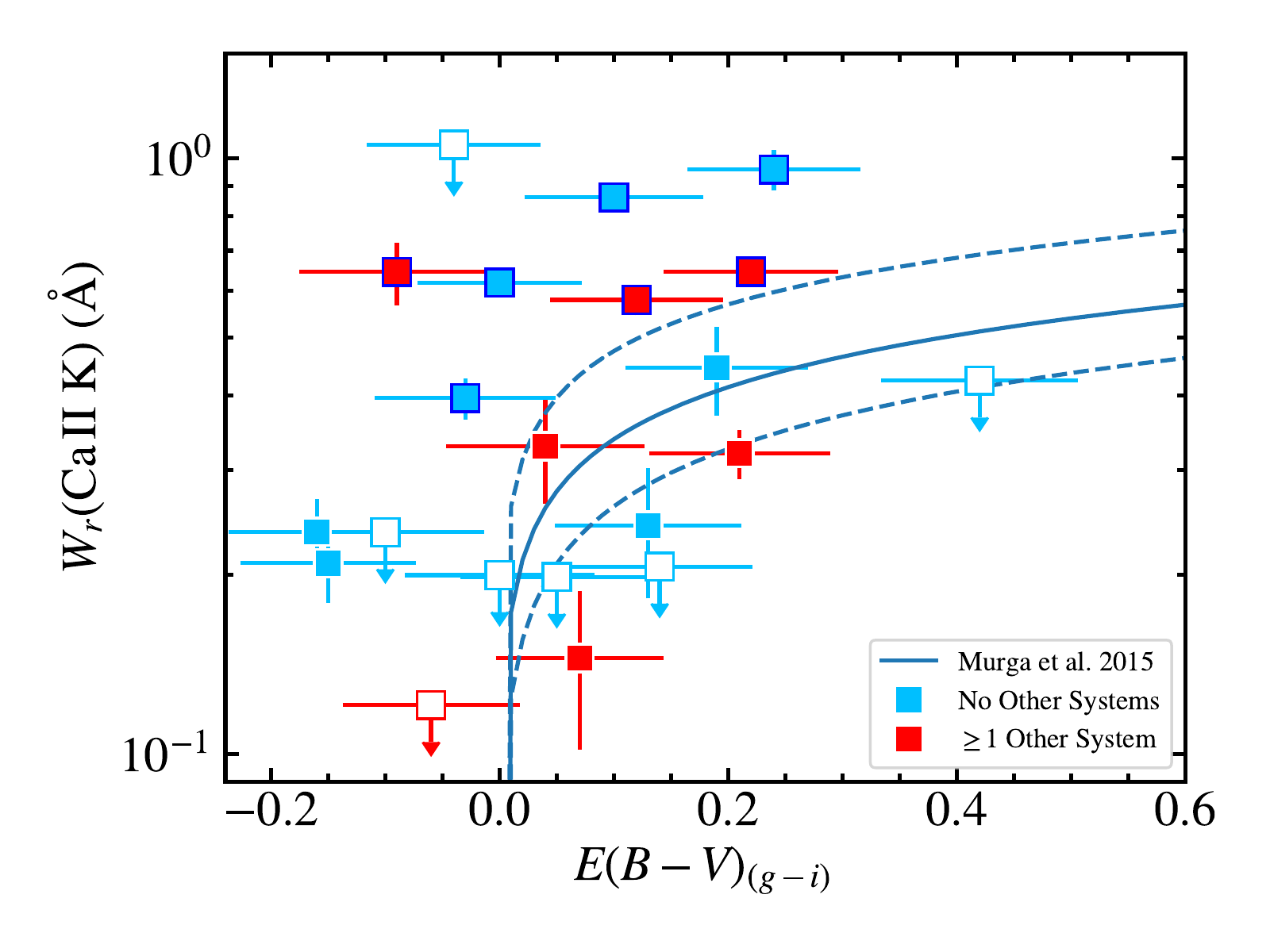}
  \includegraphics[width=0.5\textwidth]{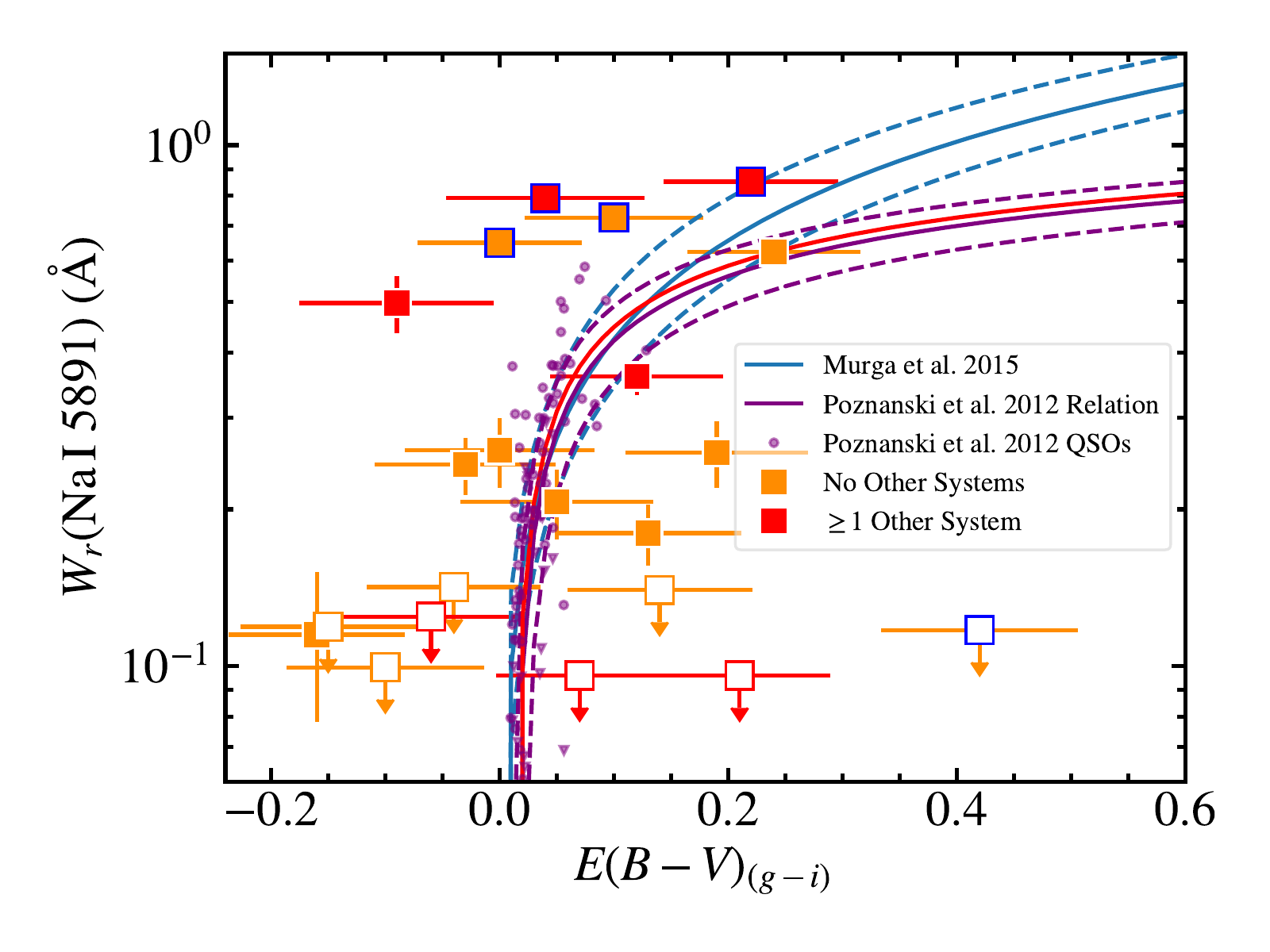}
\caption{ Total system $W_r$(\ion{Ca}{2} K) (left) and $W_r$(\ion{Na}{1} 5891) (right) vs.\ the dust reddening measured along the QSO sightline, $E(B-V)_{(g-i)}$. Upper limits, indicated with open squares, are shown in cases in which $W_r < 3\sigma_{W_r}$, and represent 3$\sigma$ limits. Sightlines shown in light blue and orange have no intervening systems that are unassociated with the known foreground galaxy.  Sightlines indicated in red exhibit one or more unassociated intervening systems in their SDSS spectra.  The solid blue curves show the best-fit relations between the $W_r$ of these ions due to the Milky Way's ISM/halo and dust reddening measured by \citet{Murga2015}.  The dashed blue curves represent the $\pm1\sigma$ uncertainties in these fits.  The purple curves show the best-fit relation (and the $\pm 1\sigma$ uncertainty in the relation) between $W_r$(\ion{Na}{1} 5891) and dust reddening in the Milky Way measured by \citet{Poznanski2012}.  
The small purple circles/triangles show $W_r(\mbox{\ion{Na}{1} 5891})$ values/3$\sigma$ upper limits measured from high-resolution QSO spectra by \citet{Poznanski2012}, plotted vs.\ the reddening toward that coordinate in the \citet{Planck2016} dust map.
Data points outlined in dark blue are offset by $>3\sigma$ from the closest point on the best-fit \citet[][left]{Murga2015} and \citet[][right]{Poznanski2012} relations.
\label{fig:ew_EBV}}
\end{figure*}

Figure~\ref{fig:ew_EBV} shows  $E(B-V)_{(g-i)}$ estimates for our sample with error bars indicating $\sigma_{\rm tot}(E(B-V)_{(g-i)})$ vs.\ the  total $W_r$ of \ion{Ca}{2} K and \ion{Na}{1} 5891 for each system.
Light blue and orange points indicate sightlines lacking any intervening absorbers (other than the system associated with \zha).  Red points indicate sightlines along which between one and nine unassociated intervening absorbers were detected in their SDSS spectra by \citet{Straka2015}.  These seven QSOs may be subject to some additional reddening from these intervening absorbers, although \citet{Straka2015} found that dust in the GOTOQs themselves is likely the dominant source of attenuation for these systems.

We first note that there is no relationship between $E(B-V)_{(g-i)}$ and either $W_r$(\ion{Ca}{2} K) or $W_r$ (\ion{Na}{1} 5891) evident among our GOTOQ sample.  The distribution of $W_r$ values in subsamples having $E(B-V)_{(g-i)} < 0.05$ and $E(B-V)_{(g-i)} > 0.05$ have medians of $W_r$(\ion{Ca}{2} K) $=0.28$ \AA\ and $0.38$ \AA, respectively, with dispersions of 0.18--0.31 \AA, and medians of $W_r$(\ion{Na}{1} 5891) $=0.18$\,\AA\ and $0.22$ \AA, with dispersions of 0.29--0.31 \AA\ (adopting the measured values of $W_r$ for all sightlines, rather than upper limits for nondetections).  We therefore are not sensitive to any significant shift in these distributions between low and high reddening values.  

We also assess the degree to which our dataset is consistent with the average relationships between dust reddening and \ion{Ca}{2}/\ion{Na}{1} absorption strength in the Milky Way.  These relationships have been investigated both in works using high-resolution spectroscopy of samples of $<100$ QSOs or early-type stars \citep[e.g.,][]{Richmond1994,MunariZwitter1997}, and more recently in studies
taking advantage of the ${>}100,000$ QSO spectra and ${>}800,000$ galaxy spectra obtained over the course of the SDSS \citep{Abazajian2009}.  These latter works \citep{Poznanski2012,Murga2015} grouped these spectra into bins based on the dust reddening of each source implied by the \citet{SFD98} map of the dust distribution across the sky.  They then constructed the median stack of the spectra in each bin and measured the $W_r$ of \ion{Ca}{2} H \& K (in the case of \citealt{Murga2015}) and the $W_r$ for both \ion{Na}{1} doublet transitions in each stack.  The best-fit relations between $E(B-V)$ and $W_r$ of the relevant transition reported in these studies are included as solid curves in Figure~\ref{fig:ew_EBV}.  Dashed curves show the same relations with the best-fit parameters offset by their $\pm1\sigma$ uncertainties.
Also included in the right-hand panel of Figure~\ref{fig:ew_EBV} are $W_r$(\ion{Na}{1}) measurements reported by \citet{Poznanski2012} for a small sample of high-resolution QSO spectra.  We estimate the reddening of these sources by querying the \citet{Planck2016} dust map available with the \texttt{dustmaps} Python package \citep{Green2018}.

Most of the measurements for our GOTOQ sample are formally consistent with these relationships, given the large uncertainties in our $E(B-V)_{(g-i)}$ estimates.  However, their distribution appears to exhibit significant scatter around these relationships, and indeed more dispersion than the \citet{Poznanski2012} sample of individual $W_r$(\ion{Na}{1}) measurements.
To quantitatively identify outliers in our sample, we first determine the closest point on each best-fit relation ($x_j$, $y_j$) to that of each data point (i.e., such that the Euclidean distance $d_j = \sqrt{(E(B-V)_{(g-i), j} - x_j)^2 + (W_{r,j} - y_j)^2}$ is minimized).  For sightlines that did not yield significant detections of a given ion, we use the formally measured value of $W_r$ (rather than its upper limit) to compute $d_j$.  We then determine the significance of the distance $d_j$ by computing 

\[
    \mathcal{N}(\sigma_{d, j}) = \sqrt{\left (\frac{E(B-V)_{(g-i), j} - x_j}{\sigma_{\rm tot}(E(B-V)_{(g-i), j})} \right )^2 + \left (\frac{W_{r,j} - y_j}{\sigma_{W_{r,j}}} \right )^2}.
\]

The seven systems for which $\mathcal{N}(\sigma_{d, j}) > 3$ 
relative to the best-fit \citet{Murga2015} relation for \ion{Ca}{2}  
are outlined in dark blue in Figure~\ref{fig:ew_EBV} (left).  All of these systems lie at $W_r$ values $\approx 0.2$--0.5 \AA\ higher than that implied by the QSO's dust reddening level.  We outline in dark blue the five \ion{Na}{1} systems for which $\mathcal{N}(\sigma_{d, j}) > 3$ relative to the best-fit \citet{Poznanski2012} relation in the right panel of Figure~\ref{fig:ew_EBV}.  Again, most of these systems have higher $W_r$(\ion{Na}{1} 5891) values than would be predicted by \citet{Poznanski2012}.  The overall high incidence of these outliers (comprising 33\% and 24\% of our \ion{Ca}{2} and \ion{Na}{1} samples, respectively), implies that these best-fit relations may underpredict the amount of low-ion metal absorption associated with low values of $E(B-V)$.  If we apply a $14\%$ recalibration to the $E(B-V)$ values used in \citet{Poznanski2012} and \citet{Murga2015} as recommended by \citet{SchlaflyFinkbeiner2011}, the number of \ion{Na}{1} outliers remains the same, and the number of \ion{Ca}{2} outliers is reduced to six (or 29\% of our sample).

Studies of dust across a range of environments, from the SMC \citep{Welty2006,Welty2012} to the ISM of QSO host galaxies \citep{Baron2016}, have likewise indicated that $E(B-V)$ values of $\gtrsim 0.05$ mag are associated with higher $W_r$(\ion{Na}{1}) than implied by \citet{Poznanski2012}.
As the \citet{Poznanski2012} relation is commonly invoked to estimate the reddening of both type I and II supernovae in combination with measurements of $W_r$(\ion{Na}{1}) in spectroscopy of these objects \citep[e.g.,][]{SmithAndrews2020,Bruch2021,Dastidar2021}, it is important to appreciate potential biases that may arise from this calibration \citep[e.g.,][]{Phillips2013}.  Moreover, given the wide range in stellar masses of our GOTOQ host galaxies, we suggest that our sample may better represent the varied dust and ISM properties of supernova host galaxies than those focused purely on the Milky Way, SMC, or QSO host systems.






\subsection{Relations between Absorption Strength and Host Galaxy Properties}\label{subsec:ew_dv_Mstar_SFR}

\begin{figure*}[ht]
 \includegraphics[width=0.5\textwidth]{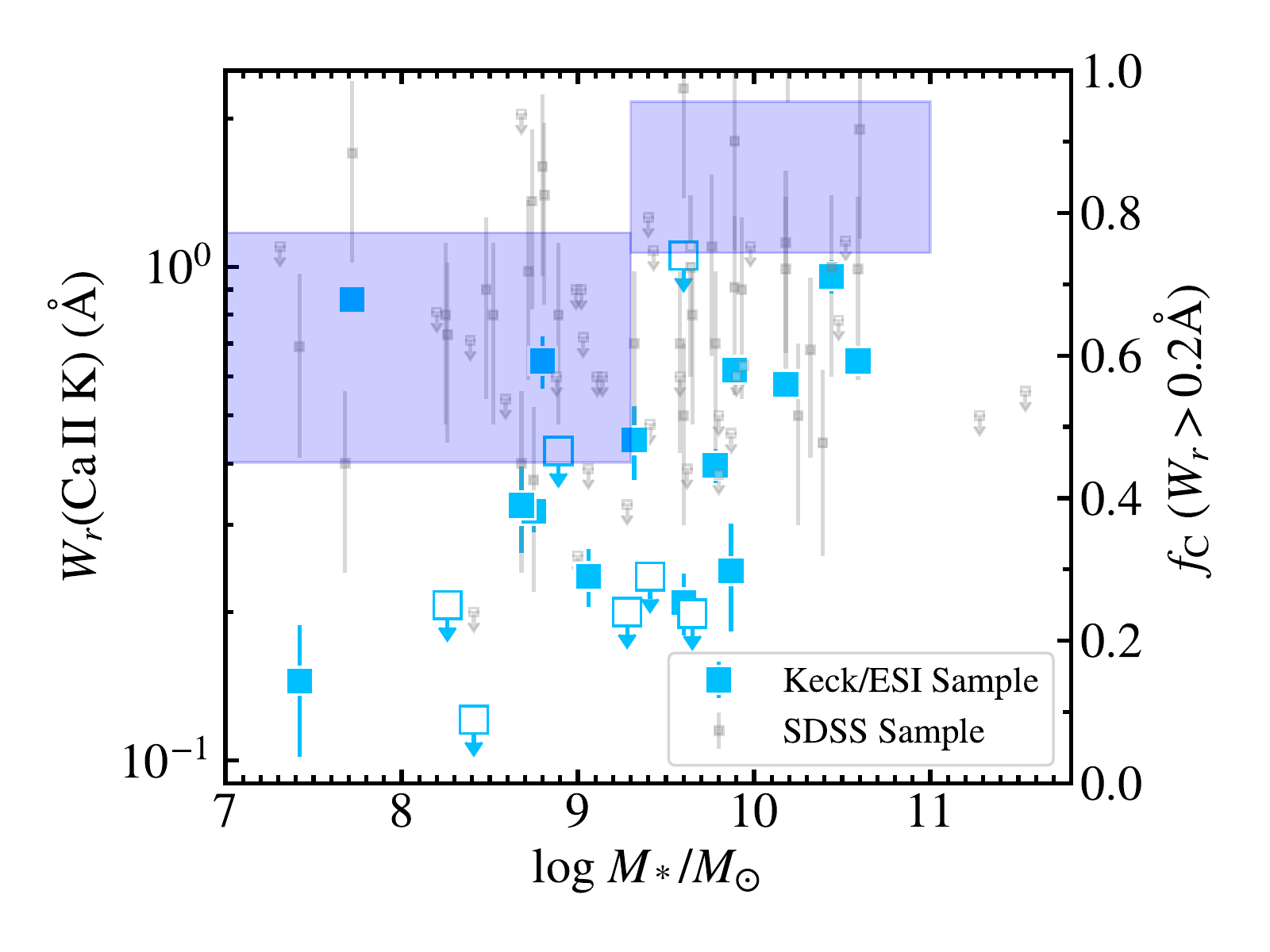}
  \includegraphics[width=0.5\textwidth]{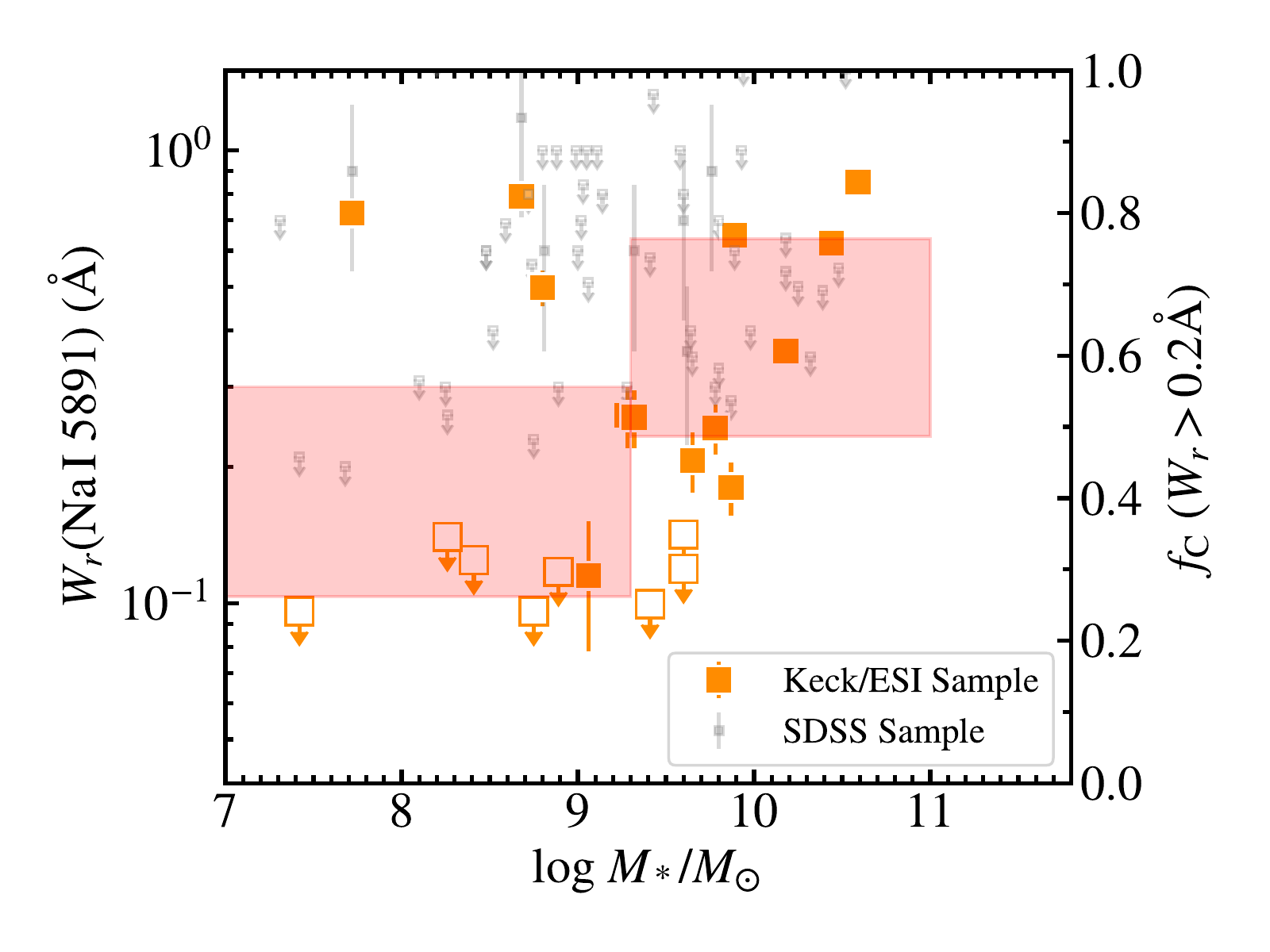}
  \includegraphics[width=0.5\textwidth]{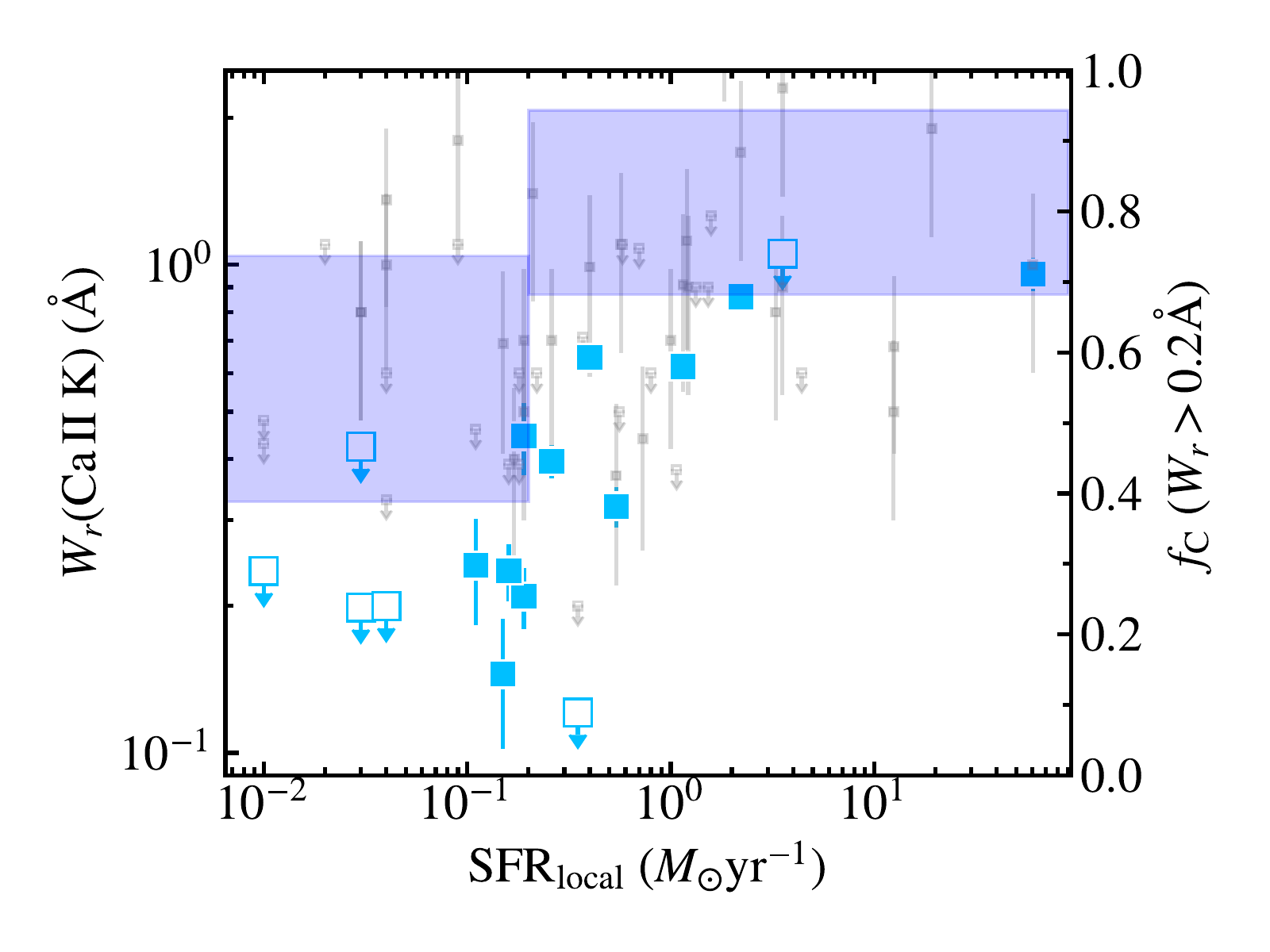}
  \includegraphics[width=0.5\textwidth]{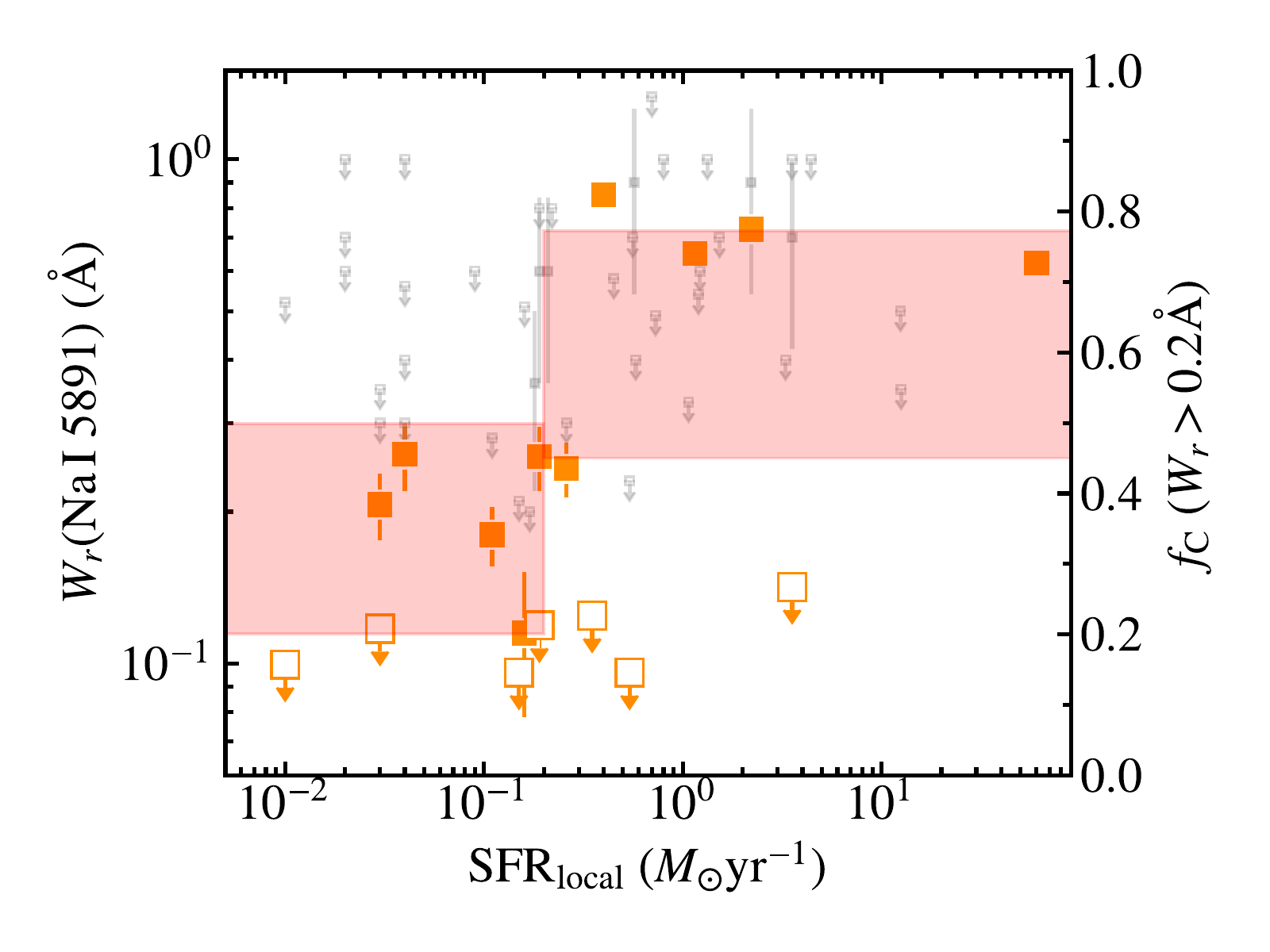}
\caption{ Total system $W_r$(\ion{Ca}{2} K) (left) and $W_r$(\ion{Na}{1} 5891) (right) vs.\ $\log M_*/M_{\odot}$ (top row) and $\rm SFR_{local}$ (bottom row) measured for the foreground host system.  Large colored points indicate constraints from our ESI spectroscopy.  Upper limits, indicated with open squares, are shown in cases in which $W_r < 3\sigma_{W_r}$ and represent 3$\sigma$ limits.  Gray points show measurements reported in \citet{Straka2015} for the parent GOTOQ sample. The filled boxes indicate the $\pm34$th percentile Wilson score confidence intervals, with respect to the right axes, for the covering fraction of absorbers having $W_r > 0.2$ \AA\ in our ESI sample.
\label{fig:ew_SFR_Mstar}}
\end{figure*}

Here we investigate the relationships between the $W_r$ of  \ion{Ca}{2} and \ion{Na}{1} absorption and the stellar masses and local star formation activity of the associated foreground host galaxies. 
Figure~\ref{fig:ew_SFR_Mstar} shows our   total system $W_r$(\ion{Ca}{2} K) and $W_r$(\ion{Na}{1} 5891) measurements vs.\ $\log M_*/M_{\odot}$ (top row) and {\bf $\rm SFR_{\rm local}$} (bottom row).  
Our $W_r$(\ion{Ca}{2} K) values appear to exhibit correlations with both  $\rm SFR_{\rm local}$ and $M_*$.  The Pearson correlation coefficient for the relationship between our directly measured $W_r$(\ion{Ca}{2} K) values and foreground galaxy local SFR is $\rho_{\rm P} = 0.61$ with a $P$-value $=0.009$, indicative of a relation that is close to linear and a very low probability that these variables are uncorrelated.  
 If we exclude the system with the highest-$\rm SFR_{local}$ value (of $60.3~M_{\odot}~\rm yr^{-1}$) from this analysis, we find a $\rho_{\rm P} = 0.50$ with a $P$-value $=0.05$, confirming that this correlation is not driven solely by a single extreme system.
For the relationship between $W_r$(\ion{Ca}{2} K) and $\log M_*/M_{\odot}$, we find $\rho_{\rm P} = 0.35$ with $P=0.12$, which does not rule out the null hypothesis that these variables are uncorrelated.   
Our $W_r$(\ion{Na}{1} 5891) measurements, shown in the right panels of Figure~\ref{fig:ew_SFR_Mstar}, 
yield correlation coefficients of $\rho_{\rm P} = 0.35$ and $0.27$ when considered vs.\ {\bf $\rm SFR_{local}$} and $\log M_*/M_{\odot}$, respectively, with associated $P$-values in the range $0.17 \le P \le 0.24$.  These values likewise do not rule out a lack of correlation between these quantities.  

We also assess the covering fraction of strong \ion{Ca}{2} and \ion{Na}{1} absorbers as a function of $\rm SFR_{local}$ and $M_*$.  Here, we consider strong systems to have $W_r > 0.2$\,\AA\ and divide our sample into bins at the median values $\log M_*/M_{\odot} = 9.3$ and ${\rm SFR_{local}} = 0.2~M_{\odot}~\rm yr^{-1}$.  
We calculate the incidence and corresponding uncertainty intervals of strong absorbers in each bin as described in Section~\ref{subsec:results_cf} and show the results with filled boxes in Figure~\ref{fig:ew_SFR_Mstar}.  
Our $f_{\rm C}$ estimates do not differ significantly at low vs.\ high $\rm SFR_{\rm local}$ or stellar mass.  Instead, we find that even systems having $\log M_*/M_{\odot} < 9.3$ have $f_{\rm C}(W_r($\ion{Ca}{2} K$)>0.2~\mathrm{\AA}) = 0.63^{+0.14}_{-0.18}$ and $f_{\rm C}(W_r($\ion{Na}{1} 5891$)>0.2~\mathrm{\AA}) = 0.40^{+0.16}_{-0.14}$.  We measure similar covering fractions for systems with $\mathrm{SFR_{local}}<0.2~M_{\odot}~\rm yr^{-1}$: $f_{\rm C}(W_r($\ion{Ca}{2} K$)>0.2~\mathrm{\AA}) = 0.57^{+0.17}_{-0.18}$ and $f_{\rm C}(W_r($\ion{Na}{1} 5891$)>0.2~\mathrm{\AA}) = 0.33^{+0.17}_{-0.13}$.  These fractions suggest that both transitions may be utilized to trace ISM kinematics in down-the-barrel spectroscopy across the galaxy population, including in systems with $\log M_{*}/M_{\odot} \lesssim 9.0$ \citep[e.g.,][]{SchwartzMartin2004}.

\begin{figure}[h]
 \includegraphics[width=\columnwidth]{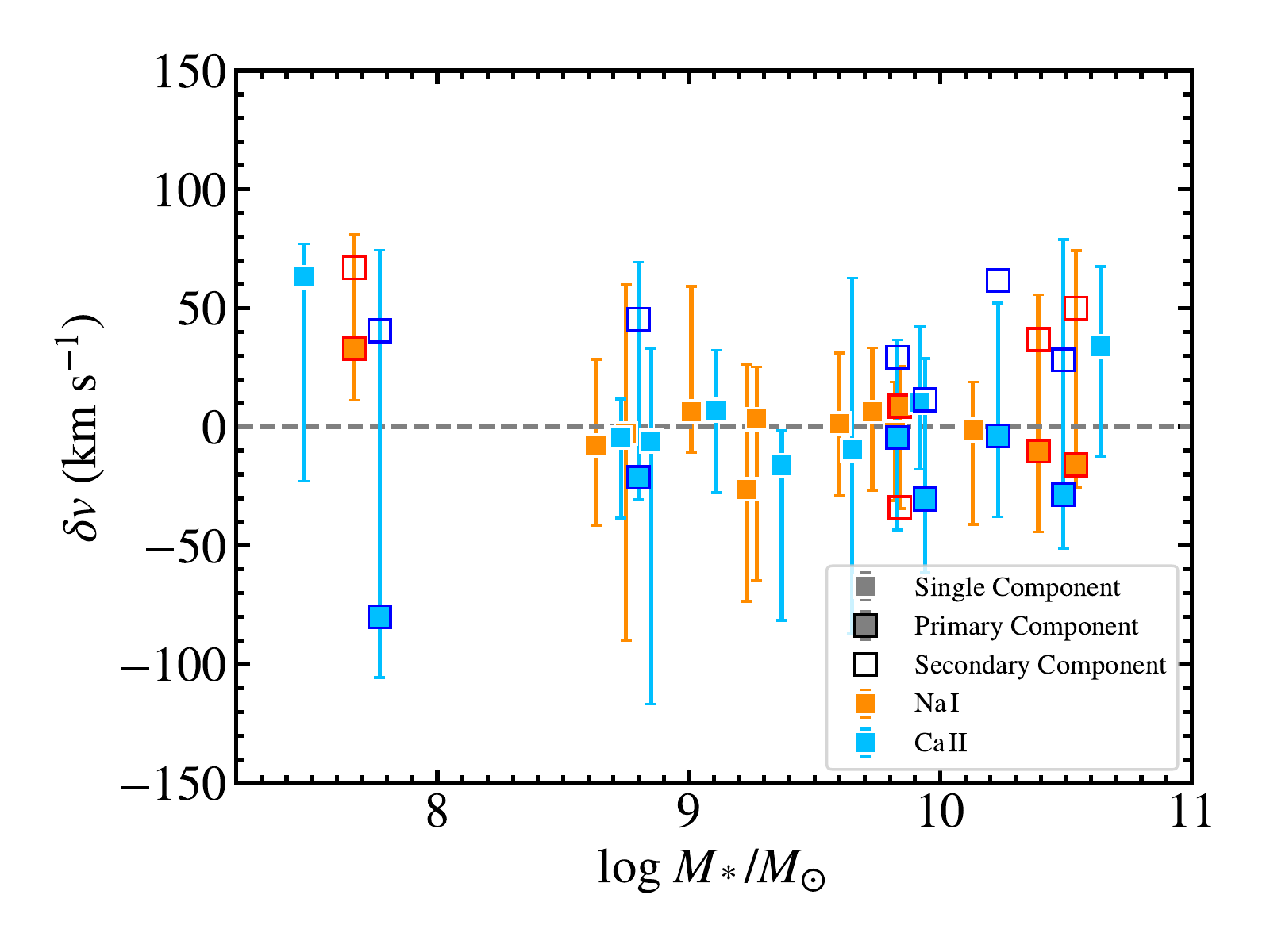}
\caption{Fitted component velocity offsets relative to \zha\ for \ion{Ca}{2} (light blue) and \ion{Na}{1} (orange) absorbers.  Systems fit with single velocity components are shown with solid light blue and orange squares.  The primary and secondary components of systems fit with two components are shown with filled and open squares outlined in a complementary color.  Error bars show the 
span of velocities included in the $\Delta v_{90}$ measurement for the \ion{Ca}{2} K and \ion{Na}{1} 5891 lines.  Symbols have been offset horizontally by $\pm 0.1$ for clarity.  
\label{fig:vel_Mstar}}
\end{figure}

Finally, we investigate the relationship between our $\delta v$ measurements for individual absorption components (presented in Section~\ref{subsec:results_kinematics}) and both $\log M_*/M_{\odot}$ and $\rm SFR_{local}$.  We show the former in Figure~\ref{fig:vel_Mstar}.  While we do not uncover notable trends in either of these relations, this figure highlights the relatively high velocity offsets ($\delta v \sim 33$--$80\mkms$) of all primary and secondary components detected close to the two lowest-$M_*$ foreground systems in our sample (having $\log M_*/M_{\odot} < 8$).  Among the 27 single/primary component velocities shown, only one other system has a primary component velocity offset $>33\mkms$.  Because such large $\delta v$ values are unusual at $\log M_*/M_{\odot} > 8.5$, and given the high equivalent widths of the absorption associated with one of these sightlines (GOTOQJ1238+6448 has $W_r$(\ion{Ca}{2} K) $= 0.86 \pm 0.04$ \AA\ and $W_r$(\ion{Na}{1} 5891) $= 0.73\pm 0.03$ \AA), we
speculate that these absorbers may in fact be associated with other nearby systems that failed to give rise to line emission that could be detected in the SDSS or ESI spectra. 
Alternatively, this absorption may be tracing either outflowing material or ongoing accretion.  

Regardless of whether we exclude these very low-$M_*$ systems from our sample, we measure a statistically significant correlation between the local SF activity in our foreground galaxies and $W_r$(\ion{Ca}{2} K) (i.e., the subsample having $\log M_{*}/M_{\odot} > 8$ yields $\rho_{\rm P} = 0.72$ and $P=0.002$).  This finding is reminiscent of the positive correlation between H$\alpha$ flux and $W_r$(\ion{Ca}{2} K) identified among the GOTOQ parent sample by \citet{Straka2015} and is suggestive of a physical link between star formation activity and the strength/velocity spread of \ion{Ca}{2} absorption in the ISM and halo.  We discuss the implications of this finding in Section~\ref{subsec:discussion_CaIISFR}.




\section{A Simple Model of the ISM Contribution to GOTOQ \ion{Ca}{2} and \ion{Na}{1} Column Densities and Kinematics}\label{sec:model}

Our QSO sightline sample is unusual in the context of CGM studies for its close impact parameters (over the range $R_{\perp} = 1$--13 kpc).  A minority of these sightlines lie within the estimated half-light radius of the foreground host, and, as a consequence of our selection technique, all of our sample sightlines lie within the extent of emission from \ion{H}{2} regions and/or an ionized gas layer.   Moreover, it is well known that the \ion{H}{1} component of disk galaxies is greater in radial extent than that of the stellar or ionized gas component (e.g., the ratio $R_{\rm HI}/R_{25}\gtrsim1.5$--2; \citealt{BroeilsRhee1997, Swaters2002,Begum2008,Wang2013,Wang2016}).  
Each of our GOTOQ sightlines is therefore very likely to be probing the warm and/or cold neutral medium within this disk, along with any outflowing or infalling material along the line of sight.  Here we consider the extent to which (1) the column densities we measure are consistent with those of a neutral gas disk having a \ion{Ca}{2} and \ion{Na}{1} distribution similar to that observed in the Milky Way; and (2) the kinematics of our absorber sample are consistent with those predicted for the ISM of galaxies with similar stellar masses.

\begin{figure*}[ht]
 \includegraphics[width=0.5\textwidth]{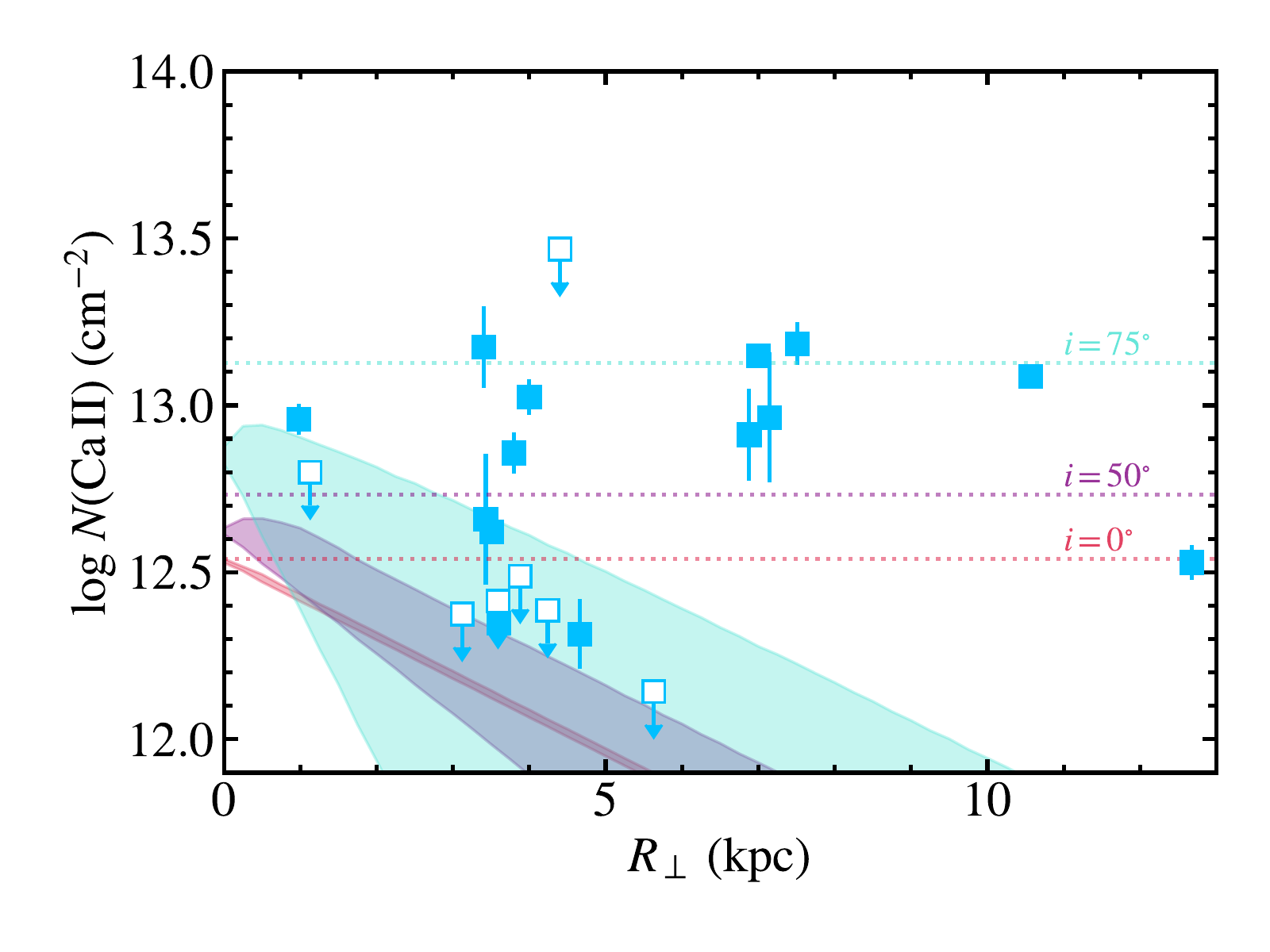}
  \includegraphics[width=0.5\textwidth]{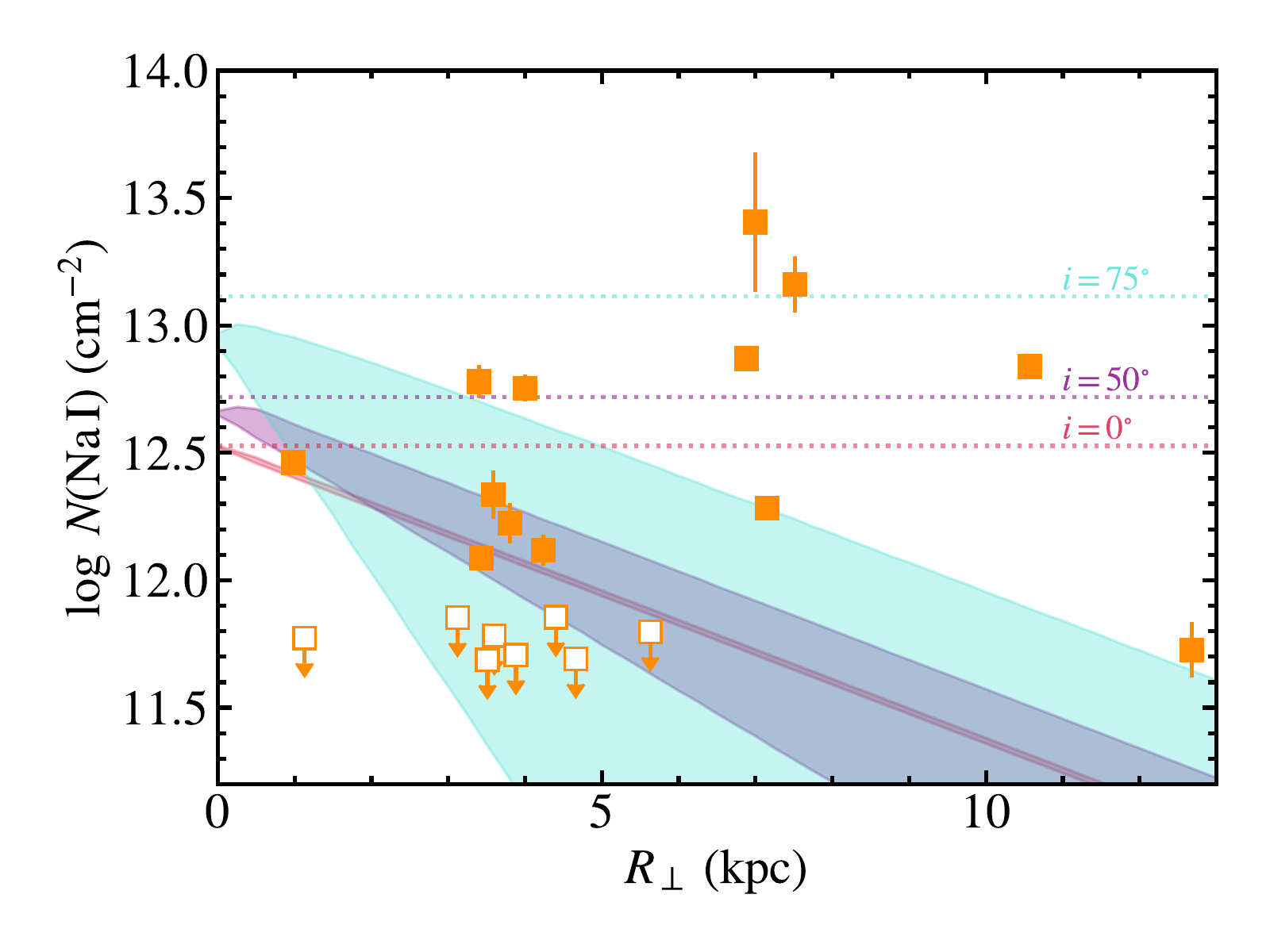}
  \includegraphics[width=0.5\textwidth]{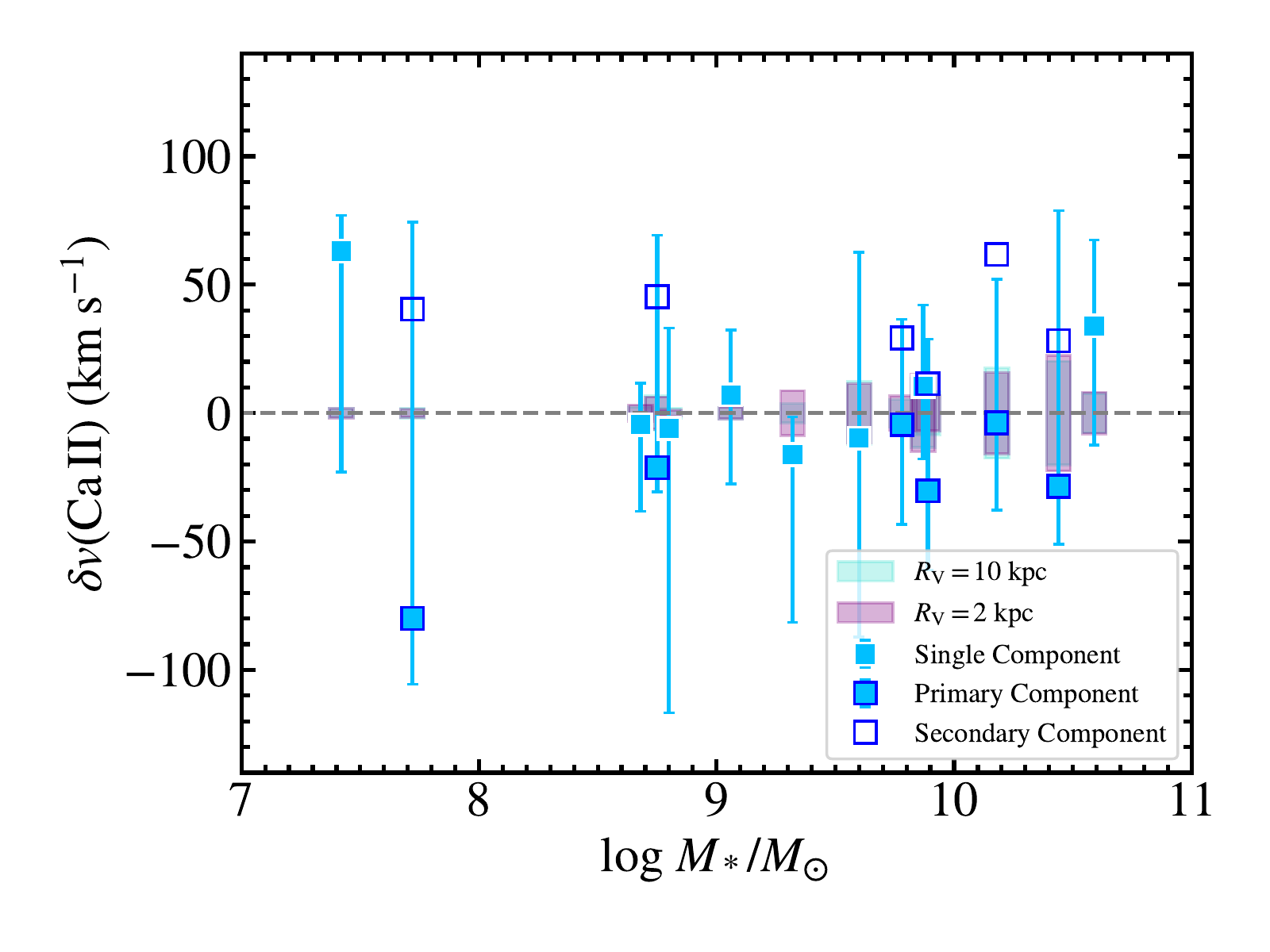}
  \includegraphics[width=0.5\textwidth]{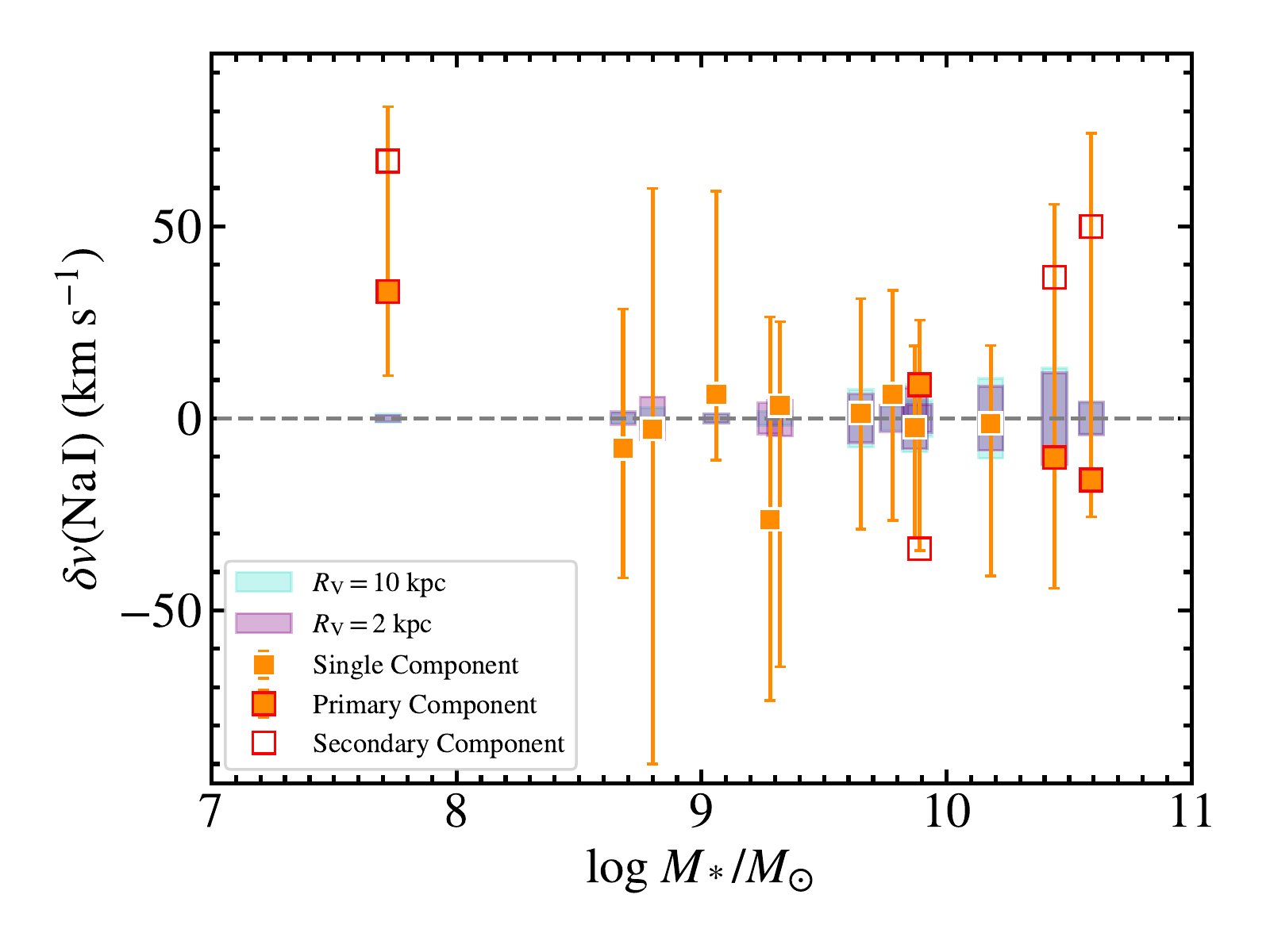}
\caption{{\it Top Row:} Total  system column density of \ion{Ca}{2} (left) and \ion{Na}{1} (right) vs.\ projected distance from the associated GOTOQs.  
Symbols correspond to those used in Figure~\ref{fig:logN_rperp}.  The filled regions indicate the range in column densities predicted for a Milky Way-like ISM observed from an external viewpoint using the simple model described in Section~\ref{subsec:modeling_columndensities}, and assuming three different disk inclinations ($i=0^{\circ}, 50^{\circ}$, and $75^{\circ}$, shown in red, purple, and turquoise, respectively).  The horizontal dotted lines show the value of the central perpendicular column density of the Milky Way disk model, adjusted by a factor $1 / \cos i$. 
{\it Bottom Row:}  Fitted component velocity offsets relative to \zha\ for \ion{Ca}{2} (left) and \ion{Na}{1} (right) absorbers.  Systems fit with single velocity components are shown with solid light blue and orange squares.  The primary and secondary components of systems fit with two components are shown with filled and open squares outlined in a complementary color.  Error bars show the 
span of velocities included in the $\Delta v_{90}$ measurement for the \ion{Ca}{2} K and \ion{Na}{1} 5891 lines.  Colored boxes indicate the maximum projected line-of-sight velocity width predicted using simple tilted-ring models with extraplanar layers placed at $z = \pm 0.82$ kpc (for \ion{Ca}{2}) and $\pm0.43$~kpc (for \ion{Na}{2}).  The maximum rotation velocity ($V_{\infty}$) of each model is set by the stellar mass Tully-Fisher Relation, and the $R_{\rm V}$ parameter is varied to model both steeply rising (purple boxes) and gradually increasing (turquoise boxes) rotation curves.
 \label{fig:logN_rperp_MWmodel}}
\end{figure*}

\subsection{Column Densities}\label{subsec:modeling_columndensities}

It is common in the literature to describe the interstellar density distribution of a given ion as an exponential function that decreases with height $\vert z\vert$ above the Milky Way disk plane: $n(z) = n_0 \mathrm{e}^{-\vert z \vert/h}$ \citep[e.g.,][]{Jenkins1978,Bohlin1978,EdgarSavage1989,Sembach1994,Savage2003,SavageWakker2009}.  The scale height, $h$, and the mid-plane density, $n_0$, may then be constrained by fitting this function to observations of ionic column densities toward samples of Milky Way disk and halo stars (and/or quasars).  \citet{Sembach1993} and \citet{Sembach1994} carried out such a study focusing on \ion{Ca}{2} and \ion{Na}{1}, finding $n_0(\mbox{\ion{Ca}{2}}) =6.85_{-0.41}^{+0.76} \times 10^{-10}~\rm cm^{-3}$, $h(\mbox{\ion{Ca}{2}}) = 0.82_{-0.09}^{+0.07}$\,kpc, 
$n_0(\mbox{\ion{Na}{1}}) =1.27_{-0.18}^{+0.20} \times 10^{-9}~\rm cm^{-3}$, and $h(\mbox{\ion{Na}{1}}) = 0.43_{-0.08}^{+0.12}$\,kpc.
We adopt these values to build our ISM model.  We further assume that the disk density declines exponentially with radius, with the scale radius measured from 21 cm mapping of the Milky Way \ion{H}{1} distribution ($R_{\rm S} = 3.75$ kpc; \citealt{KalberlaKerp2009}).  We may therefore write our adopted disk density distribution as 
\begin{equation}\label{eq:disk_density}
    n(r, z) = n_0 \exp \left [ -\frac{r}{R_{\rm S}} - \frac{\vert z \vert}{h}\right ].
\end{equation}
Given this density distribution, the total column density observed along a quasar sightline passing through the disk oriented at an inclination $i$ at a location ($x$, $y$) may be calculated via the integral 
$N(x, y) = \int n(r, z) ds$, with $ds$ representing the differential length element along the line of sight, and with $r$ and $z$ being dependent on $s$ \citep[e.g.,][]{ProchaskaWolfe1997}.

To compute this integral, we adopt a simplified version of the tilted-ring model framework that is commonly used to model \ion{H}{1} kinematics and surface brightnesses in disk galaxies \citep[e.g.,][]{Rogstad1974,Bosma1978,deBlok2008,Oh2011,Kamphuis2015,Oh2018}.  In the standard, two-dimensional approach, a galaxy's disk is modeled as a series of concentric ellipses.  Each ellipse has an independent central coordinate ($x_{\rm C}$, $y_{\rm C}$), position angle ($\phi$), and inclination ($i$).  Here, we set these parameters to the same value for every ellipse $j$.  To create a three-dimensional model, we replicate this initial set of rings, assigning each set $k$ a thickness $\Delta z = 0.05$ kpc and height $z$ such that the model extends to $z = \pm 10$ kpc.  We assign each ring an ionic volume density according to Equation~\ref{eq:disk_density} and compute the corresponding column density $N_{j, k} = n_{j, k} \Delta z / \cos i$.  We then calculate the ($x$, $y$) coordinates of each ring, interpolating the values of $N_{j, k}$ onto a fixed Cartesian grid.  Finally, we sum
these column densities over all layers to compute $N(x,y)$.

We generate three such models at inclinations $i = 0^{\circ}$, $50^{\circ}$, and $75^{\circ}$.  We then compute the range in $N(x,y)$ values predicted at a given $R_{\perp} = \sqrt{x^2 + y^2}$ for $0~\mathrm{kpc} < R_{\perp} < 15~\mathrm{kpc}$.  We show the resulting column density distributions in the upper panels of Figure~\ref{fig:logN_rperp_MWmodel}, along with the total system column density measurements for our sample (described in Section~\ref{subsec:results_cf}).  For reference, we also show the value $N = 2 n_0 h / \cos i$ with horizontal dotted lines.  
For sightlines in which \ion{Ca}{2} is securely detected, our measurements are typically well above the maximum column densities predicted for a moderately inclined disk (with $i=50^{\circ}$).  Even in the extreme case of a disk inclined to $75^{\circ}$, all six of our sightlines at $R_{\perp} > 6$ kpc yield \ion{Ca}{2} measurements significantly above the projected range of column densities at similarly large impact parameters.  
Our \ion{Na}{1} column densities overall exhibit somewhat greater consistency with our model predictions over the full range of $R_{\perp}$ of our sample; nevertheless, several of our measurements lie well above those predicted for $i = 50^{\circ}$.

Given the simplicity of this modeling, as well as our lack of knowledge of the orientation of our foreground galaxy sample, we cannot use this approach to estimate in detail the contribution of an ISM component to the column densities measured along each sightline.
Indeed, the numerous upper limits we place on $N$(\ion{Na}{1}) within $R_{\perp} < 5$ kpc suggest that our simple model likely overpredicts the \ion{Na}{1} column density in some of our foreground systems and/or does not properly capture the patchiness of \ion{Na}{1} absorption in the ISM.  Moreover, we have assumed here that the volume densities and scale heights of these ions do not vary with overall galaxy stellar mass or SFR.  If, for example, volume density is correlated with mass (as obliquely suggested by the findings presented in Section~\ref{subsec:ew_dv_Mstar_SFR}), our models would tend to overpredict the ISM contribution to the observed column densities, given the stellar mass distribution of our sample.  If the volume density of these ions is instead strongly correlated with global SFR, our modeling may underpredict their ISM column densities in light of the analysis presented in Appendix~\ref{sec:appendix_SFRfrac}.  However, we emphasize that our model predictions for moderately inclined disks lie well below ($>0.9$ dex) every measured $N$(\ion{Ca}{2}) value in our sample at $R_{\perp} > 6$ kpc.
We furthermore consider the former scenario to be more likely, given that our empirical constraints on $M_*$ are significantly more secure than those on the global SFRs of our sample.  

In view of this likelihood, 
we interpret the failure of our ISM model to reproduce the large \ion{Ca}{2} column densities (as well as the largest \ion{Na}{1} column densities) we observe as an indication that  there is a significant contribution to these columns from material that is not interstellar.  These systems must instead arise at least in part from an extraplanar, or circumgalactic, component.  Such absorbers are known to arise in the Milky Way in association with intermediate- and high-velocity \ion{H}{1} clouds, which are understood to lie at distances $\sim 0.5$--20 kpc from the disk \citep{KuntzDanly1996,Wakker2001,Thom2006,Wakker2007,Wakker2008}.  We infer that the phenomena giving rise to these extraplanar or halo clouds are active across our foreground galaxy sample.


\subsection{Kinematics}\label{subsec:modeling_kinematics}

We may also use this framework to predict the distribution of line-of-sight velocities exhibited by the neutral gas disk component of our foreground galaxy sample.  
We again begin with a single set of tilted rings, assigning each ring a rotation velocity
\begin{equation}\label{eq:vrot_r}
    V_{\rm rot} (r) = V_{\infty} \tanh (r / R_{\rm V}),
\end{equation}
with $V_{\infty}$ equal to the maximum rotation velocity, and $R_{\rm V}$ setting the steepness of the rotation curve in the central regions of the disk.  As described in \citet{Rogstad1974} and \citet{Begeman1989}, the line-of-sight component of this velocity is $V_{\rm LOS} (x, y) = V_{\rm sys} + V_{\rm rot}(r) \sin i \cos \theta$, with $\theta$ representing the azimuthal angle counterclockwise from the major axis in the disk plane, and $V_{\rm sys}$ representing the recession velocity of the system.
We then generate two additional, equivalent sets of tilted rings, placing them at heights $z = \pm h$ above and below the first set.  
This placement ensures that the map of line-of-sight velocity differences ($\Delta V_{\rm LOS}$) between these two layers is representative of the maximum velocity offsets that can be produced by a thick galactic disk exhibiting solid-body rotation.  

To generate a kinematic model for each foreground galaxy in our sample, we use the stellar mass Tully-Fisher relation derived by \citet{Bloom2017} 
from spatially resolved H$\alpha$ kinematics of nearby galaxies over the stellar mass range $8.0 < \log M_*/M_{\odot} < 11.5$ in the SAMI Galaxy Survey \citep{Allen2015}:
\begin{equation}\label{eq.TFR}
    \log (V_{\rm rot,TF}/\mathrm{km~s^{-1}}) = 0.31\log (M_*/M_{\odot}) - 0.93.\footnote{This relation is derived from stellar masses calculated by \citet{Taylor2011} for the GAMA Survey.  This work adopted the same cosmology and the same stellar population synthesis models as used in \citet{Straka2015} for stellar mass estimation.}
\end{equation}
Here, $V_{\rm rot,TF}$ is the velocity measured at $2.2R_{\rm eff}$.  This relation was determined from a fit to kinematic data for galaxies with low values of a quantitative asymmetry indicator, and thus may be considered an upper limit on the rotation velocity for lower-$M_*$, dispersion-dominated systems \citep{Bloom2017}.
We calculate the $V_{\rm rot,TF}$ implied by this relation for each foreground galaxy, and then set
$V_{\infty} = V_{\rm rot,TF}$.  
Because the $R_{\rm V}$ parameter in Equation~\ref{eq:vrot_r} is unconstrained for our sample, we generate two models for each system, one with $R_{\rm V} = 2$~kpc (creating a steeply rising rotation curve) and one with $R_{\rm V} = 10$ kpc (creating a gradually increasing rotation curve).
We compute the distribution of $\Delta V_{\rm LOS}$ for both of these models, assuming $i=75^{\circ}$.

Finally, we determine the maximum value of $\Delta V_{\rm LOS}$ predicted at the $R_{\perp}$ of the corresponding GOTOQ (max[$\Delta V_{\rm LOS}$]).  We have indicated these values with colored vertical bars in the bottom panels of Figure~\ref{fig:logN_rperp_MWmodel}.  Each bar is centered at $\delta v = 0\mkms$ and extends to $\pm {\rm max}[\Delta V_{\rm LOS}]/2$.  Note that these bars do not indicate the absolute velocity offset of the material in the layers from $V_{\rm sys}$ (which would extend to many tens of kilometers per second).  Instead, because our \zha\ measurements assess $V_{\rm LOS}(x, y)$ (rather than $V_{\rm sys}$), we are concerned only with the maximum potential velocity offset of extraplanar layers from the former quantity.

As is evident from Figure~\ref{fig:logN_rperp_MWmodel}, the magnitude of max[$\Delta V_{\rm LOS}$] increases with increasing $M_*$ and is larger for \ion{Ca}{2} relative to \ion{Na}{1} due to its larger scale height.  This quantity is also to some extent dependent on $R_{\perp}$, as sightlines that probe locations at which the rotation velocity is increasing steeply with radius are predicted to trace overall larger values of $\Delta V_{\rm LOS}$ (although we find that our predictions are not significantly affected by our choice of $R_{\rm V}$).  However,  regardless of the mass or $R_{\perp}$ of the system, we observe both \ion{Ca}{2} and \ion{Na}{1} absorption over a broader range of velocities than is predicted in this simple framework along nearly every sightline in our sample.  The eight sightlines fit with a single \ion{Ca}{2} component all exhibit $\Delta v_{90}$ values (i.e., the span of the error bars in the bottom panels of Figure~\ref{fig:logN_rperp_MWmodel})  larger than max[$\Delta V_{\rm LOS}$] by $\ge 30\mkms$. Similarly, the nine sightlines fit with a single \ion{Na}{1} component exhibit $\Delta v_{90}$(\ion{Na}{1}) values greater than the corresponding max[$\Delta V_{\rm LOS}$] by $\ge 32\mkms$.
The vast majority of sightlines fit with two \ion{Ca}{2} components or two \ion{Na}{1} components exhibit component velocity differences ($|\delta v_1 - \delta v_2|$) greater than the predicted max[$\Delta V_{\rm LOS}$] by $\ge 20\mkms$. 

The foregoing discussion does not account for the artificial broadening of our observed line profiles due to the finite resolution of our spectrograph (with FWHM $\approx 37.3\mkms$).  \citet{Prochaska2008} performed a detailed comparison of $\Delta v_{90}$ values measured from both ESI and Keck/HIRES spectra of the same QSO sightlines probing foreground damped Ly$\alpha$ systems, finding that $\Delta v_{90}$ measurements obtained from the ESI spectra were larger than those measured with HIRES by about half the FWHM spectral resolution element.  We therefore expect that our $\Delta v_{90}$ measurements may be biased high by $\approx 19\mkms$; however, this level of bias does not reconcile our measurements with the max[$\Delta V_{\rm LOS}$] predictions described above.

In light of the failure of this simple model to reproduce the broad absorption profiles observed, we conclude that the gas kinematics must be broadened by ongoing gas outflow from and/or infall onto the galaxy disks.  Moreover, given that the analysis presented in Section~\ref{subsec:results_kinematics} demonstrated that the bulk of the absorbing material remains within the gravitational potential well of each host, we ascribe the observed motions to Galactic Fountain-like activity.  We discuss the novelty and implications of this conclusion in Section~\ref{subsec:gf}.



\section{Discussion}\label{sec:discussion}

\subsection{The Relationship between Absorption Detected along GOTOQ Sightlines and in Galaxy Spectroscopy}

The rest-frame optical wavelengths of the \ion{Ca}{2} and \ion{Na}{1} transitions studied here have historically made them signatures of choice for studies of the Milky Way ISM \citep[e.g.,][]{Hobbs1969,Hobbs1974,Sembach1993,Welty1996,BenBekhti2012} and the CGM of nearby galaxies \citep{BoksenbergSargent1978,Boksenberg1980,Bergeron1987,Zych2007,Richter2011,ZhuMenard2013}.  Analysis of the \ion{Na}{1} D doublet in nearby galaxy spectroscopy has also provided some of the most important evidence for the ubiquity of cold gas outflows among star-forming systems \citep[e.g.,][]{Heckman2000,SchwartzMartin2004,Rupke2005,Martin2005,ChenTremonti2010,RobertsBorsani2020}.
Much of the literature focusing on this signature targeted galaxies known to be undergoing starburst activity \citep[e.g., by using an infrared luminosity selection criterion;][]{Heckman2000,Rupke2005,Martin2005}, establishing that outflows occur with an incidence that increases with IR luminosity (to $\approx 80\%$ among low-redshift ULIRGs; \citealt{Rupke2005}), and that their typical velocities increase from $10$ to $30\mkms$ among starbursting dwarfs to $100-1000\mkms$ among ULIRGs \citep{Martin2005}.  

Study of \ion{Na}{1} outflow signatures in more typical star-forming galaxies was facilitated by the galaxy spectroscopy obtained over the course of the SDSS
\citep[e.g.,][]{ChenTremonti2010}.  While these spectra typically lack the S/N required for analyses of \ion{Na}{1} kinematics in individual galaxies, multiple studies have taken the approach of coadding many tens or hundreds of spectra to constrain the mean outflow absorption profile as a function of, e.g., stellar mass, inclination, or specific SFR \citep[e.g.,][]{ChenTremonti2010,Concas2019}.

\begin{figure}[h]
 \includegraphics[width=\columnwidth]{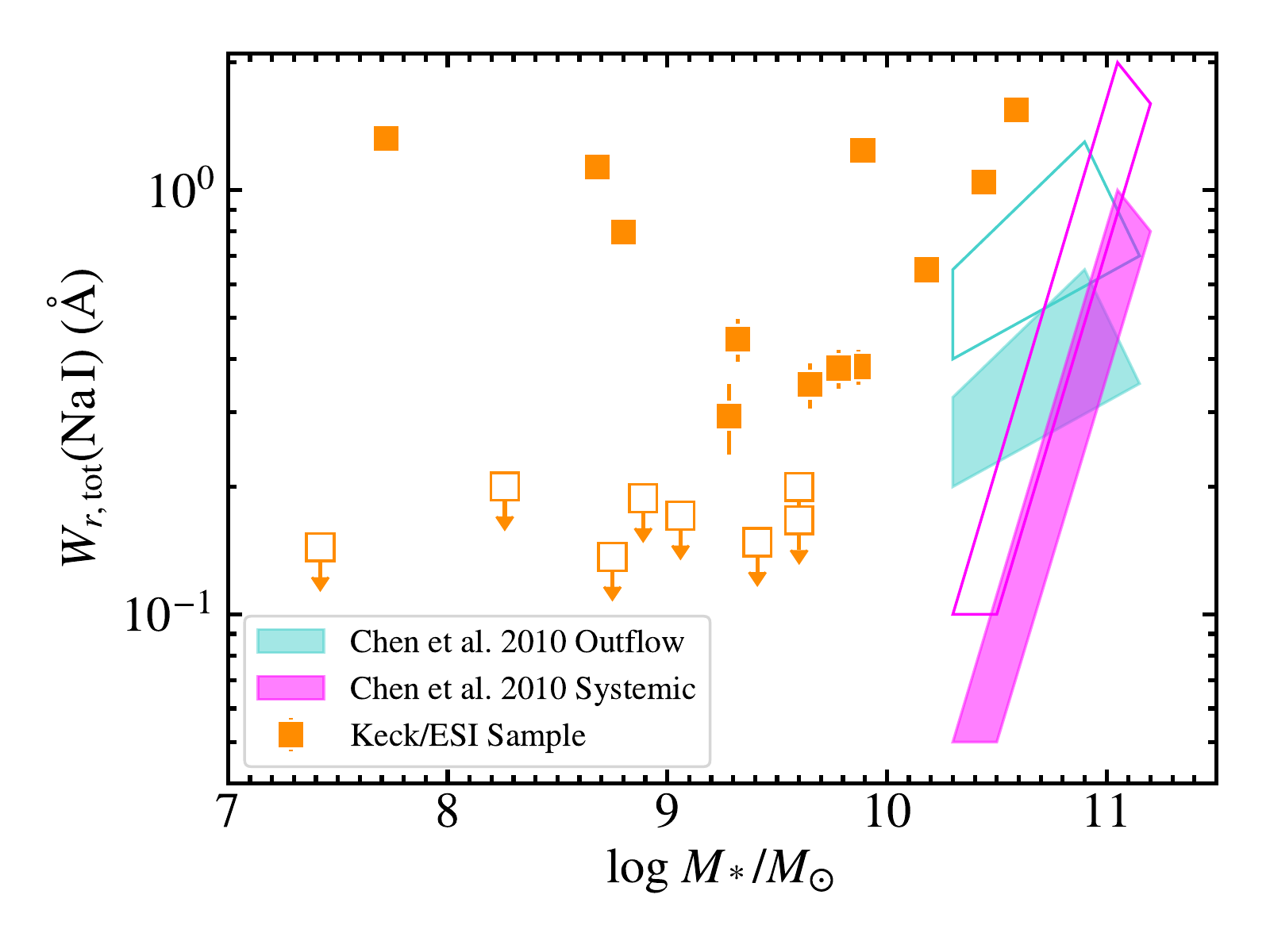}
\caption{The total  system \ion{Na}{1} equivalent width ($W_{r, \rm tot}$(\ion{Na}{1}) = $W_r$(\ion{Na}{1} 5891) $+$ $W_r$(\ion{Na}{1} 5897)) vs. $\log M_*/M_{\odot}$ measured in our ESI spectroscopy.  Upper limits, indicated with open squares, are shown in cases in which $W_{r, \rm tot} < 3\sigma_{W_{r, \rm tot}}$, and represent $3\sigma$ limits.  Measurements of GOTOQJ0851+0791 have been excluded, as the \ion{Na}{1} 5897 profile in that sightline is contaminated by blending.  The turquoise and magenta filled regions show the distribution of $W_r$ values measured by \citet{ChenTremonti2010} in coadded SDSS galaxy spectra for the blueshifted and systemic components of the \ion{Na}{1} absorption profile, respectively. The open regions show the locus of values covered by the \citet{ChenTremonti2010} measurements if they are corrected upward by a factor of two to account for the potential contribution of material located on the far side of the stellar continua.
\label{fig:ewboth_Mstar}}
\end{figure}

Figure~\ref{fig:ewboth_Mstar} compares a subset of these findings with some of the results of our GOTOQ study.  We focus here on measurements reported by \citet{ChenTremonti2010}, as they are most directly comparable to the $W_r$ measured along our GOTOQ sightlines.  In detail, \citet{ChenTremonti2010} divided their $z\sim0.1$ galaxy sample into face-on (with inclinations $i < 60^{\circ}$) and edge-on ($i > 60^{\circ}$) subsamples, then binned each of these subsamples by stellar mass over the range $10.3 \lesssim \log M_* / M_{\odot} < 11.2$.  After coadding the spectra in each of these bins, they performed stellar continuum modeling to remove the component of the \ion{Na}{1} absorption profile arising in stellar atmospheres.  They then modeled the residual \ion{Na}{1} absorption with two velocity components: a ``systemic" component with a central velocity fixed to that of the system and an ``outflow" component with a central velocity that was allowed to vary freely.  They reported the total $W_r$ (including both doublet lines) 
of the systemic components ($W_{r, \rm systemic}$) fit to their edge-on subsamples, and reported the total $W_r$ of the outflow components ($W_{r, \rm outflow}$) fit to their face-on subsamples.  The approximate parameter space covered by these measurements as a function of $M_*$ is indicated in Figure~\ref{fig:ewboth_Mstar} with filled magenta and turquoise shapes, respectively.  \citet{ChenTremonti2010} note that both $W_{r, \rm systemic}$ and $W_{r, \rm outflow}$ increase strongly with $M_*$, and these trends are reflected in the overall slopes of these regions.  Here we compare these values with the total \ion{Na}{1} rest equivalent width $W_{r, \rm tot}(\mbox{\ion{Na}{1}}) = W_r(\mbox{\ion{Na}{1} 5891}) + W_r(\mbox{\ion{Na}{1} 5897})$ measured along each of our sightlines (excluding GOTOQJ0851+0791, for which one of the doublet lines is severely blended).

This comparison reveals that all foreground galaxies in our sample having stellar masses within or close to the range studied by \citet{ChenTremonti2010} exhibit higher values of $W_{r, \rm tot}$(\ion{Na}{1}) than were measured in either the outflowing or systemic components of systems with comparable $M_*$ values.  In detail, we consider here the five GOTOQs having $\log M_*/M_{\odot} > 9.8$. Our measurements for four of these systems are close to a factor of 10 higher than $W_{r, \rm systemic}$ at approximately equivalent stellar masses,  and are $\approx 0.4$--1.0 \AA\ higher than $W_{r, \rm outflow}$.  
This offset may be due in part to the different experimental designs of these two studies: our QSO sightlines probe all absorption along the line of sight, both behind and in front of the foreground host; whereas down-the-barrel spectroscopy is sensitive only to material foreground to the galaxy's stellar populations.  We posit that were we able to use down-the-barrel spectroscopy to probe material arising on the {\it far} side of the galaxies' stellar components  (i.e., the gas along the line of sight that is beyond the galaxies' stars from the point of view of the observer), this would result in a potential increase in the observed $W_r$ by a factor of two (as indicated with the open regions in Figure~\ref{fig:ewboth_Mstar}).  This is likely an overestimate, particularly for $W_{r, \rm systemic}$, as saturated absorbing components with velocities that overlap those observed on the front side would not add to the observed total equivalent width.  Even so, our GOTOQ $W_{r, \rm tot}$ values still lie well above the predicted equivalent widths implied by the \citet{ChenTremonti2010} measurements.

The $3\arcsec$ diameter fibers used for the SDSS spectroscopy extend to radii of $R_{\perp} = 2.5$ kpc at the median redshift of the \citet{ChenTremonti2010} sample ($z=0.09$), whereas the five GOTOQs relevant to this comparison are being probed at impact parameters $R_{\perp} = 3.4, 3.4, 4.0, 7.5$, and $10.6$ kpc, or $R_{\perp}/R_{\rm eff, est} = 0.5, 1.5, 0.7, 0.7$, and 1.6.  While these sightlines are not passing through the galaxy centers, they are nevertheless likely probing star-forming regions in their disks.  The elevated absorption strengths we observe may imply either that (1) there is a significant contribution to the GOTOQ \ion{Na}{1} profiles 
from inner halo material distributed toward the galaxy outskirts; (2) our GOTOQ sightlines do not fully sample the distribution of $W_{r, \rm tot}$(\ion{Na}{1}) values for the galaxy population as a whole due to their small numbers; or (3) the \citet{ChenTremonti2010} absorption strengths are suppressed due to resonantly-scattered \ion{Na}{1} emission or to  overestimation of the contribution of stellar atmospheres to the coadded line profiles.

Galactic fountain models invoking feedback-driven condensation and cooling of material from hot coronal gas may provide a theoretical explanation for the putative detection of excess cold clouds located close to the disk but at projected separations of $>0.5 R_{\perp}/R_{\rm eff}$ from a galaxy's axis of symmetry \citep{Marasco2012,Fraternali2013}.  However, given our relatively small sample, this dataset cannot distinguish between the three scenarios laid out above.  We simply note here that \ion{Na}{1} emission from scattering has been found to be weakest in edge-on galaxies and thus should have a minimal effect on $W_{r, \rm systemic}$ \citep{ChenTremonti2010,RobertsBorsani2020}.  In addition, some degree of ``contamination" of the spectral libraries used to model the stellar continuum by interstellar material in the Milky Way remains a distinct possibility. Comparison to continuum models constructed from purely theoretical stellar spectra suggests this can result in an underestimate of the $W_{r, \rm tot}$(\ion{Na}{1}) arising from the ISM of $\approx 0.5\!-\!0.7$ \AA\  (K. S.\ Parker et al, {\it in prep}).  

Finally, we comment that the opposite effect has been found in comparisons of the $W_r$ of outflows traced by \ion{Mg}{2} $\lambda \lambda 2796, 2803$ absorption in galaxy spectra \citep{Rubin2014} to the strength of circumgalactic \ion{Mg}{2} absorption at impact parameters $10~\mathrm{kpc} \lesssim R_{\perp} \lesssim 170~\mathrm{kpc}$ \citep{Chen2010a}.  The enhanced $W_r$(\ion{Mg}{2}) detected down-the-barrel relative to those typically detected along QSO sightlines suggests that the bulk of the outflowing material does not reach distances of more than ${\sim}10$ kpc.  The $W_r$(\ion{Mg}{2}) values measured along GOTOQ sightlines at impact parameters $R_{\perp} < 6$ kpc by \citet{Kacprzak2013} are closer in strength to those observed in galaxy spectra ($1.75~\mathrm{\AA}<W_r$(\ion{Mg}{2} 2796) $<3.11~\mathrm{\AA}$), implying that the region $\sim6$--10 kpc from a galaxy's center may be an important interface for the stalling of \ion{Mg}{2}-absorbing wind material.  Comparison of our $W_r$(\ion{Na}{1}) measurements to the \citet{ChenTremonti2010} results places no such constraint on the potential extent of \ion{Na}{1}-absorbing winds.

\subsection{The Relationship between $W_r$(\ion{Ca}{2} K) and Star Formation}\label{subsec:discussion_CaIISFR}

We identified a strong, statistically significant correlation between $W_r$(\ion{Ca}{2} K) and the local SFR of the absorber host measured within the SDSS fiber (see Figure~\ref{fig:ew_SFR_Mstar}).  To our knowledge, our study presents the first evidence for such a relationship (although a correlation between H$\alpha$ flux and $W_r$(\ion{Ca}{2} K) was noted in \citealt{Straka2015}).  However, evidence that strong \ion{Ca}{2} absorbers have significant dust content was uncovered more than a decade ago \citep{Wild2005,Zych2009}.  By coadding SDSS QSO spectra in the rest frame of strong \ion{Ca}{2} systems, \citet{Wild2007} detected and characterized associated [\ion{O}{2}] $\lambda\lambda 3727, 3729$ emission arising within the SDSS fibers, measuring an average SFR of 0.1--$0.5~M_{\odot}~\rm yr^{-1}$.  Additional evidence for a connection with star formation was contributed by \citet{ZhuMenard2013}, whose stacking analysis  detected a stronger mean \ion{Ca}{2} absorption signal in the halos of star-forming galaxies relative to red-sequence hosts at fixed stellar mass.

The relation is reminiscent of the now well-established evidence for a correlation between \ion{Mg}{2} absorption strength and SFR.  This is seen both in down-the-barrel studies \citep{Rubin2014,Bordoloi2014}, as well as in QSO absorption-line systems and in QSO-galaxy pair experiments.  \citet{Menard2011} assessed the [\ion{O}{2}] luminosity of \ion{Mg}{2} absorbers as a function of their $W_r$, reporting a 15$\sigma$ correlation between these quantities, and showing that the distribution function of $W_r$(\ion{Mg}{2}) can be related to the [\ion{O}{2}] luminosity function using a simple scaling.  \citet{Lan2014} studied the host galaxy properties of \ion{Mg}{2} absorbers detected in the SDSS QSO sample as a function of their $W_r$, finding that stronger absorbers are surrounded by higher numbers of star-forming galaxies within $R_{\perp} < 50$ kpc.  \citet{LanMo2018} confirmed this connection, measuring larger average $W_r$(\ion{Mg}{2}) within $50$ kpc of emission-line galaxies selected from SDSS-IV/eBOSS with higher SFRs.  

Taken together, these studies provide strong evidence for star formation activity as a primary origin of strong \ion{Mg}{2} absorption.  Our findings, together with the literature reviewed above, are suggestive of a similarly strong link between star formation and \ion{Ca}{2}-absorbing material, which in turn implies that \ion{Ca}{2} may prove an effective tracer of winds in down-the-barrel studies.  Very few such works have made use of \ion{Ca}{2} H \& K for this purpose, due to the blue spectral coverage required, as well as to the strength of these transitions in stellar atmospheres and the potential for confusion from H$\epsilon$ $\lambda 3971$ line emission.  However, if these systematics could be successfully mitigated via detailed stellar population and emission-line modeling \citep[e.g.,][]{Westfall2019}, \ion{Ca}{2} may prove a useful probe of the spatially resolved outflow kinematics of warm, neutral gas in advance of the availability of UV-sensitive IFUs that will map the motions of more highly ionized material in absorption \citep[e.g.,][]{Tumlinson2019}.



\subsection{Galactic Fountains in External Galaxies}\label{subsec:gf}

The Galactic Fountain model was originally introduced to explain the origin of high-velocity clouds (HVCs) of \ion{H}{1} in the Milky Way halo \citep{ShapiroField1976}.  In this picture, a dynamic, hot corona is continually fed and heated by supernova ejecta.  Material lofted above the disk rises and cools, moving outward radially along the pressure gradient of the corona.  Thermal instabilities trigger the condensation of neutral clouds from the hot gas, which purportedly fall back toward the disk on ballistic trajectories.  There have been numerous theoretical investigations exploring the implications of this model for the ionized component of Milky Way HVCs \citep[e.g.,][]{Marasco2013}; the metallicities of HVCs \citep{MarascoFraternali2017}; and the X-ray emitting properties of the Milky Way's coronal plasma \citep{JoungMacLow2006,Henley2015}.  

A now significant body of theoretical work has also invoked this model to predict the properties of the gaseous components of external galaxies.  Surveys of 21 cm \ion{H}{1} emission in nearby star-forming systems having rotation velocities $\gtrsim 80\mkms$ have revealed ubiquitous extraplanar layers of neutral gas that extend to $\gtrsim 1$ kpc above the disk plane  \citep{vanderHulstSancisi1988,Fraternali2002,Oosterloo2007,Marasco2019}, and that typically lag behind the disk rotation speed \citep{Fraternali2002,Barbieri2005}.  This ``anomalous" component arises within the inner few kiloparsecs of the disk, and has 
typical masses of $\sim10^{8-9} M_{\odot}$  \citep{Marasco2019}.  
\citet{FraternaliBinney2008} argued against 
an external (or circumgalactic) origin for this material, given that if it were to fall 
onto the host galaxy disks over a freefall time, the implied accretion rates would be orders of magnitude larger than the SFRs of nearby spirals.  This inconsistency in itself is strongly suggestive of a fountain origin for extraplanar \ion{H}{1} gas \citep{FraternaliBinney2008,Marasco2019}.  While models that adopt purely ballistic trajectories for supernova ejecta fail to reproduce the observed lag in the rotation of extraplanar layers
\citep{FraternaliBinney2006}, a modification of these models that accounts for the interaction between feedback-driven outflow and cool accretion flows successfully explains both the surface brightness and kinematics of the extraplanar material in the two well-studied spirals NGC 891 and NGC 2403 \citep{FraternaliBinney2008}.

The models described above adopt a Gaussian distribution for the velocities of clouds ejected from the galactic disk, with the dispersion adjusted to achieve the closest match between the predicted and observed \ion{H}{1} surface brightnesses.  Another common approach to the modeling of extraplanar \ion{H}{1} layers is to adopt a single value for the layer velocity perpendicular to the disk \citep[e.g.,][]{Marasco2019}.  Both approaches predict velocity widths of $\gtrsim50\!-\!150\mkms$ along individual lines of sight through 21 cm emission-line maps of moderately inclined galaxies \citep{Marasco2019}.  
The $\Delta v_{90}$ widths we have measured imply similar velocity spreads of $>50 - 180\mkms$ across our sample, over the full range of impact parameters we probe ($R_{\perp} = 1\!-\!13$~kpc).  

We have seen from the analysis presented in Section~\ref{subsec:modeling_kinematics} that these velocity widths far exceed those predicted for the interstellar component of these galaxies.  Instead, they are qualitatively consistent with the kinematics predicted by commonly invoked fountain models. Our line profile modeling additionally demonstrates that these velocity widths frequently arise from multiple, distinct structures (i.e., components) with velocity offsets of $>40\mkms$, and arise from both the warm neutral material traced by \ion{Ca}{2} and a colder phase traced by \ion{Na}{1}.
Furthermore, whereas extraplanar \ion{H}{1} is typically well fit by models that adopt exponential scale lengths of $R_{\rm g} = 1\!-\!7$ kpc for the surface density of the layer (i.e., $\Sigma(R)\propto  e^{R/R_{\rm g}}$; \citealt{Marasco2019}), our absorption velocity widths suggest that the physical processes driving galactic fountain flows persist to $R_{\perp} \sim 7$ kpc and beyond.

Finally, we emphasize the novel stellar mass distribution of our sample in this context: six of the 14 foreground galaxies giving rise to securely-detected \ion{Ca}{2}  have $M_*$ values that imply rotation velocities in the range $V_{\rm rot,TF} = 20\!-\!80\mkms$.  The consistently high $\Delta v_{90}$ values we measure across this parameter space provide novel evidence for galactic fountain activity in such low-mass systems. 

A handful of alternative observational approaches have offered additional evidence for galactic fountain flows in external galaxies.  The H$\alpha$ and radio continuum emission from the extraplanar layers of the nearby edge-on spiral NGC 891, along with the properties of the dust complexes that pervade them, have long been interpreted as consistent with galactic fountain model predictions \citep{Rand1990,Dettmar1990,BregmanHouck1997,HowkSavage1997,Kamphuis2007}.  These studies also offer direct evidence for the multiphase nature of the galactic fountain material, with dust-bearing clouds likely tracing a similar phase to that giving rise to \ion{Na}{1} \citep{HowkSavage2000}.  Such multiphase ``interstellar thick disks" are now known to be ubiquitous \citep{ZschaechnerRand2015,Boettcher2016,Bizyaev2017,Li2021} and are observed to exhibit metallicities ranging from a factor of two lower than the host galaxy disk to slightly above that observed in the host
\citep[e.g.,][]{Howk2018}.  Most recently, \citet{Rupke2021} took advantage of echellette-resolution optical IFU spectroscopy of eight nearby AGN-dominated galaxies to trace the down-the-barrel kinematics of these layers, identifying ongoing outflow and inflow in nearly every system via the Doppler shift of \ion{Na}{1}, with the projected areas subtended by these flows covering up to 25\% of the optically bright stellar disks.

Modern theoretical studies of galactic fountain flows have used high-resolution numerical simulations to make detailed predictions for the temperature distribution and kinematics of extraplanar material, drawing physical links between recent-past star formation activity and the launch of expanding superbubbles \citep[e.g.,][]{Creasey2013,Martizzi2016,Girichidis2016,KimOstriker2018,Vijayan2020,Kado-Fong2020}.  While the vast majority of these studies do not simulate material in the coolest phases traced by \ion{Na}{1}, recent work by \citet{Girichidis2021} and \citet{FarberGronke2021} investigated the formation and survival of such cool, dust-enshrouded material explicitly.  The former study found that magnetized, hot wind material can effectively trigger the condensation of a molecular phase from a high-density ($n \gtrsim 0.5~\rm cm^{-3}$), warm ($T\sim10^{3-4}$ K) cloud \citep{Girichidis2021}; while the latter found that this phase can survive over numerous cloud-crushing times if the cloud is sufficiently large, and that dust grains can likewise survive in $\gtrsim 100$ pc clouds if the temperature for dust destruction $T_{\rm dest} > 10^4$ K \citep{FarberGronke2021}.  
These theoretical advances, along with ongoing efforts to link the results of parsec-resolution numerical simulations of galactic disks to simulations encompassing dark matter halo scales \citep[e.g., SMAUG;][]{Kim2020}, will enable detailed comparison of the predictions of these models to the observed kinematics and absorption/emission-line strengths of extraplanar material.
Such comparisons are crucial to affirming these theoretical efforts, as the associated predictions have not yet been rigorously compared to the numerous in-hand observational constraints.

 




\section{Summary and Conclusions}

We have analyzed medium-resolution optical spectroscopy of 21 bright quasars known \emph{a priori} to lie exceptionally close to foreground galaxies having redshifts $0.03 < z <0.20$ with the purpose of assessing the strength and kinematics of \ion{Ca}{2} H \& K and \ion{Na}{1} $\lambda\lambda 5891, 5897$ absorption arising in their ISM and disk-halo interface.
 The foreground systems were identified serendipitously via intervening nebular emission lines in SDSS spectra of the quasars by \citet{Straka2013,Straka2015}, who located their photometric counterparts in SDSS imaging and measured impact parameters in the range  $1~\mathrm{kpc} < R_{\perp} < 13~\mathrm{kpc}$. 
The foreground galaxies span a broad range of stellar masses ($7.4 \le \log M_*/M_{\odot} \le 10.6$), and 
the strength of the H$\alpha$ emission detected in the SDSS fibers implies that their global SFRs lie both within and well above the star-forming sequence at $z\sim 0$.
Our spectroscopy, with a velocity resolution $\rm FWHM\approx37.3\mkms$, is sensitive to absorbers having $W_r(\mbox{\ion{Ca}{2} K}) \gtrsim 0.2$ \AA\ and $W_r(\mbox{\ion{Na}{1} 5891}) \gtrsim 0.15$~\AA.  We used Voigt profile modeling to derive column densities, Doppler parameters, and component velocities for each securely detected system.  We also calculated a nonparametric measure of the profile velocity widths ($\Delta v_{90}$).  Our analysis has revealed the following:

\begin{itemize}
    \item We find no evidence for an anticorrelation between the $W_r$ values we measure and either $R_{\perp}$ or $R_{\perp}/R_{\rm eff, est}$ (i.e., the impact parameter normalized by an estimate of the effective radius of the foreground host galaxy).  
    Modeling of the relation between
    $\log W_r(\mbox{\ion{Ca}{2} K})$ (or $\log W_r(\mbox{\ion{Na}{1} 5891})$) and either measure of projected separation as linear yields slopes that do not significantly differ from zero.  This is unusual in the context of the QSO-galaxy pair studies literature, which in the vast majority of cases report statistically significant anticorrelations between $W_r$ and $R_{\perp}$ at larger impact parameters than we probe ($15~\mathrm{kpc} \lesssim R_{\perp} \lesssim 100~\mathrm{kpc}$).
    \item Strong absorption with column densities $N(\mbox{\ion{Ca}{2}})> 10^{12.5}~\rm cm^{-2}$ ($N(\mbox{\ion{Na}{1}})> 10^{12.0}~\rm cm^{-2}$) occurs with an incidence $f_{\rm C}(\mbox{\ion{Ca}{2}})=0.63^{+0.10}_{-0.11}$ ($f_{\rm C}(\mbox{\ion{Na}{1}})=0.57^{+0.10}_{-0.11}$) within our sample.  We find no evidence for a dependence of these covering fractions on $R_{\perp}$ or $R_{\perp}/R_{\rm eff, est}$.   These $f_{\rm C}$ values are consistent with the incidence of significantly weaker intermediate- and high-velocity \ion{Ca}{2} and \ion{Na}{1} absorbers 
    (with $N(\mbox{\ion{Ca}{2}})> 10^{11.4}~\rm cm^{-2}$ and $N(\mbox{\ion{Na}{1}})> 10^{10.9}~\rm cm^{-2}$) 
    detected in the Milky Way \citep{BenBekhti2012}. This implies that our sightlines exhibit overall stronger absorption than those probing Milky Way extraplanar/halo clouds, likely due to their longer path lengths through both the ISM and CGM.
    \item The velocities of our \ion{Ca}{2} and \ion{Na}{1} component samples exhibit overall small offsets relative to the H$\alpha$ emission velocities measured along the same sightlines (\zha).  Among 20 \ion{Ca}{2} (and 17 \ion{Na}{1}) components, only three (one) have fitted relative velocities $|\delta  v| > 50\mkms$.  The portions of each line profile contributing $90\%$ of the apparent optical depth all extend to a maximum $\delta v < 120\mkms$ and, thus, trace material that must remain gravitationally bound to even the lowest-$M_*$ system in the sample.  However, the corresponding $\Delta v_{90}$ widths lie in the range $50-180\mkms$, indicating the absorption has contributions from both interstellar and extraplanar material.
    \item We find no evidence for a correlation between the dust reddening measured along our QSO sightlines and the $W_r$ of the \ion{Ca}{2} K or \ion{Na}{1} 5891 transitions.  Between a quarter and a third of our absorber sample are 3$\sigma$ outliers from the best-fit relations between these quantities measured toward extragalactic probes of the Milky Way halo.
    \item   We find no evidence for a strong dependence of the $W_r$ of either ion on the $M_*$  of our foreground galaxies.  Instead, we measure an overall high incidence of $W_r > 0.2$ \AA\ absorbers ($f_{\rm C} \sim 0.4\!-\!0.6$) across the full $M_*$ range of our sample.
    We additionally report a significant ($>3\sigma$) correlation between $W_r(\mbox{\ion{Ca}{2} K})$ and the  local SFR implied by the H$\alpha$ emission-line luminosity measured from SDSS fiber spectra of the sightlines.  These findings suggest that (1) \ion{Na}{1} is an effective probe of disk-halo gas kinematics across the full $M_*$ range of our sample; and that (2) down-the-barrel spectroscopy of the \ion{Ca}{2} transition will be sensitive to star formation-driven outflows of warm, neutral gas.
\end{itemize}

The \ion{Na}{1} absorption strengths we measured along our sample sightlines are significantly larger than the $W_r$ of either outflowing or interstellar material close to the systemic velocity measured in coadded SDSS spectra of galaxies with similar stellar masses.  In addition, our measured column densities of both \ion{Ca}{2} and \ion{Na}{1} are too large to arise from a Milky Way-like ISM.  Instead, the columns and large velocity widths ($\Delta v_{90} = 50\!-\!180\mkms$) of these absorbers require a significant contribution from material with velocities offset by $\delta v >20\mkms$ from the galaxies' \ion{H}{2} regions, but which is gravitationally bound to each system.  

Galactic Fountain models provide a natural explanation for these kinematics and column densities
at least in a qualitative sense.  Assuming this interpretation is apt, our analysis provides novel evidence for Galactic Fountain activity in low-$M_*$, nearby galaxies.  It further suggests that fountain-driven gas motions arise at large projected separations from the nuclei of the host galaxies ($R_{\perp} \gtrsim 7$ kpc).  While some groups are now pursuing important, direct comparison between fountain flows as observed in 21 cm emission and \ion{H}{1} emission-line kinematics predicted in cosmological simulations \citep[e.g.,][]{El-Badry2018,Oman2019,Watts2020,Manuwal2021}, the QSO absorption-line measurements we present here offer a complementary, and in some ways simpler, point of comparison for Galactic Fountain model predictions.  Such comparisons are crucial to improving our understanding of the cycling of multiphase gas flows through galaxy disks.

\begin{acknowledgments}

The authors are grateful for support for this project from NSF grants AST-1715630 and AST-2009417.
K.L.C. acknowledges partial support from NSF grant AST-1615296 and appreciates the observational support of K.~Emerson and T.~Wells, University of Hawai`i at Hilo undergraduate students at the time.  V.P.K. acknowledges partial support from NSF grant AST-2009811.
J.X.P. acknowledges support from  NSF grant AST-1911140. J.K.W. acknowledges support from NSF-AST 1812521 and an RCSA Cottrell Scholar grant, ID number 26842.  The authors also wish to thank the anonymous referee, whose suggestions helped to improve this work.

The data presented herein were obtained at the W. M. Keck Observatory, which is operated as a scientific partnership among the California Institute of Technology, the University of California, and the National Aeronautics and Space Administration.  The Observatory was made possible by the generous financial support of the W. M.~Keck Foundation.

The authors wish to recognize and acknowledge the very significant cultural role and reverence that the summit of Maunakea has always had within the indigenous Hawaiian community.  We are most fortunate to have the opportunity to conduct observations from this mountain.

Funding for the SDSS and SDSS-II has been provided by the Alfred P. Sloan Foundation, the Participating Institutions, the National Science Foundation, the U.S. Department of Energy, the National Aeronautics and Space Administration, the Japanese Monbukagakusho, the Max Planck Society, and the Higher Education Funding Council for England. The SDSS Web Site is \url{http://www.sdss.org}.

The SDSS is managed by the Astrophysical Research Consortium for the Participating Institutions. The Participating Institutions are the American Museum of Natural History, Astrophysical Institute Potsdam, University of Basel, University of Cambridge, Case Western Reserve University, University of Chicago, Drexel University, Fermilab, the Institute for Advanced Study, the Japan Participation Group, Johns Hopkins University, the Joint Institute for Nuclear Astrophysics, the Kavli Institute for Particle Astrophysics and Cosmology, the Korean Scientist Group, the Chinese Academy of Sciences (LAMOST), Los Alamos National Laboratory, the Max-Planck-Institute for Astronomy (MPIA), the Max-Planck-Institute for Astrophysics (MPA), New Mexico State University, Ohio State University, University of Pittsburgh, University of Portsmouth, Princeton University, the United States Naval Observatory, and the University of Washington.
    
\end{acknowledgments}

\vspace{5mm}
\facilities{Keck(ESI), SDSS}

\software{astropy \citep{astropy2013,astropy2018}, linetools \citep{linetools2016}, veeper, MPFIT}



\clearpage
\appendix




\section{ On the Effects of Fiber Losses\label{sec:appendix_SFRfrac}}

 Our sample of foreground galaxies was identified from the serendipitous overlap of associated \ion{H}{2} regions with a nearby SDSS fiber pointing.  This implies that the nebular emission flux captured by the fiber may be a small fraction of the total line flux emitted by the host galaxy.  Moreover, the fibers with the largest angular separations from the centers of the associated hosts are likely to include smaller fractions of the total line flux than their counterparts at the closest angular separations.  Here we build simple models of the \ion{H}{2} region emission arising from each galaxy to assess the potential impacts of these limitations on our analysis.

We begin by assuming that the nebular emission from each galaxy is uniformly distributed in an exponentially declining disk with a half-light radius equal to the $R_{\rm eff,est}$ we derived in Section~\ref{sec:fg_galaxies}, and with a disk thickness $dz = 50$ pc.  This model does not account for morphological disturbances, nor for the small-scale variation in \ion{H}{2} region properties observed at very high spatial resolutions \citep[e.g.,][]{Kreckel2018}. However, given the relatively low spatial resolution of the fiber observations (with typical seeing $\rm FWHM \sim 1.32\arcsec$\footnote{Assessment of the spatial resolution of the final SDSS imaging data is provided at https://www.sdss.org/dr14/imaging/other\_info/.}), we posit that these assumptions will provide a useful 
assessment of systematics in spite of their simplicity.  We draw on the tilted-ring modeling framework discussed in Section~\ref{sec:model} to construct each disk with a single layer of rings, adopting a scale radius $R_{\rm S} = R_{\rm eff,est}/1.678$ \citep{CiottiBertin1999}.  We further assume that the surface brightness of the nebular emission (e.g., from H$\alpha$) is directly proportional to the local star formation rate surface density ($\Sigma_{\rm SFR}$), so that we may write the radial profile of this quantity as $\Sigma_{\rm SFR}(r) = \Sigma_{\rm SFR, 0} e^{-r/R_{\rm S}}$.  The total SFR of this system can then be written ${\rm SFR_{tot}} = \Sigma_{\rm SFR, 0} R_{\rm eff,est} \frac{2\pi}{1.678^2}\Gamma(2)$, as described in \citet{CiottiBertin1999}.

For each foreground galaxy, we set $\Sigma_{\rm SFR, 0}$ to an arbitrary value ($0.1~M_{\odot}~\rm yr^{-1}~kpc^{-2}$), and generate the resulting $\Sigma_{\rm SFR}$ distribution as observed both face-on (with $i=0^{\circ}$) and at an inclination of $i=75^{\circ}$ (using the same approach described in Section~\ref{subsec:modeling_columndensities}).  We then smooth these distributions with a two-dimensional Gaussian kernel having $\rm FWHM=1.32\arcsec$ to simulate the effects of seeing.  Finally, we ``observe" these distributions with a $3\arcsec$ diameter fiber placed at a distance $R_{\perp}$ from the center of the model, adopting position angles spanning between $\rm PA=0^{\circ}$ and $170^{\circ}$ at intervals of $10^{\circ}$.  We sum the model star formation activity falling within each fiber pointing, and normalize this quantity by the corresponding $\rm SFR_{\rm tot}$.

\begin{figure*}[ht]
 \includegraphics[width=0.5\textwidth]{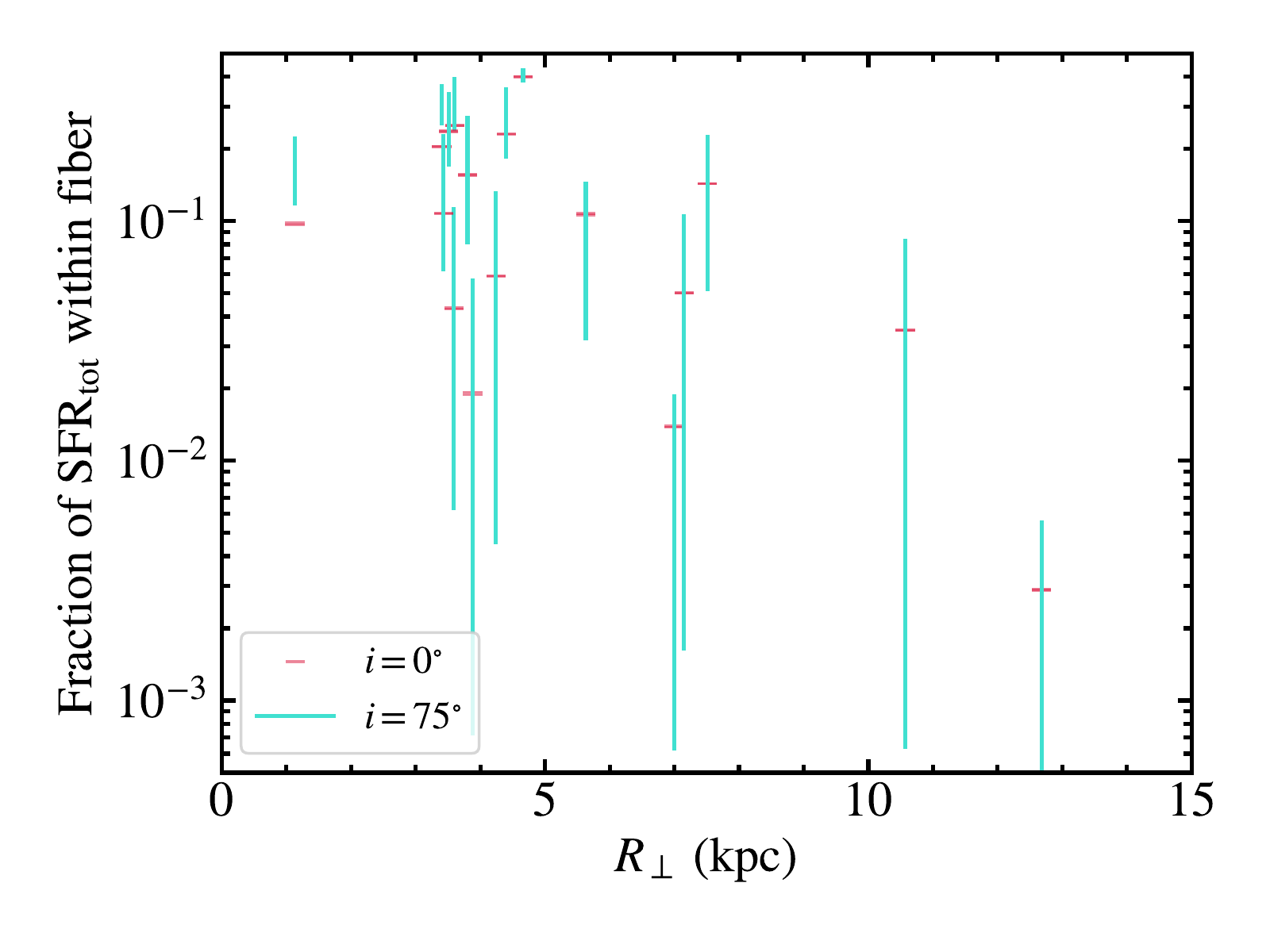}
  \includegraphics[width=0.5\textwidth]{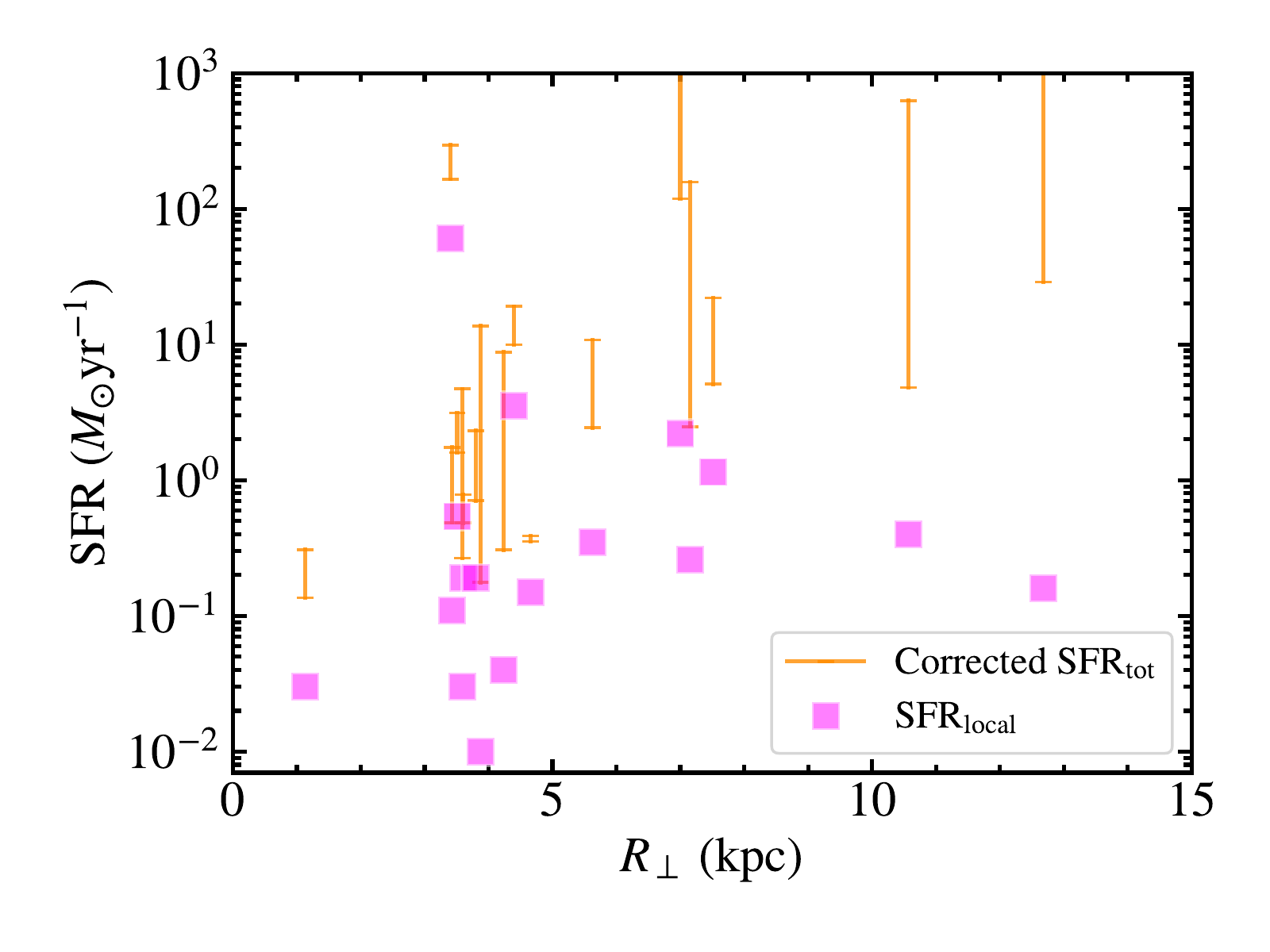}
\caption{{\it Left:} The fraction of the total SFR of each foreground galaxy probed by the corresponding SDSS fiber, as implied by our exponential disk modeling.  The rose-colored horizontal dashes indicate the fraction of \ion{H}{2} region emission observed if the galaxy is viewed face-on, while the vertical turquoise lines show the range in this observed fraction in the case that the galaxy is oriented close to edge-on with $i=75^{\circ}$ (accounting for uncertainty in the positional angle of the fiber placement).  {\it Right:}  The $\rm SFR_{local}$ measured in each fiber observation (magenta squares) vs.\ $R_{\perp}$.  The ranges indicated in orange show the span of $\rm SFR_{tot}$ values implied by the fractions shown at left.  
\label{fig:SFRfrac}}
\end{figure*}

This fraction is shown in the left panel of Figure~\ref{fig:SFRfrac} vs.\ $R_{\perp}$.  The rose-colored horizontal dash indicates the fraction of $\rm SFR_{tot}$ observed in these fibers in the face-on ($i=0^{\circ}$) case, and the vertical turquoise lines mark the range of values observed around systems oriented with $i=75^{\circ}$.  The right-hand panel of this figure shows the value of $\rm SFR_{local}$ for each system (with magenta squares), as well as the range of $\rm SFR_{tot}$ values implied by the fractions shown at left.  For systems having $R_{\perp} < 5$ kpc, our modeling suggests that the SDSS fibers include $\gtrsim 10\%$ of the total line emission associated with star formation in a majority of cases (eight out of 11).  For systems having $R_{\perp} > 5$ kpc, the SDSS fibers may exclude $\gtrsim 90\!-\!99\%$ of the total nebular line emission.  This implies that the $\rm SFR_{tot}$ values of a handful of our foreground systems ($\approx 7$ of the 17 shown here) may be quite large ($\gtrsim 10\!-\!100~M_{\odot}~\rm yr^{-1}$).  Such high global SFRs would place them well above the star-forming sequence at $z\sim0$ shown in Figure~\ref{fig:rperp_mstar_sfr}.  Moreover, under the assumption that highly star-forming galaxies have stronger interstellar and circumgalactic \ion{Na}{1} and \ion{Ca}{2} absorption overall, the higher frequency of such systems observed at larger impact parameters may drive the $\log W_r$-$R_{\perp}$ relations we fit in Section~\ref{subsec:Wr_Rperp} to have shallower slopes.

Because of the significant degree of uncertainty in these estimates (due to, e.g., uncertainties in viewing angle; uncertainties in the position angle of the fiber observations relative to the major axis of each galaxy; and the simplicity of our assumptions regarding the spatial distribution of star formation activity), we do not explicitly fold them into our analyses of the relationships between absorber properties and star formation rate.  As emphasized in Section~\ref{subsec:ew_dv_Mstar_SFR},  we have focused solely on the relation between $\rm SFR_{local}$,  which is well measured by the SDSS fiber spectroscopy at our disposal, and absorber strength and kinematics.  The analysis presented here serves to better contextualize our sample within the broader galaxy population.

\section{Sample Spectroscopy}\label{sec:appendix_spectra}

Figure~\ref{fig:velplot2} shows our spectroscopy of GOTOQ sightlines that are not included in Figure~\ref{fig:velplot1}.

\begin{figure*}[th]
  \includegraphics[trim={1cm 0cm 1cm 1cm},clip,width=1.0\textwidth]{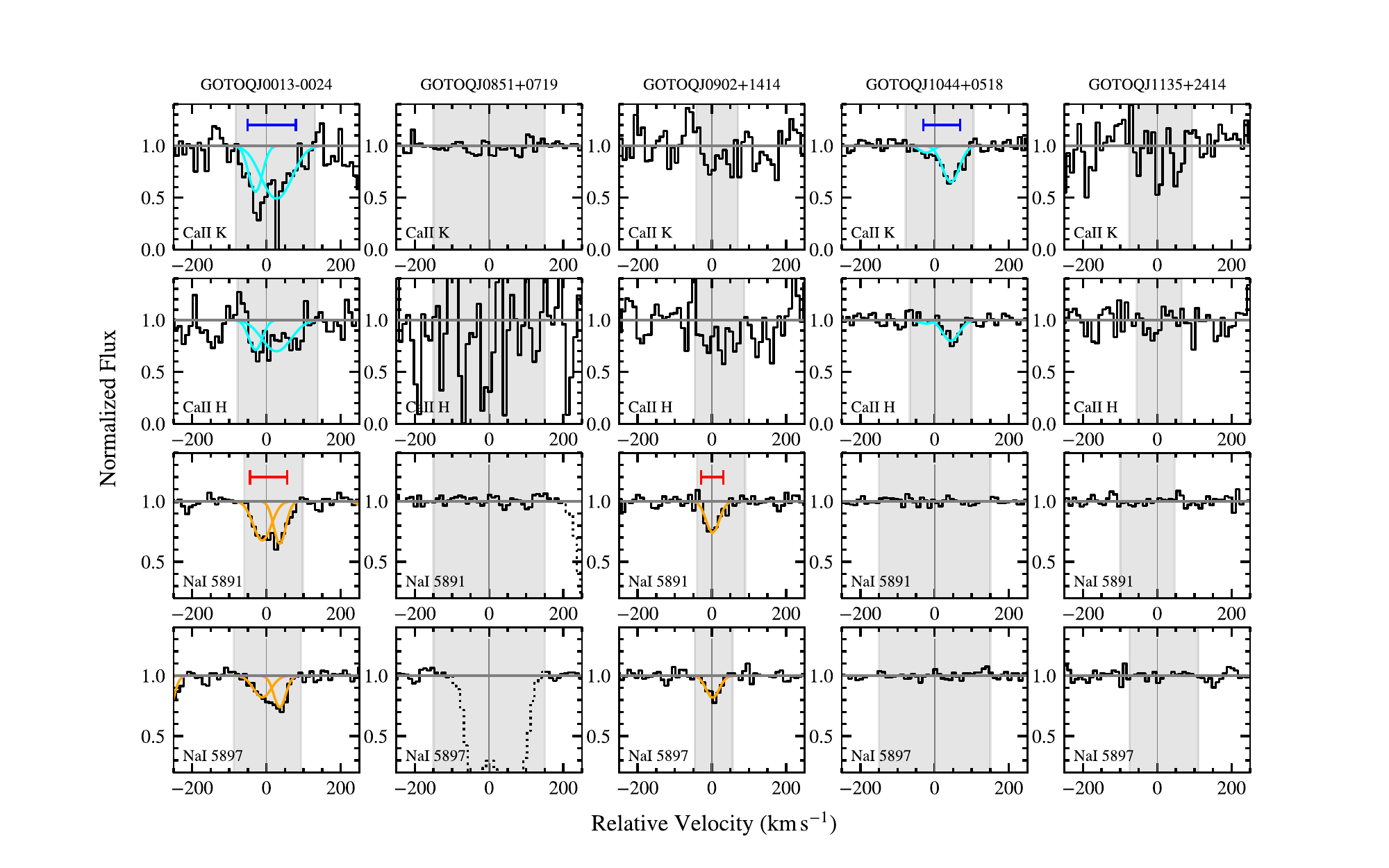}
\caption{ESI GOTOQ spectra showing the locations of the \ion{Ca}{2} H \& K and \ion{Na}{1} $\lambda \lambda 5891, 5897$ transitions in the rest frame of the foreground galaxy.  The velocity is defined relative to the GOTOQ redshift estimated from a Gaussian fit to its H$\alpha$ emission as described in Section~\ref{subsec:redshifts} (\zha).  The gray horizontal line indicates the continuum level, and the gray shaded region shows the velocity window selected for computation of $W_r$ and $\Delta v_{90}$.  Absorption from unassociated blends is shown with a dotted histogram.  The blue and red bars show the pixels that contain $>5\%$ of the total apparent optical depth of the line (determined by stepping inward from the profile edges), and the length of these bars corresponds to $\Delta v_{90}$.  Best-fit profile models are shown with cyan (for \ion{Ca}{2}) and orange (for \ion{Na}{1}) curves for systems with significantly detected absorption (see Section~\ref{subsec:abs_modeling} for details).  \label{fig:velplot2}}
\end{figure*}
\setcounter{figure}{15}
\begin{figure*}[th]
  \includegraphics[trim={1cm 0cm 1cm 1cm},clip,width=1.0\textwidth]{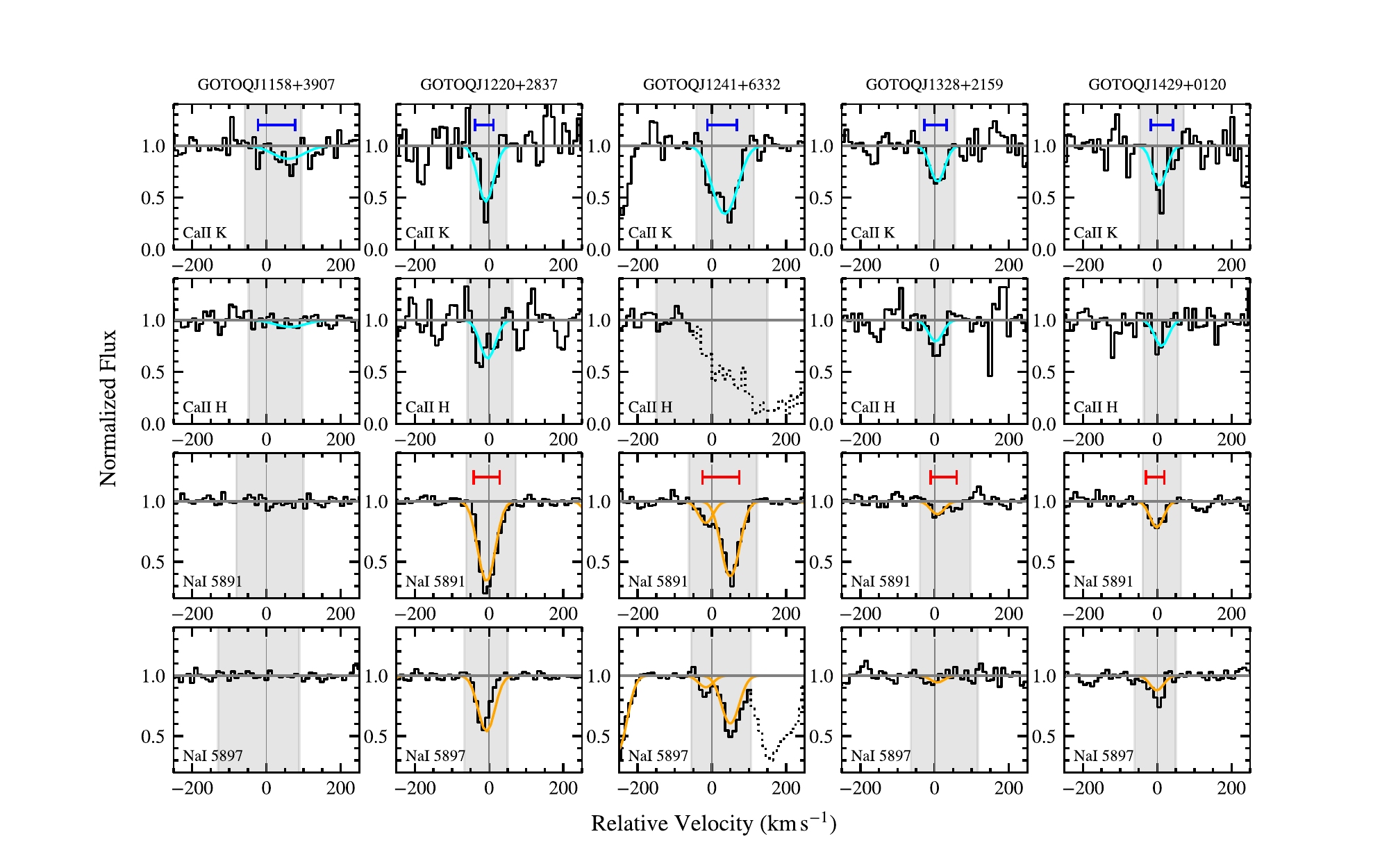}
\caption{-- continued \label{fig:velplot3}}
\end{figure*}
\setcounter{figure}{15}
\begin{figure*}[th]
  \includegraphics[trim={1cm 0cm 1cm 1cm},clip,width=1.0\textwidth]{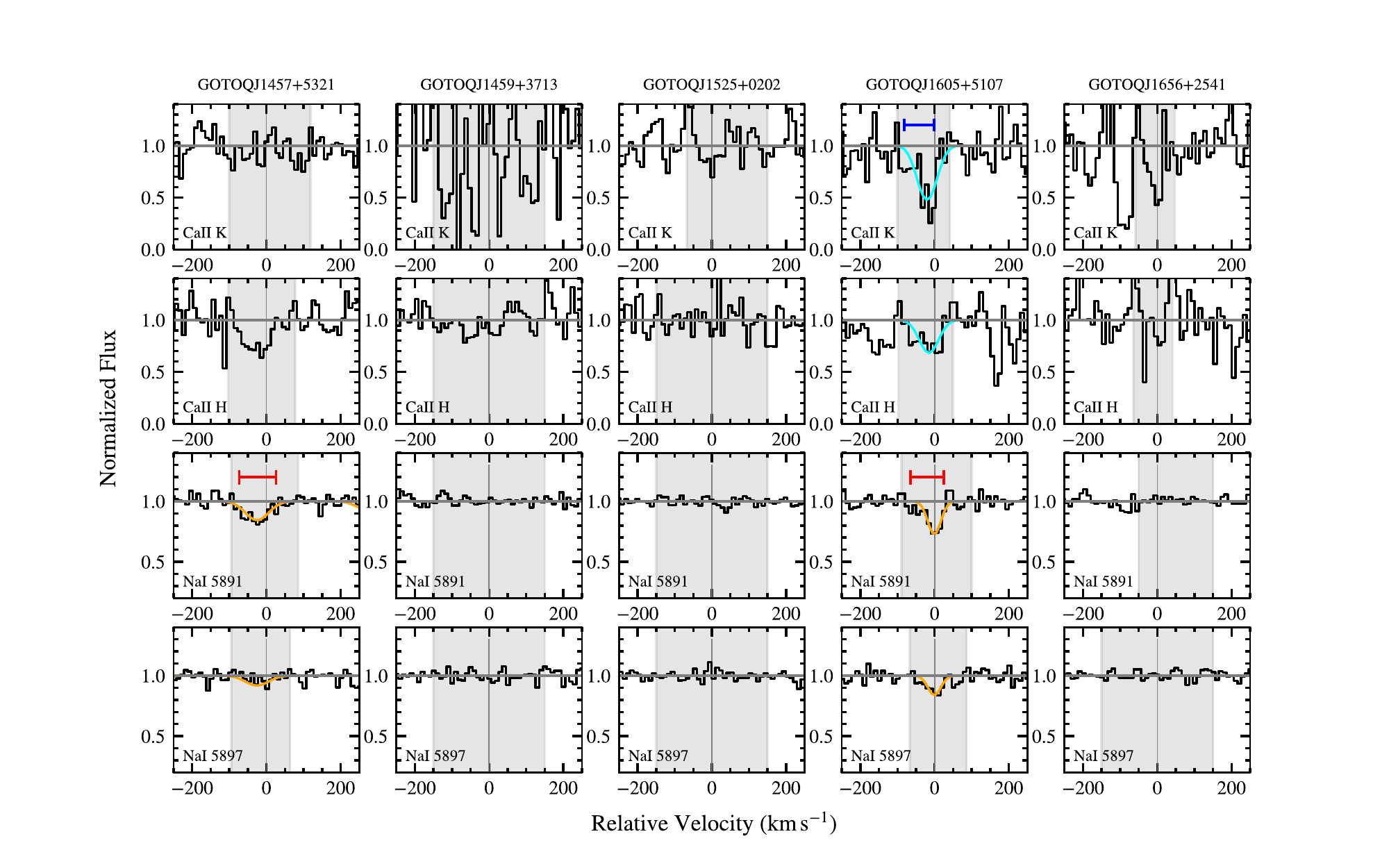}
\caption{-- continued \label{fig:velplot4}}
\end{figure*}
\setcounter{figure}{15}
\begin{figure*}[th]
  \includegraphics[trim={1cm 0cm 1cm 1cm},clip,width=1.0\textwidth]{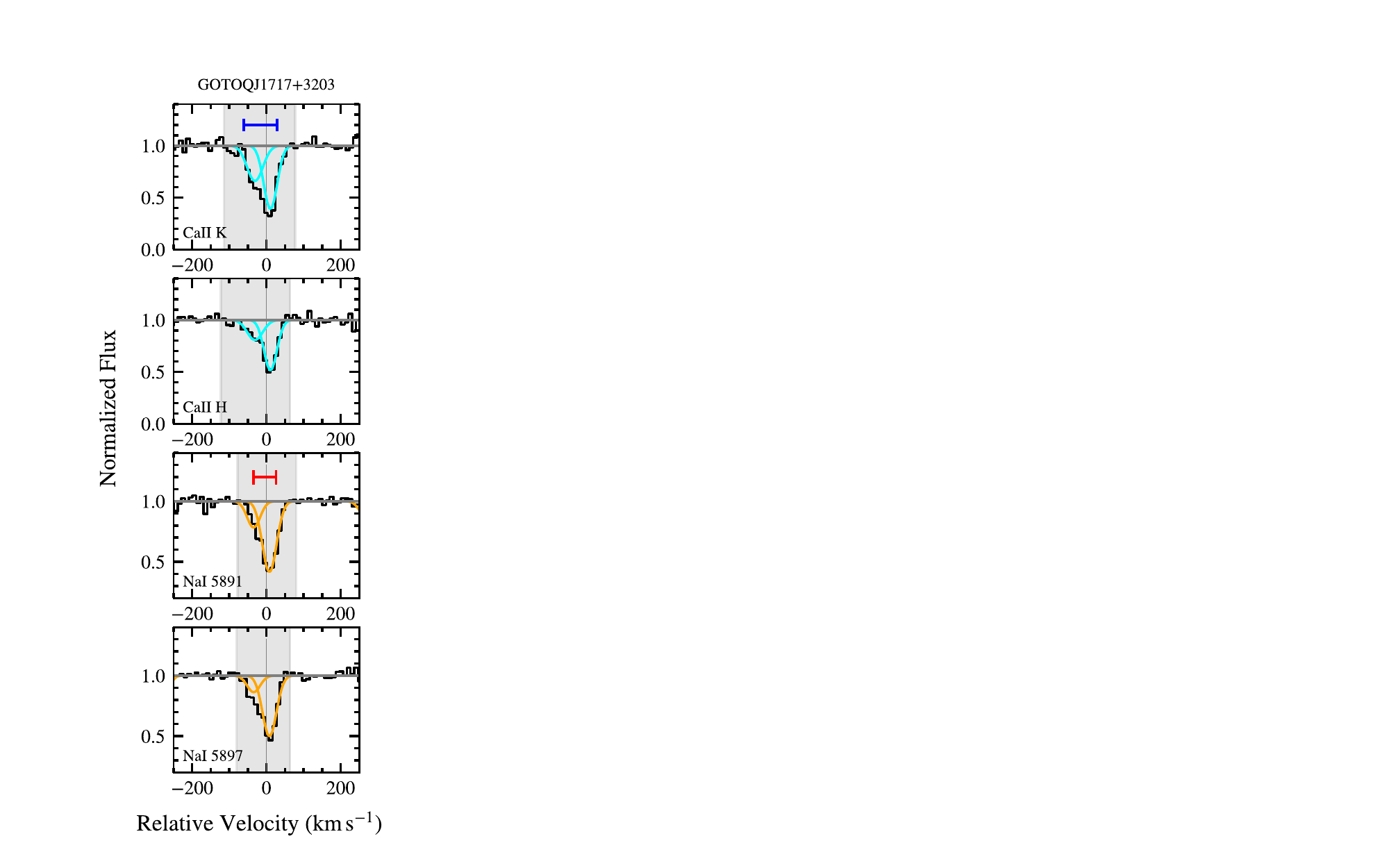}
\caption{-- continued \label{fig:velplot5}}
\end{figure*}

\bibliography{gotoqs-esi}{}

\begin{thebibliography}{}
\expandafter\ifx\csname natexlab\endcsname\relax\def\natexlab#1{#1}\fi
\providecommand{\url}[1]{\href{#1}{#1}}
\providecommand{\dodoi}[1]{doi:~\href{http://doi.org/#1}{\nolinkurl{#1}}}
\providecommand{\doeprint}[1]{\href{http://ascl.net/#1}{\nolinkurl{http://ascl.net/#1}}}
\providecommand{\doarXiv}[1]{\href{https://arxiv.org/abs/#1}{\nolinkurl{https://arxiv.org/abs/#1}}}

\bibitem[{{Abazajian} {et~al.}(2009){Abazajian}, {Adelman-McCarthy},
  {Ag{\"u}eros}, {Allam}, {Allende Prieto}, {An}, {Anderson}, {Anderson},
  {Annis}, {Bahcall}, {Bailer-Jones}, {Barentine}, {Bassett}, {Becker},
  {Beers}, {Bell}, {Belokurov}, {Berlind}, {Berman}, {Bernardi}, {Bickerton},
  {Bizyaev}, {Blakeslee}, {Blanton}, {Bochanski}, {Boroski}, {Brewington},
  {Brinchmann}, {Brinkmann}, {Brunner}, {Budav{\'a}ri}, {Carey}, {Carliles},
  {Carr}, {Castander}, {Cinabro}, {Connolly}, {Csabai}, {Cunha}, {Czarapata},
  {Davenport}, {de Haas}, {Dilday}, {Doi}, {Eisenstein}, {Evans}, {Evans},
  {Fan}, {Friedman}, {Frieman}, {Fukugita}, {G{\"a}nsicke}, {Gates},
  {Gillespie}, {Gilmore}, {Gonzalez}, {Gonzalez}, {Grebel}, {Gunn},
  {Gy{\"o}ry}, {Hall}, {Harding}, {Harris}, {Harvanek}, {Hawley}, {Hayes},
  {Heckman}, {Hendry}, {Hennessy}, {Hindsley}, {Hoblitt}, {Hogan}, {Hogg},
  {Holtzman}, {Hyde}, {Ichikawa}, {Ichikawa}, {Im}, {Ivezi{\'c}}, {Jester},
  {Jiang}, {Johnson}, {Jorgensen}, {Juri{\'c}}, {Kent}, {Kessler}, {Kleinman},
  {Knapp}, {Konishi}, {Kron}, {Krzesinski}, {Kuropatkin}, {Lampeitl},
  {Lebedeva}, {Lee}, {Lee}, {French Leger}, {L{\'e}pine}, {Li}, {Lima}, {Lin},
  {Long}, {Loomis}, {Loveday}, {Lupton}, {Magnier}, {Malanushenko},
  {Malanushenko}, {Mandelbaum}, {Margon}, {Marriner}, {Mart{\'\i}nez-Delgado},
  {Matsubara}, {McGehee}, {McKay}, {Meiksin}, {Morrison}, {Mullally}, {Munn},
  {Murphy}, {Nash}, {Nebot}, {Neilsen}, {Newberg}, {Newman}, {Nichol},
  {Nicinski}, {Nieto-Santisteban}, {Nitta}, {Okamura}, {Oravetz}, {Ostriker},
  {Owen}, {Padmanabhan}, {Pan}, {Park}, {Pauls}, {Peoples}, {Percival}, {Pier},
  {Pope}, {Pourbaix}, {Price}, {Purger}, {Quinn}, {Raddick}, {Re Fiorentin},
  {Richards}, {Richmond}, {Riess}, {Rix}, {Rockosi}, {Sako}, {Schlegel},
  {Schneider}, {Scholz}, {Schreiber}, {Schwope}, {Seljak}, {Sesar}, {Sheldon},
  {Shimasaku}, {Sibley}, {Simmons}, {Sivarani}, {Allyn Smith}, {Smith},
  {Smol{\v{c}}i{\'c}}, {Snedden}, {Stebbins}, {Steinmetz}, {Stoughton},
  {Strauss}, {SubbaRao}, {Suto}, {Szalay}, {Szapudi}, {Szkody}, {Tanaka},
  {Tegmark}, {Teodoro}, {Thakar}, {Tremonti}, {Tucker}, {Uomoto}, {Vanden
  Berk}, {Vandenberg}, {Vidrih}, {Vogeley}, {Voges}, {Vogt}, {Wadadekar},
  {Watters}, {Weinberg}, {West}, {White}, {Wilhite}, {Wonders}, {Yanny},
  {Yocum}, {York}, {Zehavi}, {Zibetti}, \& {Zucker}}]{Abazajian2009}
{Abazajian}, K.~N., {Adelman-McCarthy}, J.~K., {Ag{\"u}eros}, M.~A., {et~al.}
  2009, \apjs, 182, 543, \dodoi{10.1088/0067-0049/182/2/543}

\bibitem[{{Ahumada} {et~al.}(2020){Ahumada}, {Prieto}, {Almeida}, {Anders},
  {Anderson}, {Andrews}, {Anguiano}, {Arcodia}, {Armengaud}, {Aubert}, {Avila},
  {Avila-Reese}, {Badenes}, {Balland}, {Barger}, {Barrera-Ballesteros}, {Basu},
  {Bautista}, {Beaton}, {Beers}, {Benavides}, {Bender}, {Bernardi}, {Bershady},
  {Beutler}, {Bidin}, {Bird}, {Bizyaev}, {Blanc}, {Blanton}, {Boquien},
  {Borissova}, {Bovy}, {Brandt}, {Brinkmann}, {Brownstein}, {Bundy}, {Bureau},
  {Burgasser}, {Burtin}, {Cano-D{\'\i}az}, {Capasso}, {Cappellari}, {Carrera},
  {Chabanier}, {Chaplin}, {Chapman}, {Cherinka}, {Chiappini}, {Doohyun Choi},
  {Chojnowski}, {Chung}, {Clerc}, {Coffey}, {Comerford}, {Comparat}, {da
  Costa}, {Cousinou}, {Covey}, {Crane}, {Cunha}, {Ilha}, {Dai}, {Damsted},
  {Darling}, {Davidson}, {Davies}, {Dawson}, {De}, {de la Macorra}, {De Lee},
  {Queiroz}, {Deconto Machado}, {de la Torre}, {Dell'Agli}, {du Mas des
  Bourboux}, {Diamond-Stanic}, {Dillon}, {Donor}, {Drory}, {Duckworth},
  {Dwelly}, {Ebelke}, {Eftekharzadeh}, {Davis Eigenbrot}, {Elsworth},
  {Eracleous}, {Erfanianfar}, {Escoffier}, {Fan}, {Farr},
  {Fern{\'a}ndez-Trincado}, {Feuillet}, {Finoguenov}, {Fofie},
  {Fraser-McKelvie}, {Frinchaboy}, {Fromenteau}, {Fu}, {Galbany}, {Garcia},
  {Garc{\'\i}a-Hern{\'a}ndez}, {Oehmichen}, {Ge}, {Maia}, {Geisler}, {Gelfand},
  {Goddy}, {Gonzalez-Perez}, {Grabowski}, {Green}, {Grier}, {Guo}, {Guy},
  {Harding}, {Hasselquist}, {Hawken}, {Hayes}, {Hearty}, {Hekker}, {Hogg},
  {Holtzman}, {Horta}, {Hou}, {Hsieh}, {Huber}, {Hunt}, {Chitham}, {Imig},
  {Jaber}, {Angel}, {Johnson}, {Jones}, {J{\"o}nsson}, {Jullo}, {Kim},
  {Kinemuchi}, {Kirkpatrick}, {Kite}, {Klaene}, {Kneib}, {Kollmeier}, {Kong},
  {Kounkel}, {Krishnarao}, {Lacerna}, {Lan}, {Lane}, {Law}, {Le Goff}, {Leung},
  {Lewis}, {Li}, {Lian}, {Lin}, {Long}, {Longa-Pe{\~n}a}, {Lundgren}, {Lyke},
  {Ted Mackereth}, {MacLeod}, {Majewski}, {Manchado}, {Maraston}, {Martini},
  {Masseron}, {Masters}, {Mathur}, {McDermid}, {Merloni}, {Merrifield},
  {M{\'e}sz{\'a}ros}, {Miglio}, {Minniti}, {Minsley}, {Miyaji}, {Mohammad},
  {Mosser}, {Mueller}, {Muna}, {Mu{\~n}oz-Guti{\'e}rrez}, {Myers}, {Nadathur},
  {Nair}, {Nandra}, {do Nascimento}, {Nevin}, {Newman}, {Nidever}, {Nitschelm},
  {Noterdaeme}, {O'Connell}, {Olmstead}, {Oravetz}, {Oravetz}, {Osorio},
  {Pace}, {Padilla}, {Palanque-Delabrouille}, {Palicio}, {Pan}, {Pan},
  {Parker}, {Paviot}, {Peirani}, {Ram{\'r}ez}, {Penny}, {Percival},
  {Perez-Fournon}, {P{\'e}rez-R{\`a}fols}, {Petitjean}, {Pieri},
  {Pinsonneault}, {Poovelil}, {Povick}, {Prakash}, {Price-Whelan}, {Raddick},
  {Raichoor}, {Ray}, {Rembold}, {Rezaie}, {Riffel}, {Riffel}, {Rix}, {Robin},
  {Roman-Lopes}, {Rom{\'a}n-Z{\'u}{\~n}iga}, {Rose}, {Ross}, {Rossi},
  {Rowlands}, {Rubin}, {Salvato}, {S{\'a}nchez}, {S{\'a}nchez-Menguiano},
  {S{\'a}nchez-Gallego}, {Sayres}, {Schaefer}, {Schiavon}, {Schimoia},
  {Schlafly}, {Schlegel}, {Schneider}, {Schultheis}, {Schwope}, {Seo},
  {Serenelli}, {Shafieloo}, {Shamsi}, {Shao}, {Shen}, {Shetrone}, {Shirley},
  {Aguirre}, {Simon}, {Skrutskie}, {Slosar}, {Smethurst}, {Sobeck}, {Sodi},
  {Souto}, {Stark}, {Stassun}, {Steinmetz}, {Stello}, {Stermer},
  {Storchi-Bergmann}, {Streblyanska}, {Stringfellow}, {Stutz}, {Su{\'a}rez},
  {Sun}, {Taghizadeh-Popp}, {Talbot}, {Tayar}, {Thakar}, {Theriault}, {Thomas},
  {Thomas}, {Tinker}, {Tojeiro}, {Toledo}, {Tremonti}, {Troup}, {Tuttle},
  {Unda-Sanzana}, {Valentini}, {Vargas-Gonz{\'a}lez}, {Vargas-Maga{\~n}a},
  {V{\'a}zquez-Mata}, {Vivek}, {Wake}, {Wang}, {Weaver}, {Weijmans}, {Wild},
  {Wilson}, {Wilson}, {Wolthuis}, {Wood-Vasey}, {Yan}, {Yang}, {Y{\`e}che},
  {Zamora}, {Zarrouk}, {Zasowski}, {Zhang}, {Zhao}, {Zhao}, {Zheng}, {Zheng},
  {Zhu}, \& {Zou}}]{Ahumada2020}
{Ahumada}, R., {Prieto}, C.~A., {Almeida}, A., {et~al.} 2020, \apjs, 249, 3,
  \dodoi{10.3847/1538-4365/ab929e}

\bibitem[{{Allen} {et~al.}(2015){Allen}, {Croom}, {Konstantopoulos}, {Bryant},
  {Sharp}, {Cecil}, {Fogarty}, {Foster}, {Green}, {Ho}, {Owers}, {Schaefer},
  {Scott}, {Bauer}, {Baldry}, {Barnes}, {Bland-Hawthorn}, {Bloom}, {Brough},
  {Colless}, {Cortese}, {Couch}, {Drinkwater}, {Driver}, {Goodwin},
  {Gunawardhana}, {Hampton}, {Hopkins}, {Kewley}, {Lawrence}, {Leon-Saval},
  {Liske}, {L{\'o}pez-S{\'a}nchez}, {Lorente}, {McElroy}, {Medling}, {Mould},
  {Norberg}, {Parker}, {Power}, {Pracy}, {Richards}, {Robotham}, {Sweet},
  {Taylor}, {Thomas}, {Tonini}, \& {Walcher}}]{Allen2015}
{Allen}, J.~T., {Croom}, S.~M., {Konstantopoulos}, I.~S., {et~al.} 2015,
  \mnras, 446, 1567, \dodoi{10.1093/mnras/stu2057}

\bibitem[{{Astropy Collaboration} {et~al.}(2013){Astropy Collaboration},
  {Robitaille}, {Tollerud}, {Greenfield}, {Droettboom}, {Bray}, {Aldcroft},
  {Davis}, {Ginsburg}, {Price-Whelan}, {Kerzendorf}, {Conley}, {Crighton},
  {Barbary}, {Muna}, {Ferguson}, {Grollier}, {Parikh}, {Nair}, {Unther},
  {Deil}, {Woillez}, {Conseil}, {Kramer}, {Turner}, {Singer}, {Fox}, {Weaver},
  {Zabalza}, {Edwards}, {Azalee Bostroem}, {Burke}, {Casey}, {Crawford},
  {Dencheva}, {Ely}, {Jenness}, {Labrie}, {Lim}, {Pierfederici}, {Pontzen},
  {Ptak}, {Refsdal}, {Servillat}, \& {Streicher}}]{astropy2013}
{Astropy Collaboration}, {Robitaille}, T.~P., {Tollerud}, E.~J., {et~al.} 2013,
  \aap, 558, A33, \dodoi{10.1051/0004-6361/201322068}

\bibitem[{{Astropy Collaboration} {et~al.}(2018){Astropy Collaboration},
  {Price-Whelan}, {Sip{\H{o}}cz}, {G{\"u}nther}, {Lim}, {Crawford}, {Conseil},
  {Shupe}, {Craig}, {Dencheva}, {Ginsburg}, {VanderPlas}, {Bradley},
  {P{\'e}rez-Su{\'a}rez}, {de Val-Borro}, {Aldcroft}, {Cruz}, {Robitaille},
  {Tollerud}, {Ardelean}, {Babej}, {Bach}, {Bachetti}, {Bakanov}, {Bamford},
  {Barentsen}, {Barmby}, {Baumbach}, {Berry}, {Biscani}, {Boquien}, {Bostroem},
  {Bouma}, {Brammer}, {Bray}, {Breytenbach}, {Buddelmeijer}, {Burke},
  {Calderone}, {Cano Rodr{\'\i}guez}, {Cara}, {Cardoso}, {Cheedella}, {Copin},
  {Corrales}, {Crichton}, {D'Avella}, {Deil}, {Depagne}, {Dietrich}, {Donath},
  {Droettboom}, {Earl}, {Erben}, {Fabbro}, {Ferreira}, {Finethy}, {Fox},
  {Garrison}, {Gibbons}, {Goldstein}, {Gommers}, {Greco}, {Greenfield},
  {Groener}, {Grollier}, {Hagen}, {Hirst}, {Homeier}, {Horton}, {Hosseinzadeh},
  {Hu}, {Hunkeler}, {Ivezi{\'c}}, {Jain}, {Jenness}, {Kanarek}, {Kendrew},
  {Kern}, {Kerzendorf}, {Khvalko}, {King}, {Kirkby}, {Kulkarni}, {Kumar},
  {Lee}, {Lenz}, {Littlefair}, {Ma}, {Macleod}, {Mastropietro}, {McCully},
  {Montagnac}, {Morris}, {Mueller}, {Mumford}, {Muna}, {Murphy}, {Nelson},
  {Nguyen}, {Ninan}, {N{\"o}the}, {Ogaz}, {Oh}, {Parejko}, {Parley}, {Pascual},
  {Patil}, {Patil}, {Plunkett}, {Prochaska}, {Rastogi}, {Reddy Janga},
  {Sabater}, {Sakurikar}, {Seifert}, {Sherbert}, {Sherwood-Taylor}, {Shih},
  {Sick}, {Silbiger}, {Singanamalla}, {Singer}, {Sladen}, {Sooley},
  {Sornarajah}, {Streicher}, {Teuben}, {Thomas}, {Tremblay}, {Turner},
  {Terr{\'o}n}, {van Kerkwijk}, {de la Vega}, {Watkins}, {Weaver}, {Whitmore},
  {Woillez}, {Zabalza}, \& {Astropy Contributors}}]{astropy2018}
{Astropy Collaboration}, {Price-Whelan}, A.~M., {Sip{\H{o}}cz}, B.~M., {et~al.}
  2018, \aj, 156, 123, \dodoi{10.3847/1538-3881/aabc4f}

\bibitem[{{Bajaja} {et~al.}(1985){Bajaja}, {Cappa de Nicolau}, {Cersosimo},
  {Martin}, {Loiseau}, {Morras}, {Olano}, \& {Poeppel}}]{Bajaja1985}
{Bajaja}, E., {Cappa de Nicolau}, C.~E., {Cersosimo}, J.~C., {et~al.} 1985,
  \apjs, 58, 143, \dodoi{10.1086/191033}

\bibitem[{{Barbieri} {et~al.}(2005){Barbieri}, {Fraternali}, {Oosterloo},
  {Bertin}, {Boomsma}, \& {Sancisi}}]{Barbieri2005}
{Barbieri}, C.~V., {Fraternali}, F., {Oosterloo}, T., {et~al.} 2005, \aap, 439,
  947, \dodoi{10.1051/0004-6361:20042395}

\bibitem[{{Baron} {et~al.}(2016){Baron}, {Stern}, {Poznanski}, \&
  {Netzer}}]{Baron2016}
{Baron}, D., {Stern}, J., {Poznanski}, D., \& {Netzer}, H. 2016, \apj, 832, 8,
  \dodoi{10.3847/0004-637X/832/1/8}

\bibitem[{{Begeman}(1989)}]{Begeman1989}
{Begeman}, K.~G. 1989, \aap, 223, 47

\bibitem[{{Begum} {et~al.}(2008){Begum}, {Chengalur}, {Karachentsev},
  {Sharina}, \& {Kaisin}}]{Begum2008}
{Begum}, A., {Chengalur}, J.~N., {Karachentsev}, I.~D., {Sharina}, M.~E., \&
  {Kaisin}, S.~S. 2008, \mnras, 386, 1667,
  \dodoi{10.1111/j.1365-2966.2008.13150.x}

\bibitem[{{Ben Bekhti} {et~al.}(2008){Ben Bekhti}, {Richter}, {Westmeier}, \&
  {Murphy}}]{BenBekhti2008}
{Ben Bekhti}, N., {Richter}, P., {Westmeier}, T., \& {Murphy}, M.~T. 2008,
  \aap, 487, 583, \dodoi{10.1051/0004-6361:20079067}

\bibitem[{{Ben Bekhti} {et~al.}(2012){Ben Bekhti}, {Winkel}, {Richter}, {Kerp},
  {Klein}, \& {Murphy}}]{BenBekhti2012}
{Ben Bekhti}, N., {Winkel}, B., {Richter}, P., {et~al.} 2012, \aap, 542, A110,
  \dodoi{10.1051/0004-6361/201118673}

\bibitem[{{Bergeron} {et~al.}(1987){Bergeron}, {D'Odorico}, \&
  {Kunth}}]{Bergeron1987}
{Bergeron}, J., {D'Odorico}, S., \& {Kunth}, D. 1987, \aap, 180, 1

\bibitem[{{Bergeron} \& {Stasi{\'n}ska}(1986)}]{BergeronStasinska1986}
{Bergeron}, J., \& {Stasi{\'n}ska}, G. 1986, \aap, 169, 1

\bibitem[{{Bish} {et~al.}(2021){Bish}, {Werk}, {Peek}, {Zheng}, \&
  {Putman}}]{Bish2021}
{Bish}, H.~V., {Werk}, J.~K., {Peek}, J., {Zheng}, Y., \& {Putman}, M. 2021,
  \apj, 912, 8, \dodoi{10.3847/1538-4357/abeb6b}

\bibitem[{{Bish} {et~al.}(2019){Bish}, {Werk}, {Prochaska}, {Rubin}, {Zheng},
  {O'Meara}, \& {Deason}}]{Bish2019}
{Bish}, H.~V., {Werk}, J.~K., {Prochaska}, J.~X., {et~al.} 2019, \apj, 882, 76,
  \dodoi{10.3847/1538-4357/ab3414}

\bibitem[{{Bizyaev} {et~al.}(2017){Bizyaev}, {Walterbos}, {Yoachim}, {Riffel},
  {Fern{\'a}ndez-Trincado}, {Pan}, {Diamond-Stanic}, {Jones}, {Thomas},
  {Cleary}, \& {Brinkmann}}]{Bizyaev2017}
{Bizyaev}, D., {Walterbos}, R.~A.~M., {Yoachim}, P., {et~al.} 2017, \apj, 839,
  87, \dodoi{10.3847/1538-4357/aa6979}

\bibitem[{{Blades} {et~al.}(1981){Blades}, {Hunstead}, \&
  {Murdoch}}]{Blades1981}
{Blades}, J.~C., {Hunstead}, R.~W., \& {Murdoch}, H.~S. 1981, \mnras, 194, 669,
  \dodoi{10.1093/mnras/194.3.669}

\bibitem[{{Bland-Hawthorn} {et~al.}(1998){Bland-Hawthorn}, {Veilleux}, {Cecil},
  {Putman}, {Gibson}, \& {Maloney}}]{Bland-Hawthorn1998}
{Bland-Hawthorn}, J., {Veilleux}, S., {Cecil}, G.~N., {et~al.} 1998, \mnras,
  299, 611, \dodoi{10.1046/j.1365-8711.1998.01902.x}

\bibitem[{{Bloom} {et~al.}(2017){Bloom}, {Croom}, {Bryant}, {Callingham},
  {Schaefer}, {Cortese}, {Hopkins}, {D'Eugenio}, {Scott}, {Glazebrook},
  {Tonini}, {McElroy}, {Clark}, {Catinella}, {Allen}, {Bland-Hawthorn},
  {Goodwin}, {Green}, {Konstantopoulos}, {Lawrence}, {Lorente}, {Medling},
  {Owers}, {Richards}, \& {Sharp}}]{Bloom2017}
{Bloom}, J.~V., {Croom}, S.~M., {Bryant}, J.~J., {et~al.} 2017, \mnras, 472,
  1809, \dodoi{10.1093/mnras/stx1701}

\bibitem[{{Boettcher} {et~al.}(2016){Boettcher}, {Zweibel}, {Gallagher}, \&
  {Benjamin}}]{Boettcher2016}
{Boettcher}, E., {Zweibel}, E.~G., {Gallagher}, J.~S., I., \& {Benjamin}, R.~A.
  2016, \apj, 832, 118, \dodoi{10.3847/0004-637X/832/2/118}

\bibitem[{{Bohlin} {et~al.}(1978){Bohlin}, {Savage}, \& {Drake}}]{Bohlin1978}
{Bohlin}, R.~C., {Savage}, B.~D., \& {Drake}, J.~F. 1978, \apj, 224, 132,
  \dodoi{10.1086/156357}

\bibitem[{{Boksenberg} {et~al.}(1980){Boksenberg}, {Danziger}, {Fosbury}, \&
  {Goss}}]{Boksenberg1980}
{Boksenberg}, A., {Danziger}, I.~J., {Fosbury}, R.~A.~E., \& {Goss}, W.~M.
  1980, \apjl, 242, L145, \dodoi{10.1086/183420}

\bibitem[{{Boksenberg} \& {Sargent}(1978)}]{BoksenbergSargent1978}
{Boksenberg}, A., \& {Sargent}, W.~L.~W. 1978, \apj, 220, 42,
  \dodoi{10.1086/155880}

\bibitem[{{Bolzonella} {et~al.}(2000){Bolzonella}, {Miralles}, \&
  {Pell{\'o}}}]{Bolzonella2000}
{Bolzonella}, M., {Miralles}, J.~M., \& {Pell{\'o}}, R. 2000, \aap, 363, 476.
\newblock \doarXiv{astro-ph/0003380}

\bibitem[{{Bordoloi} {et~al.}(2014){Bordoloi}, {Lilly}, {Hardmeier}, {Contini},
  {Kneib}, {Le Fevre}, {Mainieri}, {Renzini}, {Scodeggio}, {Zamorani},
  {Bardelli}, {Bolzonella}, {Bongiorno}, {Caputi}, {Carollo}, {Cucciati}, {de
  la Torre}, {de Ravel}, {Garilli}, {Iovino}, {Kampczyk}, {Kova{\v{c}}},
  {Knobel}, {Lamareille}, {Le Borgne}, {Le Brun}, {Maier}, {Mignoli}, {Oesch},
  {Pello}, {Peng}, {Perez Montero}, {Presotto}, {Silverman}, {Tanaka}, {Tasca},
  {Tresse}, {Vergani}, {Zucca}, {Cappi}, {Cimatti}, {Coppa}, {Franzetti},
  {Koekemoer}, {Moresco}, {Nair}, \& {Pozzetti}}]{Bordoloi2014}
{Bordoloi}, R., {Lilly}, S.~J., {Hardmeier}, E., {et~al.} 2014, \apj, 794, 130,
  \dodoi{10.1088/0004-637X/794/2/130}

\bibitem[{{Bosma}(1978)}]{Bosma1978}
{Bosma}, A. 1978, PhD thesis, Groningen University

\bibitem[{{Bregman}(1980)}]{Bregman1980}
{Bregman}, J.~N. 1980, \apj, 236, 577, \dodoi{10.1086/157776}

\bibitem[{{Bregman} \& {Houck}(1997)}]{BregmanHouck1997}
{Bregman}, J.~N., \& {Houck}, J.~C. 1997, \apj, 485, 159,
  \dodoi{10.1086/304397}

\bibitem[{{Brinchmann} {et~al.}(2004){Brinchmann}, {Charlot}, {White},
  {Tremonti}, {Kauffmann}, {Heckman}, \& {Brinkmann}}]{Brinchmann2004}
{Brinchmann}, J., {Charlot}, S., {White}, S.~D.~M., {et~al.} 2004, \mnras, 351,
  1151, \dodoi{10.1111/j.1365-2966.2004.07881.x}

\bibitem[{{Broeils} \& {Rhee}(1997)}]{BroeilsRhee1997}
{Broeils}, A.~H., \& {Rhee}, M.~H. 1997, \aap, 324, 877

\bibitem[{{Bruch} {et~al.}(2021){Bruch}, {Gal-Yam}, {Schulze}, {Yaron}, {Yang},
  {Soumagnac}, {Rigault}, {Strotjohann}, {Ofek}, {Sollerman}, {Masci},
  {Barbarino}, {Ho}, {Fremling}, {Perley}, {Nordin}, {Cenko}, {Adams},
  {Adreoni}, {Bellm}, {Blagorodnova}, {Bulla}, {Burdge}, {De}, {Dhawan},
  {Drake}, {Duev}, {Dugas}, {Graham}, {Graham}, {Irani}, {Jencson},
  {Karamehmetoglu}, {Kasliwal}, {Kim}, {Kulkarni}, {Kupfer}, {Liang},
  {Mahabal}, {Miller}, {Prince}, {Riddle}, {Sharma}, {Smith}, {Taddia},
  {Taggart}, {Walters}, \& {Yan}}]{Bruch2021}
{Bruch}, R.~J., {Gal-Yam}, A., {Schulze}, S., {et~al.} 2021, \apj, 912, 46,
  \dodoi{10.3847/1538-4357/abef05}

\bibitem[{{Burchett} {et~al.}(2016){Burchett}, {Tripp}, {Bordoloi}, {Werk},
  {Prochaska}, {Tumlinson}, {Willmer}, {O'Meara}, \& {Katz}}]{Burchett2016}
{Burchett}, J.~N., {Tripp}, T.~M., {Bordoloi}, R., {et~al.} 2016, \apj, 832,
  124, \dodoi{10.3847/0004-637X/832/2/124}

\bibitem[{{Chen} {et~al.}(2010{\natexlab{a}}){Chen}, {Helsby}, {Gauthier},
  {Shectman}, {Thompson}, \& {Tinker}}]{Chen2010a}
{Chen}, H.-W., {Helsby}, J.~E., {Gauthier}, J.-R., {et~al.} 2010{\natexlab{a}},
  \apj, 714, 1521, \dodoi{10.1088/0004-637X/714/2/1521}

\bibitem[{{Chen} {et~al.}(2010{\natexlab{b}}){Chen}, {Tremonti}, {Heckman},
  {Kauffmann}, {Weiner}, {Brinchmann}, \& {Wang}}]{ChenTremonti2010}
{Chen}, Y.-M., {Tremonti}, C.~A., {Heckman}, T.~M., {et~al.}
  2010{\natexlab{b}}, \aj, 140, 445, \dodoi{10.1088/0004-6256/140/2/445}

\bibitem[{{Cherinka} \& {Schulte-Ladbeck}(2011)}]{Cherinka2011}
{Cherinka}, B., \& {Schulte-Ladbeck}, R.~E. 2011, \aj, 142, 122,
  \dodoi{10.1088/0004-6256/142/4/122}

\bibitem[{{Ciotti} \& {Bertin}(1999)}]{CiottiBertin1999}
{Ciotti}, L., \& {Bertin}, G. 1999, \aap, 352, 447.
\newblock \doarXiv{astro-ph/9911078}

\bibitem[{{Concas} {et~al.}(2019){Concas}, {Popesso}, {Brusa}, {Mainieri}, \&
  {Thomas}}]{Concas2019}
{Concas}, A., {Popesso}, P., {Brusa}, M., {Mainieri}, V., \& {Thomas}, D. 2019,
  \aap, 622, A188, \dodoi{10.1051/0004-6361/201732152}

\bibitem[{{Cooper} {et~al.}(2008){Cooper}, {Bicknell}, {Sutherland}, \&
  {Bland-Hawthorn}}]{Cooper2008}
{Cooper}, J.~L., {Bicknell}, G.~V., {Sutherland}, R.~S., \& {Bland-Hawthorn},
  J. 2008, \apj, 674, 157, \dodoi{10.1086/524918}

\bibitem[{{Crawford}(1992)}]{Crawford1992}
{Crawford}, I.~A. 1992, \mnras, 259, 47, \dodoi{10.1093/mnras/259.1.47}

\bibitem[{{Creasey} {et~al.}(2013){Creasey}, {Theuns}, \&
  {Bower}}]{Creasey2013}
{Creasey}, P., {Theuns}, T., \& {Bower}, R.~G. 2013, \mnras, 429, 1922,
  \dodoi{10.1093/mnras/sts439}

\bibitem[{{Dastidar} {et~al.}(2021){Dastidar}, {Misra}, {Singh}, {Pastorello},
  {Sahu}, {Wang}, {Gangopadhyay}, {Tomasella}, {Zhang}, {Bose}, {Mo},
  {Elias-Rosa}, {Tartaglia}, {Yan}, {Kumar}, {Anupama}, {Pandey}, {Rui},
  {Zhang}, {Terreran}, {Ochner}, \& {Huang}}]{Dastidar2021}
{Dastidar}, R., {Misra}, K., {Singh}, M., {et~al.} 2021, \mnras, 504, 1009,
  \dodoi{10.1093/mnras/stab831}

\bibitem[{{de Avillez}(2000)}]{deAvillez2000}
{de Avillez}, M.~A. 2000, \mnras, 315, 479,
  \dodoi{10.1046/j.1365-8711.2000.03464.x}

\bibitem[{{de Blok} {et~al.}(2008){de Blok}, {Walter}, {Brinks},
  {Trachternach}, {Oh}, \& {Kennicutt}}]{deBlok2008}
{de Blok}, W.~J.~G., {Walter}, F., {Brinks}, E., {et~al.} 2008, \aj, 136, 2648,
  \dodoi{10.1088/0004-6256/136/6/2648}

\bibitem[{{Dettmar}(1990)}]{Dettmar1990}
{Dettmar}, R.~J. 1990, \aap, 232, L15

\bibitem[{{Edgar} \& {Savage}(1989)}]{EdgarSavage1989}
{Edgar}, R.~J., \& {Savage}, B.~D. 1989, \apj, 340, 762, \dodoi{10.1086/167435}

\bibitem[{{Egger} \& {Aschenbach}(1995)}]{Egger1995}
{Egger}, R.~J., \& {Aschenbach}, B. 1995, \aap, 294, L25.
\newblock \doarXiv{astro-ph/9412086}

\bibitem[{{El-Badry} {et~al.}(2018){El-Badry}, {Bradford}, {Quataert}, {Geha},
  {Boylan-Kolchin}, {Weisz}, {Wetzel}, {Hopkins}, {Chan}, {Fitts},
  {Kere{\v{s}}}, \& {Faucher-Gigu{\`e}re}}]{El-Badry2018}
{El-Badry}, K., {Bradford}, J., {Quataert}, E., {et~al.} 2018, \mnras, 477,
  1536, \dodoi{10.1093/mnras/sty730}

\bibitem[{{Farber} \& {Gronke}(2021)}]{FarberGronke2021}
{Farber}, R.~J., \& {Gronke}, M. 2021, arXiv e-prints, arXiv:2107.07991.
\newblock \doarXiv{2107.07991}

\bibitem[{{Fielding} {et~al.}(2018){Fielding}, {Quataert}, \&
  {Martizzi}}]{Fielding2018}
{Fielding}, D., {Quataert}, E., \& {Martizzi}, D. 2018, \mnras, 481, 3325,
  \dodoi{10.1093/mnras/sty2466}

\bibitem[{{Foreman-Mackey} {et~al.}(2013){Foreman-Mackey}, {Hogg}, {Lang}, \&
  {Goodman}}]{Foreman-Mackey2013}
{Foreman-Mackey}, D., {Hogg}, D.~W., {Lang}, D., \& {Goodman}, J. 2013, \pasp,
  125, 306, \dodoi{10.1086/670067}

\bibitem[{{Fraternali} \& {Binney}(2006)}]{FraternaliBinney2006}
{Fraternali}, F., \& {Binney}, J.~J. 2006, \mnras, 366, 449,
  \dodoi{10.1111/j.1365-2966.2005.09816.x}

\bibitem[{{Fraternali} \& {Binney}(2008)}]{FraternaliBinney2008}
---. 2008, \mnras, 386, 935, \dodoi{10.1111/j.1365-2966.2008.13071.x}

\bibitem[{{Fraternali} {et~al.}(2013){Fraternali}, {Marasco}, {Marinacci}, \&
  {Binney}}]{Fraternali2013}
{Fraternali}, F., {Marasco}, A., {Marinacci}, F., \& {Binney}, J. 2013, \apjl,
  764, L21, \dodoi{10.1088/2041-8205/764/2/L21}

\bibitem[{{Fraternali} {et~al.}(2002){Fraternali}, {van Moorsel}, {Sancisi}, \&
  {Oosterloo}}]{Fraternali2002}
{Fraternali}, F., {van Moorsel}, G., {Sancisi}, R., \& {Oosterloo}, T. 2002,
  \aj, 123, 3124, \dodoi{10.1086/340358}

\bibitem[{{Gillmon} \& {Shull}(2006)}]{Gillmon2006}
{Gillmon}, K., \& {Shull}, J.~M. 2006, \apj, 636, 908, \dodoi{10.1086/498055}

\bibitem[{{Girichidis} {et~al.}(2021){Girichidis}, {Naab}, {Walch}, \&
  {Berlok}}]{Girichidis2021}
{Girichidis}, P., {Naab}, T., {Walch}, S., \& {Berlok}, T. 2021, \mnras, 505,
  1083, \dodoi{10.1093/mnras/stab1203}

\bibitem[{{Girichidis} {et~al.}(2016){Girichidis}, {Walch}, {Naab}, {Gatto},
  {W{\"u}nsch}, {Glover}, {Klessen}, {Clark}, {Peters}, {Derigs}, \&
  {Baczynski}}]{Girichidis2016}
{Girichidis}, P., {Walch}, S., {Naab}, T., {et~al.} 2016, \mnras, 456, 3432,
  \dodoi{10.1093/mnras/stv2742}

\bibitem[{{Green}(2018)}]{Green2018}
{Green}, G. 2018, The Journal of Open Source Software, 3, 695,
  \dodoi{10.21105/joss.00695}

\bibitem[{{Haffner} {et~al.}(2003){Haffner}, {Reynolds}, {Tufte}, {Madsen},
  {Jaehnig}, \& {Percival}}]{Haffner2003}
{Haffner}, L.~M., {Reynolds}, R.~J., {Tufte}, S.~L., {et~al.} 2003, \apjs, 149,
  405, \dodoi{10.1086/378850}

\bibitem[{{Hartmann}(1904)}]{Hartmann1904}
{Hartmann}, J. 1904, \apj, 19, 268, \dodoi{10.1086/141112}

\bibitem[{{Heckman} {et~al.}(2000){Heckman}, {Lehnert}, {Strickland}, \&
  {Armus}}]{Heckman2000}
{Heckman}, T.~M., {Lehnert}, M.~D., {Strickland}, D.~K., \& {Armus}, L. 2000,
  \apjs, 129, 493, \dodoi{10.1086/313421}

\bibitem[{{Henley} {et~al.}(2015){Henley}, {Shelton}, {Kwak}, {Hill}, \& {Mac
  Low}}]{Henley2015}
{Henley}, D.~B., {Shelton}, R.~L., {Kwak}, K., {Hill}, A.~S., \& {Mac Low},
  M.-M. 2015, \apj, 800, 102, \dodoi{10.1088/0004-637X/800/2/102}

\bibitem[{{Heyer} \& {Dame}(2015)}]{Heyer2015}
{Heyer}, M., \& {Dame}, T.~M. 2015, \araa, 53, 583,
  \dodoi{10.1146/annurev-astro-082214-122324}

\bibitem[{{Hobbs}(1969)}]{Hobbs1969}
{Hobbs}, L.~M. 1969, \apj, 158, 461, \dodoi{10.1086/150210}

\bibitem[{{Hobbs}(1974)}]{Hobbs1974}
---. 1974, \apj, 191, 381, \dodoi{10.1086/152976}

\bibitem[{{Howk} {et~al.}(2018){Howk}, {Rueff}, {Lehner}, {Wotta}, {Croxall},
  \& {Savage}}]{Howk2018}
{Howk}, J.~C., {Rueff}, K.~M., {Lehner}, N., {et~al.} 2018, \apj, 856, 166,
  \dodoi{10.3847/1538-4357/aab1fa}

\bibitem[{{Howk} \& {Savage}(1997)}]{HowkSavage1997}
{Howk}, J.~C., \& {Savage}, B.~D. 1997, \aj, 114, 2463, \dodoi{10.1086/118660}

\bibitem[{{Howk} \& {Savage}(2000)}]{HowkSavage2000}
---. 2000, \aj, 119, 644, \dodoi{10.1086/301210}

\bibitem[{{Howk} {et~al.}(2003){Howk}, {Sembach}, \& {Savage}}]{Howk2003}
{Howk}, J.~C., {Sembach}, K.~R., \& {Savage}, B.~D. 2003, \apj, 586, 249,
  \dodoi{10.1086/346262}

\bibitem[{{Jenkins}(1978)}]{Jenkins1978}
{Jenkins}, E.~B. 1978, \apj, 220, 107, \dodoi{10.1086/155885}

\bibitem[{{Joung} \& {Mac Low}(2006)}]{JoungMacLow2006}
{Joung}, M.~K.~R., \& {Mac Low}, M.-M. 2006, \apj, 653, 1266,
  \dodoi{10.1086/508795}

\bibitem[{{Kacprzak} {et~al.}(2008){Kacprzak}, {Churchill}, {Steidel}, \&
  {Murphy}}]{Kacprzak2008}
{Kacprzak}, G.~G., {Churchill}, C.~W., {Steidel}, C.~C., \& {Murphy}, M.~T.
  2008, \aj, 135, 922, \dodoi{10.1088/0004-6256/135/3/922}

\bibitem[{{Kacprzak} {et~al.}(2013){Kacprzak}, {Cooke}, {Churchill},
  {Ryan-Weber}, \& {Nielsen}}]{Kacprzak2013}
{Kacprzak}, G.~G., {Cooke}, J., {Churchill}, C.~W., {Ryan-Weber}, E.~V., \&
  {Nielsen}, N.~M. 2013, \apjl, 777, L11, \dodoi{10.1088/2041-8205/777/1/L11}

\bibitem[{{Kado-Fong} {et~al.}(2020){Kado-Fong}, {Kim}, {Ostriker}, \&
  {Kim}}]{Kado-Fong2020}
{Kado-Fong}, E., {Kim}, J.-G., {Ostriker}, E.~C., \& {Kim}, C.-G. 2020, \apj,
  897, 143, \dodoi{10.3847/1538-4357/ab9abd}

\bibitem[{{Kalberla} {et~al.}(2005){Kalberla}, {Burton}, {Hartmann}, {Arnal},
  {Bajaja}, {Morras}, \& {P{\"o}ppel}}]{Kalberla2005}
{Kalberla}, P.~M.~W., {Burton}, W.~B., {Hartmann}, D., {et~al.} 2005, \aap,
  440, 775, \dodoi{10.1051/0004-6361:20041864}

\bibitem[{{Kalberla} \& {Kerp}(2009)}]{KalberlaKerp2009}
{Kalberla}, P. M.~W., \& {Kerp}, J. 2009, \araa, 47, 27,
  \dodoi{10.1146/annurev-astro-082708-101823}

\bibitem[{{Kamphuis} {et~al.}(2007){Kamphuis}, {Holwerda}, {Allen}, {Peletier},
  \& {van der Kruit}}]{Kamphuis2007}
{Kamphuis}, P., {Holwerda}, B.~W., {Allen}, R.~J., {Peletier}, R.~F., \& {van
  der Kruit}, P.~C. 2007, \aap, 471, L1, \dodoi{10.1051/0004-6361:20077951}

\bibitem[{{Kamphuis} {et~al.}(2015){Kamphuis}, {J{\'o}zsa}, {Oh}, {Spekkens},
  {Urbancic}, {Serra}, {Koribalski}, \& {Dettmar}}]{Kamphuis2015}
{Kamphuis}, P., {J{\'o}zsa}, G.~I.~G., {Oh}, S. .~H., {et~al.} 2015, \mnras,
  452, 3139, \dodoi{10.1093/mnras/stv1480}

\bibitem[{{Kennicutt}(1998)}]{Kennicutt1998}
{Kennicutt}, Robert~C., J. 1998, \araa, 36, 189,
  \dodoi{10.1146/annurev.astro.36.1.189}

\bibitem[{{Kerp} {et~al.}(1999){Kerp}, {Burton}, {Egger}, {Freyberg},
  {Hartmann}, {Kalberla}, {Mebold}, \& {Pietz}}]{Kerp1999}
{Kerp}, J., {Burton}, W.~B., {Egger}, R., {et~al.} 1999, \aap, 342, 213.
\newblock \doarXiv{astro-ph/9810307}

\bibitem[{{Kim} \& {Ostriker}(2018)}]{KimOstriker2018}
{Kim}, C.-G., \& {Ostriker}, E.~C. 2018, \apj, 853, 173,
  \dodoi{10.3847/1538-4357/aaa5ff}

\bibitem[{{Kim} {et~al.}(2020){Kim}, {Ostriker}, {Somerville}, {Bryan},
  {Fielding}, {Forbes}, {Hayward}, {Hernquist}, \& {Pandya}}]{Kim2020}
{Kim}, C.-G., {Ostriker}, E.~C., {Somerville}, R.~S., {et~al.} 2020, \apj, 900,
  61, \dodoi{10.3847/1538-4357/aba962}

\bibitem[{{Korpi} {et~al.}(1999){Korpi}, {Brandenburg}, {Shukurov}, \&
  {Tuominen}}]{Korpi1999}
{Korpi}, M.~J., {Brandenburg}, A., {Shukurov}, A., \& {Tuominen}, I. 1999,
  \aap, 350, 230

\bibitem[{{Kreckel} {et~al.}(2018){Kreckel}, {Faesi}, {Kruijssen}, {Schruba},
  {Groves}, {Leroy}, {Bigiel}, {Blanc}, {Chevance}, {Herrera}, {Hughes},
  {McElroy}, {Pety}, {Querejeta}, {Rosolowsky}, {Schinnerer}, {Sun}, {Usero},
  \& {Utomo}}]{Kreckel2018}
{Kreckel}, K., {Faesi}, C., {Kruijssen}, J.~M.~D., {et~al.} 2018, \apjl, 863,
  L21, \dodoi{10.3847/2041-8213/aad77d}

\bibitem[{{Kulkarni} {et~al.}(2022){Kulkarni}, {Bowen}, {Straka}, {York},
  {Gupta}, {Noterdaeme}, \& {Srianand}}]{Kulkarni2022}
{Kulkarni}, V.~P., {Bowen}, D.~V., {Straka}, L.~A., {et~al.} 2022, \apj, 929,
  150, \dodoi{10.3847/1538-4357/ac5fab}

\bibitem[{{Kuntz} \& {Danly}(1996)}]{KuntzDanly1996}
{Kuntz}, K.~D., \& {Danly}, L. 1996, \apj, 457, 703, \dodoi{10.1086/176765}

\bibitem[{{Kuntz} \& {Snowden}(2000)}]{KuntzSnowden2000}
{Kuntz}, K.~D., \& {Snowden}, S.~L. 2000, \apj, 543, 195,
  \dodoi{10.1086/317071}

\bibitem[{{Lan} {et~al.}(2014){Lan}, {M{\'e}nard}, \& {Zhu}}]{Lan2014}
{Lan}, T.-W., {M{\'e}nard}, B., \& {Zhu}, G. 2014, \apj, 795, 31,
  \dodoi{10.1088/0004-637X/795/1/31}

\bibitem[{{Lan} \& {Mo}(2018)}]{LanMo2018}
{Lan}, T.-W., \& {Mo}, H. 2018, \apj, 866, 36, \dodoi{10.3847/1538-4357/aadc08}

\bibitem[{{Lanzetta} \& {Bowen}(1990)}]{LanzettaBowen1990}
{Lanzetta}, K.~M., \& {Bowen}, D. 1990, \apj, 357, 321, \dodoi{10.1086/168922}

\bibitem[{{Lehner} \& {Howk}(2011)}]{LehnerHowk2011}
{Lehner}, N., \& {Howk}, J.~C. 2011, Science, 334, 955,
  \dodoi{10.1126/science.1209069}

\bibitem[{{Li} {et~al.}(2021){Li}, {Marasco}, {Fraternali}, {Trager}, \&
  {Verheijen}}]{Li2021}
{Li}, A., {Marasco}, A., {Fraternali}, F., {Trager}, S., \& {Verheijen}, M.
  A.~W. 2021, \mnras, 504, 3013, \dodoi{10.1093/mnras/stab1043}

\bibitem[{{Mac Low} \& {McCray}(1988)}]{MacLowMcCray1988}
{Mac Low}, M.-M., \& {McCray}, R. 1988, \apj, 324, 776, \dodoi{10.1086/165936}

\bibitem[{{Manuwal} {et~al.}(2021){Manuwal}, {Ludlow}, {Stevens}, {Wright}, \&
  {Robotham}}]{Manuwal2021}
{Manuwal}, A., {Ludlow}, A.~D., {Stevens}, A. R.~H., {Wright}, R.~J., \&
  {Robotham}, A. S.~G. 2021, arXiv e-prints, arXiv:2109.11214.
\newblock \doarXiv{2109.11214}

\bibitem[{{Marasco} \& {Fraternali}(2017)}]{MarascoFraternali2017}
{Marasco}, A., \& {Fraternali}, F. 2017, \mnras, 464, L100,
  \dodoi{10.1093/mnrasl/slw195}

\bibitem[{{Marasco} {et~al.}(2012){Marasco}, {Fraternali}, \&
  {Binney}}]{Marasco2012}
{Marasco}, A., {Fraternali}, F., \& {Binney}, J.~J. 2012, \mnras, 419, 1107,
  \dodoi{10.1111/j.1365-2966.2011.19771.x}

\bibitem[{{Marasco} {et~al.}(2013){Marasco}, {Marinacci}, \&
  {Fraternali}}]{Marasco2013}
{Marasco}, A., {Marinacci}, F., \& {Fraternali}, F. 2013, \mnras, 433, 1634,
  \dodoi{10.1093/mnras/stt836}

\bibitem[{{Marasco} {et~al.}(2019){Marasco}, {Fraternali}, {Heald}, {de Blok},
  {Oosterloo}, {Kamphuis}, {J{\'o}zsa}, {Vargas}, {Winkel}, {Walterbos},
  {Dettmar}, \& {Juẗte}}]{Marasco2019}
{Marasco}, A., {Fraternali}, F., {Heald}, G., {et~al.} 2019, \aap, 631, A50,
  \dodoi{10.1051/0004-6361/201936338}

\bibitem[{{Martin}(2005)}]{Martin2005}
{Martin}, C.~L. 2005, \apj, 621, 227, \dodoi{10.1086/427277}

\bibitem[{{Martizzi} {et~al.}(2016){Martizzi}, {Fielding},
  {Faucher-Gigu{\`e}re}, \& {Quataert}}]{Martizzi2016}
{Martizzi}, D., {Fielding}, D., {Faucher-Gigu{\`e}re}, C.-A., \& {Quataert}, E.
  2016, \mnras, 459, 2311, \dodoi{10.1093/mnras/stw745}

\bibitem[{{McClure-Griffiths} {et~al.}(2009){McClure-Griffiths}, {Pisano},
  {Calabretta}, {Ford}, {Lockman}, {Staveley-Smith}, {Kalberla}, {Bailin},
  {Dedes}, {Janowiecki}, {Gibson}, {Murphy}, {Nakanishi}, \&
  {Newton-McGee}}]{McClure-Griffiths2009}
{McClure-Griffiths}, N.~M., {Pisano}, D.~J., {Calabretta}, M.~R., {et~al.}
  2009, \apjs, 181, 398, \dodoi{10.1088/0067-0049/181/2/398}

\bibitem[{{M{\'e}nard} {et~al.}(2011){M{\'e}nard}, {Wild}, {Nestor}, {Quider},
  {Zibetti}, {Rao}, \& {Turnshek}}]{Menard2011}
{M{\'e}nard}, B., {Wild}, V., {Nestor}, D., {et~al.} 2011, \mnras, 417, 801,
  \dodoi{10.1111/j.1365-2966.2011.18227.x}

\bibitem[{{Moster} {et~al.}(2013){Moster}, {Naab}, \& {White}}]{Moster2013}
{Moster}, B.~P., {Naab}, T., \& {White}, S. D.~M. 2013, \mnras, 428, 3121,
  \dodoi{10.1093/mnras/sts261}

\bibitem[{{Moustakas} {et~al.}(2013){Moustakas}, {Coil}, {Aird}, {Blanton},
  {Cool}, {Eisenstein}, {Mendez}, {Wong}, {Zhu}, \& {Arnouts}}]{Moustakas2013}
{Moustakas}, J., {Coil}, A.~L., {Aird}, J., {et~al.} 2013, \apj, 767, 50,
  \dodoi{10.1088/0004-637X/767/1/50}

\bibitem[{{Munari} \& {Zwitter}(1997)}]{MunariZwitter1997}
{Munari}, U., \& {Zwitter}, T. 1997, \aap, 318, 269

\bibitem[{{M{\"u}nch} \& {Zirin}(1961)}]{MunchZirin1961}
{M{\"u}nch}, G., \& {Zirin}, H. 1961, \apj, 133, 11, \dodoi{10.1086/146999}

\bibitem[{{Murga} {et~al.}(2015){Murga}, {Zhu}, {M{\'e}nard}, \&
  {Lan}}]{Murga2015}
{Murga}, M., {Zhu}, G., {M{\'e}nard}, B., \& {Lan}, T.-W. 2015, \mnras, 452,
  511, \dodoi{10.1093/mnras/stv1277}

\bibitem[{{Nielsen} {et~al.}(2013){Nielsen}, {Churchill}, \&
  {Kacprzak}}]{Nielsen2013}
{Nielsen}, N.~M., {Churchill}, C.~W., \& {Kacprzak}, G.~G. 2013, \apj, 776,
  115, \dodoi{10.1088/0004-637X/776/2/115}

\bibitem[{{Norman} \& {Ikeuchi}(1989)}]{Norman1989}
{Norman}, C.~A., \& {Ikeuchi}, S. 1989, \apj, 345, 372, \dodoi{10.1086/167912}

\bibitem[{{Noterdaeme} {et~al.}(2010){Noterdaeme}, {Srianand}, \&
  {Mohan}}]{Noterdaeme2010}
{Noterdaeme}, P., {Srianand}, R., \& {Mohan}, V. 2010, \mnras, 403, 906,
  \dodoi{10.1111/j.1365-2966.2009.16169.x}

\bibitem[{{Oh} {et~al.}(2011){Oh}, {de Blok}, {Brinks}, {Walter}, \&
  {Kennicutt}}]{Oh2011}
{Oh}, S.-H., {de Blok}, W.~J.~G., {Brinks}, E., {Walter}, F., \& {Kennicutt},
  Robert~C., J. 2011, \aj, 141, 193, \dodoi{10.1088/0004-6256/141/6/193}

\bibitem[{{Oh} {et~al.}(2018){Oh}, {Staveley-Smith}, {Spekkens}, {Kamphuis}, \&
  {Koribalski}}]{Oh2018}
{Oh}, S.-H., {Staveley-Smith}, L., {Spekkens}, K., {Kamphuis}, P., \&
  {Koribalski}, B.~S. 2018, \mnras, 473, 3256, \dodoi{10.1093/mnras/stx2304}

\bibitem[{{Oman} {et~al.}(2019){Oman}, {Marasco}, {Navarro}, {Frenk}, {Schaye},
  \& {Ben{\'\i}tez-Llambay}}]{Oman2019}
{Oman}, K.~A., {Marasco}, A., {Navarro}, J.~F., {et~al.} 2019, \mnras, 482,
  821, \dodoi{10.1093/mnras/sty2687}

\bibitem[{{Oosterloo} {et~al.}(2007){Oosterloo}, {Fraternali}, \&
  {Sancisi}}]{Oosterloo2007}
{Oosterloo}, T., {Fraternali}, F., \& {Sancisi}, R. 2007, \aj, 134, 1019,
  \dodoi{10.1086/520332}

\bibitem[{{Phillips} {et~al.}(1984){Phillips}, {Pettini}, \&
  {Gondhalekar}}]{Phillips1984}
{Phillips}, A.~P., {Pettini}, M., \& {Gondhalekar}, P.~M. 1984, \mnras, 206,
  337, \dodoi{10.1093/mnras/206.2.337}

\bibitem[{{Phillips} {et~al.}(2013){Phillips}, {Simon}, {Morrell}, {Burns},
  {Cox}, {Foley}, {Karakas}, {Patat}, {Sternberg}, {Williams}, {Gal-Yam},
  {Hsiao}, {Leonard}, {Persson}, {Stritzinger}, {Thompson}, {Campillay},
  {Contreras}, {Folatelli}, {Freedman}, {Hamuy}, {Roth}, {Shields}, {Suntzeff},
  {Chomiuk}, {Ivans}, {Madore}, {Penprase}, {Perley}, {Pignata}, {Preston}, \&
  {Soderberg}}]{Phillips2013}
{Phillips}, M.~M., {Simon}, J.~D., {Morrell}, N., {et~al.} 2013, \apj, 779, 38,
  \dodoi{10.1088/0004-637X/779/1/38}

\bibitem[{{Planck Collaboration} {et~al.}(2016){Planck Collaboration},
  {Aghanim}, {Ashdown}, {Aumont}, {Baccigalupi}, {Ballardini}, {Banday},
  {Barreiro}, {Bartolo}, {Basak}, {Benabed}, {Bernard}, {Bersanelli},
  {Bielewicz}, {Bonavera}, {Bond}, {Borrill}, {Bouchet}, {Boulanger},
  {Burigana}, {Calabrese}, {Cardoso}, {Carron}, {Chiang}, {Colombo}, {Comis},
  {Couchot}, {Coulais}, {Crill}, {Curto}, {Cuttaia}, {de Bernardis}, {de
  Zotti}, {Delabrouille}, {Di Valentino}, {Dickinson}, {Diego}, {Dor{\'e}},
  {Douspis}, {Ducout}, {Dupac}, {Dusini}, {Elsner}, {En{\ss}lin}, {Eriksen},
  {Falgarone}, {Fantaye}, {Finelli}, {Forastieri}, {Frailis}, {Fraisse},
  {Franceschi}, {Frolov}, {Galeotta}, {Galli}, {Ganga}, {G{\'e}nova-Santos},
  {Gerbino}, {Ghosh}, {Giraud-H{\'e}raud}, {Gonz{\'a}lez-Nuevo}, {G{\'o}rski},
  {Gruppuso}, {Gudmundsson}, {Hansen}, {Helou}, {Henrot-Versill{\'e}},
  {Herranz}, {Hivon}, {Huang}, {Jaffe}, {Jones}, {Keih{\"a}nen}, {Keskitalo},
  {Kiiveri}, {Kisner}, {Krachmalnicoff}, {Kunz}, {Kurki-Suonio}, {Lamarre},
  {Langer}, {Lasenby}, {Lattanzi}, {Lawrence}, {Le Jeune}, {Levrier}, {Lilje},
  {Lilley}, {Lindholm}, {L{\'o}pez-Caniego}, {Ma}, {Mac{\'\i}as-P{\'e}rez},
  {Maggio}, {Maino}, {Mandolesi}, {Mangilli}, {Maris}, {Martin},
  {Mart{\'\i}nez-Gonz{\'a}lez}, {Matarrese}, {Mauri}, {McEwen}, {Melchiorri},
  {Mennella}, {Migliaccio}, {Miville-Desch{\^e}nes}, {Molinari}, {Moneti},
  {Montier}, {Morgante}, {Moss}, {Natoli}, {Oxborrow}, {Pagano}, {Paoletti},
  {Patanchon}, {Perdereau}, {Perotto}, {Pettorino}, {Piacentini},
  {Plaszczynski}, {Polastri}, {Polenta}, {Puget}, {Rachen}, {Racine},
  {Reinecke}, {Remazeilles}, {Renzi}, {Rocha}, {Rosset}, {Rossetti}, {Roudier},
  {Rubi{\~n}o-Mart{\'\i}n}, {Ruiz-Granados}, {Salvati}, {Sandri}, {Savelainen},
  {Scott}, {Sirignano}, {Sirri}, {Soler}, {Spencer}, {Suur-Uski}, {Tauber},
  {Tavagnacco}, {Tenti}, {Toffolatti}, {Tomasi}, {Tristram}, {Trombetti},
  {Valiviita}, {Van Tent}, {Vielva}, {Villa}, {Vittorio}, {Wandelt}, {Wehus},
  {Zacchei}, \& {Zonca}}]{Planck2016}
{Planck Collaboration}, {Aghanim}, N., {Ashdown}, M., {et~al.} 2016, \aap, 596,
  A109, \dodoi{10.1051/0004-6361/201629022}

\bibitem[{{Poznanski} {et~al.}(2012){Poznanski}, {Prochaska}, \&
  {Bloom}}]{Poznanski2012}
{Poznanski}, D., {Prochaska}, J.~X., \& {Bloom}, J.~S. 2012, \mnras, 426, 1465,
  \dodoi{10.1111/j.1365-2966.2012.21796.x}

\bibitem[{{Prevot} {et~al.}(1984){Prevot}, {Lequeux}, {Maurice}, {Prevot}, \&
  {Rocca-Volmerange}}]{Prevot1984}
{Prevot}, M.~L., {Lequeux}, J., {Maurice}, E., {Prevot}, L., \&
  {Rocca-Volmerange}, B. 1984, \aap, 132, 389

\bibitem[{{Prochaska} {et~al.}(2008){Prochaska}, {Chen}, {Wolfe},
  {Dessauges-Zavadsky}, \& {Bloom}}]{Prochaska2008}
{Prochaska}, J.~X., {Chen}, H.-W., {Wolfe}, A.~M., {Dessauges-Zavadsky}, M., \&
  {Bloom}, J.~S. 2008, \apj, 672, 59, \dodoi{10.1086/523689}

\bibitem[{{Prochaska} \& {Wolfe}(1997)}]{ProchaskaWolfe1997}
{Prochaska}, J.~X., \& {Wolfe}, A.~M. 1997, \apj, 487, 73,
  \dodoi{10.1086/304591}

\bibitem[{Prochaska {et~al.}(2016)Prochaska, Tejos, Crighton, jnburchett,
  Tuo-Ji, tiffanyhsyu, ktirimba, jhennawi, O'Meara, \& Werk}]{linetools2016}
Prochaska, J.~X., Tejos, N., Crighton, N., {et~al.} 2016, linetools/linetools:
  Second major release, v0.2,  Zenodo, \dodoi{10.5281/zenodo.168270}

\bibitem[{{Puspitarini} \& {Lallement}(2012)}]{Puspitarini2012}
{Puspitarini}, L., \& {Lallement}, R. 2012, \aap, 545, A21,
  \dodoi{10.1051/0004-6361/201219284}

\bibitem[{{Rand} {et~al.}(1990){Rand}, {Kulkarni}, \& {Hester}}]{Rand1990}
{Rand}, R.~J., {Kulkarni}, S.~R., \& {Hester}, J.~J. 1990, \apjl, 352, L1,
  \dodoi{10.1086/185679}

\bibitem[{{Richmond} {et~al.}(1994){Richmond}, {Treffers}, {Filippenko},
  {Paik}, {Leibundgut}, {Schulman}, \& {Cox}}]{Richmond1994}
{Richmond}, M.~W., {Treffers}, R.~R., {Filippenko}, A.~V., {et~al.} 1994, \aj,
  107, 1022, \dodoi{10.1086/116915}

\bibitem[{{Richter}(2017)}]{Richter2017}
{Richter}, P. 2017, {\it Gas Accretion onto Galaxies}, ed. A.~{Fox} \&
  R.~{Dav{\'e}}, Vol. 430, 15, \dodoi{10.1007/978-3-319-52512-9\_2}

\bibitem[{{Richter} {et~al.}(2011){Richter}, {Krause}, {Fechner}, {Charlton},
  \& {Murphy}}]{Richter2011}
{Richter}, P., {Krause}, F., {Fechner}, C., {Charlton}, J.~C., \& {Murphy},
  M.~T. 2011, \aap, 528, A12, \dodoi{10.1051/0004-6361/201015566}

\bibitem[{{Richter} {et~al.}(2001{\natexlab{a}}){Richter}, {Sembach}, {Wakker},
  \& {Savage}}]{Richter2001c}
{Richter}, P., {Sembach}, K.~R., {Wakker}, B.~P., \& {Savage}, B.~D.
  2001{\natexlab{a}}, \apjl, 562, L181, \dodoi{10.1086/338050}

\bibitem[{{Richter} {et~al.}(2001{\natexlab{b}}){Richter}, {Sembach}, {Wakker},
  {Savage}, {Tripp}, {Murphy}, {Kalberla}, \& {Jenkins}}]{Richter2001a}
{Richter}, P., {Sembach}, K.~R., {Wakker}, B.~P., {et~al.} 2001{\natexlab{b}},
  \apj, 559, 318, \dodoi{10.1086/322401}

\bibitem[{{Roberts-Borsani} {et~al.}(2020){Roberts-Borsani}, {Saintonge},
  {Masters}, \& {Stark}}]{RobertsBorsani2020}
{Roberts-Borsani}, G.~W., {Saintonge}, A., {Masters}, K.~L., \& {Stark}, D.~V.
  2020, \mnras, 493, 3081, \dodoi{10.1093/mnras/staa464}

\bibitem[{{Rogstad} {et~al.}(1974){Rogstad}, {Lockhart}, \&
  {Wright}}]{Rogstad1974}
{Rogstad}, D.~H., {Lockhart}, I.~A., \& {Wright}, M.~C.~H. 1974, \apj, 193,
  309, \dodoi{10.1086/153164}

\bibitem[{{R{\"o}hser} {et~al.}(2016){R{\"o}hser}, {Kerp}, {Ben Bekhti}, \&
  {Winkel}}]{Rohser2016}
{R{\"o}hser}, T., {Kerp}, J., {Ben Bekhti}, N., \& {Winkel}, B. 2016, \aap,
  592, A142, \dodoi{10.1051/0004-6361/201526801}

\bibitem[{{Routly} \& {Spitzer}(1952)}]{RoutlySpitzer1952}
{Routly}, P.~M., \& {Spitzer}, Lyman, J. 1952, \apj, 115, 227,
  \dodoi{10.1086/145535}

\bibitem[{{Rubin} {et~al.}(2018){Rubin}, {Diamond-Stanic}, {Coil}, {Crighton},
  \& {Moustakas}}]{Rubin2018a}
{Rubin}, K. H.~R., {Diamond-Stanic}, A.~M., {Coil}, A.~L., {Crighton}, N.
  H.~M., \& {Moustakas}, J. 2018, \apj, 853, 95,
  \dodoi{10.3847/1538-4357/aa9792}

\bibitem[{{Rubin} {et~al.}(2014){Rubin}, {Prochaska}, {Koo}, {Phillips},
  {Martin}, \& {Winstrom}}]{Rubin2014}
{Rubin}, K. H.~R., {Prochaska}, J.~X., {Koo}, D.~C., {et~al.} 2014, \apj, 794,
  156, \dodoi{10.1088/0004-637X/794/2/156}

\bibitem[{{Rupke} {et~al.}(2005){Rupke}, {Veilleux}, \& {Sanders}}]{Rupke2005}
{Rupke}, D.~S., {Veilleux}, S., \& {Sanders}, D.~B. 2005, \apjs, 160, 87,
  \dodoi{10.1086/432886}

\bibitem[{{Rupke} {et~al.}(2017){Rupke}, {G{\"u}ltekin}, \&
  {Veilleux}}]{Rupke2017}
{Rupke}, D. S.~N., {G{\"u}ltekin}, K., \& {Veilleux}, S. 2017, \apj, 850, 40,
  \dodoi{10.3847/1538-4357/aa94d1}

\bibitem[{{Rupke} {et~al.}(2021){Rupke}, {Thomas}, \& {Dopita}}]{Rupke2021}
{Rupke}, D. S.~N., {Thomas}, A.~D., \& {Dopita}, M.~A. 2021, \mnras, 503, 4748,
  \dodoi{10.1093/mnras/stab743}

\bibitem[{{Savage} \& {Sembach}(1991)}]{SavageSembach1991}
{Savage}, B.~D., \& {Sembach}, K.~R. 1991, \apj, 379, 245,
  \dodoi{10.1086/170498}

\bibitem[{{Savage} \& {Wakker}(2009)}]{SavageWakker2009}
{Savage}, B.~D., \& {Wakker}, B.~P. 2009, \apj, 702, 1472,
  \dodoi{10.1088/0004-637X/702/2/1472}

\bibitem[{{Savage} {et~al.}(2003){Savage}, {Sembach}, {Wakker}, {Richter},
  {Meade}, {Jenkins}, {Shull}, {Moos}, \& {Sonneborn}}]{Savage2003}
{Savage}, B.~D., {Sembach}, K.~R., {Wakker}, B.~P., {et~al.} 2003, \apjs, 146,
  125, \dodoi{10.1086/346229}

\bibitem[{{Schlafly} \& {Finkbeiner}(2011)}]{SchlaflyFinkbeiner2011}
{Schlafly}, E.~F., \& {Finkbeiner}, D.~P. 2011, \apj, 737, 103,
  \dodoi{10.1088/0004-637X/737/2/103}

\bibitem[{{Schlegel} {et~al.}(1998){Schlegel}, {Finkbeiner}, \&
  {Davis}}]{SFD98}
{Schlegel}, D.~J., {Finkbeiner}, D.~P., \& {Davis}, M. 1998, \apj, 500, 525,
  \dodoi{10.1086/305772}

\bibitem[{{Schneider} {et~al.}(2007){Schneider}, {Hall}, {Richards}, {Strauss},
  {Vanden Berk}, {Anderson}, {Brandt}, {Fan}, {Jester}, {Gray}, {Gunn},
  {SubbaRao}, {Thakar}, {Stoughton}, {Szalay}, {Yanny}, {York}, {Bahcall},
  {Barentine}, {Blanton}, {Brewington}, {Brinkmann}, {Brunner}, {Castander},
  {Csabai}, {Frieman}, {Fukugita}, {Harvanek}, {Hogg}, {Ivezi{\'c}}, {Kent},
  {Kleinman}, {Knapp}, {Kron}, {Krzesi{\'n}ski}, {Long}, {Lupton}, {Nitta},
  {Pier}, {Saxe}, {Shen}, {Snedden}, {Weinberg}, \& {Wu}}]{Schneider2007}
{Schneider}, D.~P., {Hall}, P.~B., {Richards}, G.~T., {et~al.} 2007, \aj, 134,
  102, \dodoi{10.1086/518474}

\bibitem[{{Schwartz} \& {Martin}(2004)}]{SchwartzMartin2004}
{Schwartz}, C.~M., \& {Martin}, C.~L. 2004, \apj, 610, 201,
  \dodoi{10.1086/421546}

\bibitem[{{Sembach} \& {Danks}(1994)}]{Sembach1994}
{Sembach}, K.~R., \& {Danks}, A.~C. 1994, \aap, 289, 539

\bibitem[{{Sembach} {et~al.}(1993){Sembach}, {Danks}, \&
  {Savage}}]{Sembach1993}
{Sembach}, K.~R., {Danks}, A.~C., \& {Savage}, B.~D. 1993, \aaps, 100, 107

\bibitem[{{Shapiro} \& {Field}(1976)}]{ShapiroField1976}
{Shapiro}, P.~R., \& {Field}, G.~B. 1976, \apj, 205, 762,
  \dodoi{10.1086/154332}

\bibitem[{{Sheinis} {et~al.}(2002){Sheinis}, {Bolte}, {Epps}, {Kibrick},
  {Miller}, {Radovan}, {Bigelow}, \& {Sutin}}]{Sheinis2002}
{Sheinis}, A.~I., {Bolte}, M., {Epps}, H.~W., {et~al.} 2002, \pasp, 114, 851,
  \dodoi{10.1086/341706}

\bibitem[{{Smith} \& {Andrews}(2020)}]{SmithAndrews2020}
{Smith}, N., \& {Andrews}, J.~E. 2020, \mnras, 499, 3544,
  \dodoi{10.1093/mnras/staa3047}

\bibitem[{{Smoker} {et~al.}(2015){Smoker}, {Keenan}, \& {Fox}}]{Smoker2015}
{Smoker}, J.~V., {Keenan}, F.~P., \& {Fox}, A.~J. 2015, \aap, 582, A59,
  \dodoi{10.1051/0004-6361/201425190}

\bibitem[{{Stocke} {et~al.}(1991){Stocke}, {Case}, {Donahue}, {Shull}, \&
  {Snow}}]{Stocke1991}
{Stocke}, J.~T., {Case}, J., {Donahue}, M., {Shull}, J.~M., \& {Snow}, T.~P.
  1991, \apj, 374, 72, \dodoi{10.1086/170097}

\bibitem[{{Straka} {et~al.}(2013){Straka}, {Whichard}, {Kulkarni}, {Bishof},
  {Bowen}, {Khare}, \& {York}}]{Straka2013}
{Straka}, L.~A., {Whichard}, Z.~L., {Kulkarni}, V.~P., {et~al.} 2013, \mnras,
  436, 3200, \dodoi{10.1093/mnras/stt1798}

\bibitem[{{Straka} {et~al.}(2015){Straka}, {Noterdaeme}, {Srianand},
  {Nutalaya}, {Kulkarni}, {Khare}, {Bowen}, {Bishof}, \& {York}}]{Straka2015}
{Straka}, L.~A., {Noterdaeme}, P., {Srianand}, R., {et~al.} 2015, \mnras, 447,
  3856, \dodoi{10.1093/mnras/stu2739}

\bibitem[{{Swaters} {et~al.}(2002){Swaters}, {van Albada}, {van der Hulst}, \&
  {Sancisi}}]{Swaters2002}
{Swaters}, R.~A., {van Albada}, T.~S., {van der Hulst}, J.~M., \& {Sancisi}, R.
  2002, \aap, 390, 829, \dodoi{10.1051/0004-6361:20011755}

\bibitem[{{Taylor} {et~al.}(2011){Taylor}, {Hopkins}, {Baldry}, {Brown},
  {Driver}, {Kelvin}, {Hill}, {Robotham}, {Bland-Hawthorn}, {Jones}, {Sharp},
  {Thomas}, {Liske}, {Loveday}, {Norberg}, {Peacock}, {Bamford}, {Brough},
  {Colless}, {Cameron}, {Conselice}, {Croom}, {Frenk}, {Gunawardhana},
  {Kuijken}, {Nichol}, {Parkinson}, {Phillipps}, {Pimbblet}, {Popescu},
  {Prescott}, {Sutherland}, {Tuffs}, {van Kampen}, \&
  {Wijesinghe}}]{Taylor2011}
{Taylor}, E.~N., {Hopkins}, A.~M., {Baldry}, I.~K., {et~al.} 2011, \mnras, 418,
  1587, \dodoi{10.1111/j.1365-2966.2011.19536.x}

\bibitem[{{Tenorio-Tagle} \& {Bodenheimer}(1988)}]{TenorioTagle1988}
{Tenorio-Tagle}, G., \& {Bodenheimer}, P. 1988, \araa, 26, 145,
  \dodoi{10.1146/annurev.aa.26.090188.001045}

\bibitem[{{Thom} {et~al.}(2006){Thom}, {Putman}, {Gibson}, {Christlieb},
  {Flynn}, {Beers}, {Wilhelm}, \& {Lee}}]{Thom2006}
{Thom}, C., {Putman}, M.~E., {Gibson}, B.~K., {et~al.} 2006, \apjl, 638, L97,
  \dodoi{10.1086/501005}

\bibitem[{{Tomisaka} \& {Ikeuchi}(1986)}]{TomisakaIkeuchi1986}
{Tomisaka}, K., \& {Ikeuchi}, S. 1986, \pasj, 38, 697

\bibitem[{{Tumlinson} {et~al.}(2019){Tumlinson}, {Oey}, {Henry}, {James},
  {Rubin}, {Lanz}, {Ravindranath}, {Chisholm}, {Aloisi}, {Bordoloi},
  {Burchett}, {De Rosa}, {Fox}, {France}, {French}, {Fullerton}, {Hayes},
  {Law}, {Lehner}, {Lochhaas}, {Martin}, {O'Meara}, {Peeples}, {Rafelski},
  {Rigby}, {Roman-Duval}, {Rowlands}, {Wang}, \& {Werk}}]{Tumlinson2019}
{Tumlinson}, J., {Oey}, S., {Henry}, A., {et~al.} 2019, \baas, 51, 380

\bibitem[{{Vallerga} {et~al.}(1993){Vallerga}, {Vedder}, {Craig}, \&
  {Welsh}}]{Vallerga1993}
{Vallerga}, J.~V., {Vedder}, P.~W., {Craig}, N., \& {Welsh}, B.~Y. 1993, \apj,
  411, 729, \dodoi{10.1086/172875}

\bibitem[{{van der Hulst} \& {Sancisi}(1988)}]{vanderHulstSancisi1988}
{van der Hulst}, T., \& {Sancisi}, R. 1988, \aj, 95, 1354,
  \dodoi{10.1086/114731}

\bibitem[{{van der Wel} {et~al.}(2014){van der Wel}, {Franx}, {van Dokkum},
  {Skelton}, {Momcheva}, {Whitaker}, {Brammer}, {Bell}, {Rix}, {Wuyts},
  {Ferguson}, {Holden}, {Barro}, {Koekemoer}, {Chang}, {McGrath},
  {H{\"a}ussler}, {Dekel}, {Behroozi}, {Fumagalli}, {Leja}, {Lundgren},
  {Maseda}, {Nelson}, {Wake}, {Patel}, {Labb{\'e}}, {Faber}, {Grogin}, \&
  {Kocevski}}]{vanderWel2014}
{van der Wel}, A., {Franx}, M., {van Dokkum}, P.~G., {et~al.} 2014, \apj, 788,
  28, \dodoi{10.1088/0004-637X/788/1/28}

\bibitem[{{Veilleux} {et~al.}(1995){Veilleux}, {Cecil}, \&
  {Bland-Hawthorn}}]{Veilleux1995}
{Veilleux}, S., {Cecil}, G., \& {Bland-Hawthorn}, J. 1995, \apj, 445, 152,
  \dodoi{10.1086/175681}

\bibitem[{{Veilleux} {et~al.}(2020){Veilleux}, {Maiolino}, {Bolatto}, \&
  {Aalto}}]{Veilleux2020}
{Veilleux}, S., {Maiolino}, R., {Bolatto}, A.~D., \& {Aalto}, S. 2020, \aapr,
  28, 2, \dodoi{10.1007/s00159-019-0121-9}

\bibitem[{{Vijayan} {et~al.}(2020){Vijayan}, {Kim}, {Armillotta}, {Ostriker},
  \& {Li}}]{Vijayan2020}
{Vijayan}, A., {Kim}, C.-G., {Armillotta}, L., {Ostriker}, E.~C., \& {Li}, M.
  2020, \apj, 894, 12, \dodoi{10.3847/1538-4357/ab8474}

\bibitem[{{Wakker}(2001)}]{Wakker2001}
{Wakker}, B.~P. 2001, \apjs, 136, 463, \dodoi{10.1086/321783}

\bibitem[{{Wakker} \& {van Woerden}(1991)}]{WakkervanWoerden1991}
{Wakker}, B.~P., \& {van Woerden}, H. 1991, \aap, 250, 509

\bibitem[{{Wakker} {et~al.}(2008){Wakker}, {York}, {Wilhelm}, {Barentine},
  {Richter}, {Beers}, {Ivezi{\'c}}, \& {Howk}}]{Wakker2008}
{Wakker}, B.~P., {York}, D.~G., {Wilhelm}, R., {et~al.} 2008, \apj, 672, 298,
  \dodoi{10.1086/523845}

\bibitem[{{Wakker} {et~al.}(2007){Wakker}, {York}, {Howk}, {Barentine},
  {Wilhelm}, {Peletier}, {van Woerden}, {Beers}, {Ivezi{\'c}}, {Richter}, \&
  {Schwarz}}]{Wakker2007}
{Wakker}, B.~P., {York}, D.~G., {Howk}, J.~C., {et~al.} 2007, \apjl, 670, L113,
  \dodoi{10.1086/524222}

\bibitem[{{Wang} {et~al.}(2016){Wang}, {Koribalski}, {Serra}, {van der Hulst},
  {Roychowdhury}, {Kamphuis}, \& {Chengalur}}]{Wang2016}
{Wang}, J., {Koribalski}, B.~S., {Serra}, P., {et~al.} 2016, \mnras, 460, 2143,
  \dodoi{10.1093/mnras/stw1099}

\bibitem[{{Wang} {et~al.}(2013){Wang}, {Kauffmann}, {J{\'o}zsa}, {Serra}, {van
  der Hulst}, {Bigiel}, {Brinchmann}, {Verheijen}, {Oosterloo}, {Wang}, {Li},
  {den Heijer}, \& {Kerp}}]{Wang2013}
{Wang}, J., {Kauffmann}, G., {J{\'o}zsa}, G. I.~G., {et~al.} 2013, \mnras, 433,
  270, \dodoi{10.1093/mnras/stt722}

\bibitem[{{Watts} {et~al.}(2020){Watts}, {Power}, {Catinella}, {Cortese}, \&
  {Stevens}}]{Watts2020}
{Watts}, A.~B., {Power}, C., {Catinella}, B., {Cortese}, L., \& {Stevens}, A.
  R.~H. 2020, \mnras, 499, 5205, \dodoi{10.1093/mnras/staa3200}

\bibitem[{{Weiner} \& {Williams}(1996)}]{WeinerWilliams1996}
{Weiner}, B.~J., \& {Williams}, T.~B. 1996, \aj, 111, 1156,
  \dodoi{10.1086/117860}

\bibitem[{{Welty} {et~al.}(2006){Welty}, {Federman}, {Gredel}, {Thorburn}, \&
  {Lambert}}]{Welty2006}
{Welty}, D.~E., {Federman}, S.~R., {Gredel}, R., {Thorburn}, J.~A., \&
  {Lambert}, D.~L. 2006, \apjs, 165, 138, \dodoi{10.1086/504153}

\bibitem[{{Welty} {et~al.}(1996){Welty}, {Morton}, \& {Hobbs}}]{Welty1996}
{Welty}, D.~E., {Morton}, D.~C., \& {Hobbs}, L.~M. 1996, \apjs, 106, 533,
  \dodoi{10.1086/192347}

\bibitem[{{Welty} {et~al.}(2012){Welty}, {Xue}, \& {Wong}}]{Welty2012}
{Welty}, D.~E., {Xue}, R., \& {Wong}, T. 2012, \apj, 745, 173,
  \dodoi{10.1088/0004-637X/745/2/173}

\bibitem[{{Werk} {et~al.}(2013){Werk}, {Prochaska}, {Thom}, {Tumlinson},
  {Tripp}, {O'Meara}, \& {Peeples}}]{Werk2013}
{Werk}, J.~K., {Prochaska}, J.~X., {Thom}, C., {et~al.} 2013, \apjs, 204, 17,
  \dodoi{10.1088/0067-0049/204/2/17}

\bibitem[{{Werk} {et~al.}(2019){Werk}, {Rubin}, {Bish}, {Prochaska}, {Zheng},
  {O'Meara}, {Lenz}, {Hummels}, \& {Deason}}]{Werk2019}
{Werk}, J.~K., {Rubin}, K.~H.~R., {Bish}, H.~V., {et~al.} 2019, \apj, 887, 89,
  \dodoi{10.3847/1538-4357/ab54cf}

\bibitem[{{Westfall} {et~al.}(2019){Westfall}, {Cappellari}, {Bershady},
  {Bundy}, {Belfiore}, {Ji}, {Law}, {Schaefer}, {Shetty}, {Tremonti}, {Yan},
  {Andrews}, {Brownstein}, {Cherinka}, {Coccato}, {Drory}, {Maraston},
  {Parikh}, {S{\'a}nchez-Gallego}, {Thomas}, {Weijmans}, {Barrera-Ballesteros},
  {Du}, {Goddard}, {Li}, {Masters}, {Ibarra Medel}, {S{\'a}nchez}, {Yang},
  {Zheng}, \& {Zhou}}]{Westfall2019}
{Westfall}, K.~B., {Cappellari}, M., {Bershady}, M.~A., {et~al.} 2019, \aj,
  158, 231, \dodoi{10.3847/1538-3881/ab44a2}

\bibitem[{{Wild} \& {Hewett}(2005)}]{Wild2005}
{Wild}, V., \& {Hewett}, P.~C. 2005, \mnras, 361, L30,
  \dodoi{10.1111/j.1745-3933.2005.00058.x}

\bibitem[{{Wild} {et~al.}(2007){Wild}, {Hewett}, \& {Pettini}}]{Wild2007}
{Wild}, V., {Hewett}, P.~C., \& {Pettini}, M. 2007, \mnras, 374, 292,
  \dodoi{10.1111/j.1365-2966.2006.11146.x}

\bibitem[{{Womble} {et~al.}(1990){Womble}, {Junkkarinen}, {Cohen}, \&
  {Burbidge}}]{Womble1990}
{Womble}, D.~S., {Junkkarinen}, V.~T., {Cohen}, R.~D., \& {Burbidge}, E.~M.
  1990, \aj, 100, 1785, \dodoi{10.1086/115636}

\bibitem[{{Yao} {et~al.}(2009){Yao}, {Tripp}, {Wang}, {Danforth}, {Canizares},
  {Shull}, {Marshall}, \& {Song}}]{Yao2009}
{Yao}, Y., {Tripp}, T.~M., {Wang}, Q.~D., {et~al.} 2009, \apj, 697, 1784,
  \dodoi{10.1088/0004-637X/697/2/1784}

\bibitem[{{York} {et~al.}(2000){York}, {Adelman}, {Anderson}, {Anderson},
  {Annis}, {Bahcall}, {Bakken}, {Barkhouser}, {Bastian}, {Berman}, {Boroski},
  {Bracker}, {Briegel}, {Briggs}, {Brinkmann}, {Brunner}, {Burles}, {Carey},
  {Carr}, {Castander}, {Chen}, {Colestock}, {Connolly}, {Crocker}, {Csabai},
  {Czarapata}, {Davis}, {Doi}, {Dombeck}, {Eisenstein}, {Ellman}, {Elms},
  {Evans}, {Fan}, {Federwitz}, {Fiscelli}, {Friedman}, {Frieman}, {Fukugita},
  {Gillespie}, {Gunn}, {Gurbani}, {de Haas}, {Haldeman}, {Harris}, {Hayes},
  {Heckman}, {Hennessy}, {Hindsley}, {Holm}, {Holmgren}, {Huang}, {Hull},
  {Husby}, {Ichikawa}, {Ichikawa}, {Ivezi{\'c}}, {Kent}, {Kim}, {Kinney},
  {Klaene}, {Kleinman}, {Kleinman}, {Knapp}, {Korienek}, {Kron}, {Kunszt},
  {Lamb}, {Lee}, {Leger}, {Limmongkol}, {Lindenmeyer}, {Long}, {Loomis},
  {Loveday}, {Lucinio}, {Lupton}, {MacKinnon}, {Mannery}, {Mantsch}, {Margon},
  {McGehee}, {McKay}, {Meiksin}, {Merelli}, {Monet}, {Munn}, {Narayanan},
  {Nash}, {Neilsen}, {Neswold}, {Newberg}, {Nichol}, {Nicinski}, {Nonino},
  {Okada}, {Okamura}, {Ostriker}, {Owen}, {Pauls}, {Peoples}, {Peterson},
  {Petravick}, {Pier}, {Pope}, {Pordes}, {Prosapio}, {Rechenmacher}, {Quinn},
  {Richards}, {Richmond}, {Rivetta}, {Rockosi}, {Ruthmansdorfer}, {Sandford},
  {Schlegel}, {Schneider}, {Sekiguchi}, {Sergey}, {Shimasaku}, {Siegmund},
  {Smee}, {Smith}, {Snedden}, {Stone}, {Stoughton}, {Strauss}, {Stubbs},
  {SubbaRao}, {Szalay}, {Szapudi}, {Szokoly}, {Thakar}, {Tremonti}, {Tucker},
  {Uomoto}, {Vanden Berk}, {Vogeley}, {Waddell}, {Wang}, {Watanabe},
  {Weinberg}, {Yanny}, {Yasuda}, \& {SDSS Collaboration}}]{York2000}
{York}, D.~G., {Adelman}, J., {Anderson}, John~E., J., {et~al.} 2000, \aj, 120,
  1579, \dodoi{10.1086/301513}

\bibitem[{{York} {et~al.}(2006){York}, {Khare}, {Vanden Berk}, {Kulkarni},
  {Crotts}, {Lauroesch}, {Richards}, {Schneider}, {Welty}, {Alsayyad}, {Kumar},
  {Lundgren}, {Shanidze}, {Smith}, {Vanlandingham}, {Baugher}, {Hall},
  {Jenkins}, {Menard}, {Rao}, {Tumlinson}, {Turnshek}, {Yip}, \&
  {Brinkmann}}]{York2006}
{York}, D.~G., {Khare}, P., {Vanden Berk}, D., {et~al.} 2006, \mnras, 367, 945,
  \dodoi{10.1111/j.1365-2966.2005.10018.x}

\bibitem[{{York} {et~al.}(2012){York}, {Straka}, {Bishof}, {Kuttruff}, {Bowen},
  {Kulkarni}, {Subbarao}, {Richards}, {Vanden Berk}, {Hall}, {Heckman},
  {Khare}, {Quashnock}, {Ghering}, \& {Johnson}}]{York2012}
{York}, D.~G., {Straka}, L.~A., {Bishof}, M., {et~al.} 2012, \mnras, 423, 3692,
  \dodoi{10.1111/j.1365-2966.2012.21166.x}

\bibitem[{{Zheng} {et~al.}(2017){Zheng}, {Peek}, {Werk}, \&
  {Putman}}]{Zheng2017}
{Zheng}, Y., {Peek}, J.~E.~G., {Werk}, J.~K., \& {Putman}, M.~E. 2017, \apj,
  834, 179, \dodoi{10.3847/1538-4357/834/2/179}

\bibitem[{{Zhu} \& {M{\'e}nard}(2013)}]{ZhuMenard2013}
{Zhu}, G., \& {M{\'e}nard}, B. 2013, \apj, 773, 16,
  \dodoi{10.1088/0004-637X/773/1/16}

\bibitem[{{Zschaechner} \& {Rand}(2015)}]{ZschaechnerRand2015}
{Zschaechner}, L.~K., \& {Rand}, R.~J. 2015, \apj, 808, 153,
  \dodoi{10.1088/0004-637X/808/2/153}

\bibitem[{{Zych} {et~al.}(2009){Zych}, {Murphy}, {Hewett}, \&
  {Prochaska}}]{Zych2009}
{Zych}, B.~J., {Murphy}, M.~T., {Hewett}, P.~C., \& {Prochaska}, J.~X. 2009,
  \mnras, 392, 1429, \dodoi{10.1111/j.1365-2966.2008.14157.x}

\bibitem[{{Zych} {et~al.}(2007){Zych}, {Murphy}, {Pettini}, {Hewett},
  {Ryan-Weber}, \& {Ellison}}]{Zych2007}
{Zych}, B.~J., {Murphy}, M.~T., {Pettini}, M., {et~al.} 2007, \mnras, 379,
  1409, \dodoi{10.1111/j.1365-2966.2007.12015.x}

\end{thebibliography}
\bibliographystyle{aasjournal}



\end{document}